%% file: skirt_paper.tex
\documentclass[a4paper,fleqn,usenatbib]{mnras}

\pdfoutput=1 
\usepackage{graphicx}
\usepackage{float}
\usepackage{acronym}
\usepackage{aas_macros}
\usepackage{amssymb}
\usepackage{amsmath}
\usepackage{amsfonts}
\usepackage{xcolor}
\usepackage{upgreek}
\usepackage{hyperref}

\usepackage[percent]{overpic}
\usepackage{gensymb}

\hypersetup{colorlinks=true,linkcolor=blue,citecolor=blue,filecolor=blue,urlcolor=blue}

\newcommand{\eagle}[0]{{\sc eagle}}
\newcommand{\skirt}[0]{{\sc skirt}}
\newcommand{\gama}{{\sc gama}}
\newcommand{\sdss}{{\sc sdss}}
\newcommand{\galaxev}[0]{{\sc galaxev}}
\newcommand{\mappings}[0]{{\sc mappings-iii}}
\newcommand{\starburst}[0]{{\sc mappings-iii}}
\newcommand{\Ha}[0]{{H$\alpha$}}
\newcommand{\Hb}[0]{{H$\beta$}}
\newcommand{\Hii}[0]{{H{\sc ii}}}
\newcommand{\Oii}[0]{{O{\sc ii}}}

\newcommand\changes[1]{#1}

\title[Observing EAGLE with SKIRT]{Optical colours and spectral indices of $z=0.1$ EAGLE galaxies with 3D dust radiative transfer code SKIRT}

\author[James W. Trayford et al.]{James W. Trayford$^{1}$\thanks{E-mail: j.w.trayford@durham.ac.uk (JWT)}, Peter Camps$^2$, Tom Theuns$^1$, Maarten Baes$^2$, \newauthor Richard G. Bower$^{1}$,
 Robert A. Crain$^3$, Madusha L. P. Gunawardhana$^{4,1}$, \newauthor Matthieu Schaller$^1$, Joop Schaye$^5$ and Carlos S. Frenk$^1$\\
$^1$Institute for Computational Cosmology, Durham University, South Road, Durham, DH1 3LE\\
$^2$Sterrenkundig Observatorium, Universiteit Gent, Krijgslaan 281, B-9000 Gent, Belgium\\
$^3$Astrophysics Research Institute, Liverpool John Moores University, 146 Brownlow Hill, Liverpool L3 5RF, UK\\
$^4$Instituto de Astrof\'isica and Centro de Astroingenier\'ia, Facultad de F\'isica, Pontificia Universidad Cat\'olica de Chile,\\
 Vicu\~na Mackenna 4860, 7820436 Macul, Santiago, Chile\\
$^5$Leiden Observatory, Leiden University, P.O. Box 9513, 2300 RA Leiden, the Netherlands}

\begin{document}

\date{Accepted . Received ; in original form }

\pagerange{\pageref{firstpage}--\pageref{lastpage}} \pubyear{2017}

\maketitle
\label{firstpage}

\begin{abstract}
We present mock optical images, broad-band and \Ha{} fluxes, and D4000 spectral indices for $30,145$ galaxies from the \eagle\ hydrodynamical simulation at redshift $z=0.1$, modelling dust with the \skirt\ Monte Carlo radiative transfer code. The modelling includes a subgrid prescription for dusty star-forming regions, with both the subgrid obscuration of these regions and the fraction of metals in diffuse interstellar dust calibrated against far-infrared fluxes of local galaxies. The predicted optical colours as a function of stellar mass agree well with observation, with the \skirt{} model showing marked improvement over a simple dust screen model. The orientation dependence of attenuation is weaker than observed because \eagle\ galaxies are generally puffier than real galaxies, due to the pressure floor imposed on the interstellar medium. The mock H$\alpha$ luminosity function agrees reasonably well with the data, and we quantify the extent to which dust obscuration affects observed \Ha\ fluxes. The distribution of D4000 break values is bimodal, as observed. In the simulation, 20~per cent of galaxies deemed `passive' for the \skirt{} model, i.e. exhibiting D4000 $> 1.8$, are classified `active' when ISM dust attenuation is not included. The fraction of galaxies with stellar mass greater than $10^{10}$~M$_\odot$ that are deemed passive is slightly smaller than observed, which is due to low levels of residual star formation in these simulated galaxies. Colour images, fluxes and spectra of \eagle\ galaxies are to be made available through the public \eagle\ database.
\end{abstract}


\input{acronyms}

\input{Intro}

\input{Simulation}

\input{SKIRT}

\input{Orientation}

\input{SkirtPhotometry}

\input{Indices}
\input{Summary}

\bibliographystyle{mnras}
\bibliography{references}

\input{Appendices}

\label{lastpage}

\end{document}

%% file: acronyms.tex
\acrodef{CMD}{colour magnitude diagram}
\acrodef{LF}{luminosity function}
\acrodef{GAMA}{Galaxy And Mass Assembly}
\acrodef{MGC}{Millenium Galaxy Catalogue}
\acrodef{UKIDSS}{UK Infra-red Digital Sky Survey}
\acrodef{SDSS}{Sloan Digital Sky Survey}
\acrodef{RT}{Full 3-dimensional Radiative Transfer}
\acrodef{SED}{Spectral Energy Distribution}
\acrodef{T15}{\citet{Trayford15}}
\acrodef{K98}{\citet{Kennicutt98}}
\acrodef{S15}{\citet{Schaye15}}
\acrodef{C15}{\citet{Cheng15}}
\acrodef{C16}{Camps et al. ({\it in prep.})}
\acrodef{D4000}{the 4000\AA{} break}

\acrodef{Ref-100}{Ref-L100N1504}
\acrodef{RefHi-25}{Ref-L025N0754}
\acrodef{Recal-25}{Recal-L025N0754}
\acrodef{Ref-25}{Ref-L025N0376}
\acrodef{ML}{maximum-likelihood}
\acrodef{MCMC}{Markov-chain Monte Carlo}

%% file: Intro.tex
\section{Introduction}
\label{sec:intro}

Cosmological simulations are instrumental for our understanding of how competing physical processes shape galaxies. N-body simulations played a crucial role in establishing the cold dark matter paradigm, demonstrating that dark matter halos provide the potential wells into which gas can collapse, cool and form stars \citep[e.g.][]{White78, Frenk88, White91}. Simulations that also include hydrodynamics have by now matured to such an extent that they show good agreement between simulated and observed galaxies for a wide range of properties, provided feedback from forming stars and accreting black holes is implemented to be very efficient \citep[e.g.][]{Vogelsberger14, Murante15, Schaye15, Dave16}. Differences in the properties of simulated galaxies result primarily from the choices made in how to implement unresolved physical processes, particularly star formation and feedback, as shown by e.g. \citet{Schaye10, Scannapieco12, Kim14}. 

The comparisons to data are relatively indirect, however, because simulations yield intrinsic properties of galaxies, such as galaxy stellar mass or star formation rate, that are not directly observable. \lq Inverse models\rq\ attempt to infer such physical properties from the observed fluxes. The main ingredients of such models are the assumed stellar initial mass function (IMF), templates for the star formation and enrichment histories, a model for dust effects (absorption and scattering), and a model for stellar evolution encapsulated by a population synthesis description to yield fluxes. This is exemplified by the analysis of \cite{Li09} applied to the 7th data release \citep{SDSSdr7} of the Sloan Digital Sky Survey (\sdss, \citealt{York00}), or the analysis by \cite{Baldry12} applied to the Galaxy And Mass Assembly (\gama, \citealt{Driver09}) survey. Such analysis makes necessarily bold simplifications, for example assuming exponential star formation histories, uniform stellar metallicities and a dust screen model. \cite{Mitchell13} demonstrated how this methodology suffers from degeneracy between the star formation history, metallicity and dust properties of galaxies.
	 
Similar models are needed to infer star formation rates	or passive fractions. The strength of the \Ha{} recombination line is sensitive to recent star formation, as it probes UV-continuum emission from stars that are $\lesssim$10~Myr old \citep{Kennicutt98}. However, a significant fraction of the \Ha{} flux in star forming galaxies is emitted by dusty \Hii{} regions \citep[e.g.][]{Zurita00}, and therefore the conversion from flux to star formation rate requires a model to account for obscuration \citep[e.g.][]{James04, Best13, Gunawardhana13}. Similarly, the continuum strength on either side of the 4000\AA{} break depends on the relative contribution to the flux of old versus younger stars, and hence is a useful proxy for the specific star formation rate of a galaxy \citep[e.g.][]{Kauffmann03, Balogh99}. However, the amplitude of the break may also be affected by dust, and hence the observed passive fraction depends on the assumed dust properties.

\changes{In addition to} inverse modelling, it should be possible to apply the ingredients of the inverse models to the simulated galaxies instead, and compare mock {\em fluxes} to the observations. Such \lq forward models\rq\ have many potential advantages. For example, the star formation histories of simulated galaxies are more detailed and diverse than the parametric models used in inverse modelling. Similarly, the simulated - and presumably also the observed stars in any galaxy - have a considerable spread in metallicity, rather than a single uniform value. These assumed priors may introduce biases in the inferred properties of galaxies, see e.g. \cite{Trayford16}. Notwithstanding any practical considerations, surely it should be the ultimate aim of the simulations to predict observables.

\changes{In practice there are still formidable challenges with forward modelling, and inverse and forward modelling approaches are largely complementary. Inverse modelling is useful to assess how distinct physical quantities may contribute to observables and elucidate discrepancies between real and mock observations, while insights gained from a forward modelling approach can inform and improve our inverse models. For instance, generating mock galaxy observations with attenuation and re-emission by dust can demonstrate how numerous degeneracies in SED inversion can be lifted by incorporating FIR observations \citep[e.g.][]{Hayward15}. The treatment of dust, however, exemplifies a major challenge of forward modelling. While dust represents a marginal fraction of the mass in galaxies} and simulations do not typically model an explicit dust phase \citep[with exceptions, e.g.][]{Bekki15, Aoyama16, McKinnon16}, interstellar dust can play an important role in processing the light we observe from galaxies. 

On average, about a third of the UV plus optical starlight emitted in local star-forming galaxies is absorbed by dust and re-radiated at longer wavelengths \citep[e.g.][]{Popescu02,Viaene16}. The effect of dust attenuation introduces various systematics, particularly for young stellar populations:  the cross-sections for absorption and scattering generally increase with the frequency of incident light, such that the relatively blue emission from young stars is more affected. But the impact of dust also depends on the morphology and orientation of the galaxy: young stellar populations are typically found in a thin disc and near the dense, dusty ISM regions from which they formed. These regions are therefore likely to have relatively high dust obscuration, and, if this is not accounted for, may affect inferred structural measures of galaxies such as scale lengths/heights and bulge-to-disc ratios. Because young stars may be more obscured than old stars, the {\em attenuation curve} (the ratio of observed over emitted radiation as function of wavelength for the observed galaxy) of a galaxy differs in general from the {\em extinction curve} that describes the wavelength dependence of the photon-dust interaction. The modelling of dust is clearly an important aspect of comparing models to observations, and is the subject of this paper.

An idealised dust model is that of an intervening dust screen with a wavelength-dependent optical depth, the attenuation of which can be computed analytically. The geometry of the dust distribution and some effects of scattering, can be accounted for to some extent by making the attenuation curve \lq greyer\rq\ (i.e. less wavelength dependent) than the extinction curve \citep[e.g.][]{Calzetti01}, and/or by using multiple screen components \citep[e.g.][]{CF00}. \citet{Trayford15} (hereafter T15) adopted the two component screen model of \citet[][]{CF00} to represent dust absorption in the \eagle{} simulations \citep{Schaye15} when generating broad-band luminosities and colours. Absorption is boosted by a fixed factor for young ($< 30$~Myr) stars in this model, and the overall optical depths depend on the gas properties of each simulated galaxy with additional scatter to account for orientation dependence.

T15 showed that optical colours and luminosities generated for \eagle{} galaxies are broadly compatible with the \gama{} measurements, while exhibiting some notable discrepancies. In particular, the modelling resulted in a {\em more} pronounced bimodal distribution of colours at a given stellar mass than observed. In particular, model colours exhibited bimodality amongst even the most massive galaxies for which bimodality is absent in the data. A related discrepancy was that the red and blue fractions were also somewhat inconsistent between model and data, with the model yielding an excess of blue galaxies at high $M_\star$. The cause of these discrepancies was attributed to both differences between the intrinsic properties of the simulated and observed galaxy populations (higher specific star formation rates in \eagle), as well as uncertainties in the modelling of the photometry (especially dust effects). Of particular concern was that lower than observed average star formation rates in \eagle{} galaxies \citep{Furlong15} and underestimated reddening could have a compensatory effect, potentially yielding a fortuitous agreement with \gama{} colours.

The dust screen model applied by T15 may not represent attenuation in \eagle\ galaxies well. Indeed, the effects of complex geometries cannot be captured by a screen \citep[e.g.][]{Witt92}: there are degeneracies between dust geometry and dust content. Colours of galaxies where dust and stars are well mixed can be confused with dimmer dust-free galaxies if a screen is assumed \citep[e.g.][]{Calzetti01, Disney89}. Screens also neglect scattering into the line of sight, or attempt to account for it with approximative absorption. With scattering being as important as geometry at optical wavelengths, and often producing effects entirely dissimilar to absorption, a screen-based approximation is often insufficient \citep{Baes01, Byun94}.

Fully accounting for dust requires three-dimensional radiative transfer calculations \citep[e.g.][]{Steinacker13}, and, given the lack of symmetry, Monte Carlo radiative transfer (MCRT) \citep[e.g][]{Whitney11} techniques appear to be well suited. These follow the path of photons
	from sources through the dusty ISM to a camera. A variety of MCRT codes are
	publically available, such as {\sc sunrise} \citep{Jonsson06} and \skirt{} \citep{Baes03, Baes11,
	 Camps15}. {\sc sunrise} has been applied to zoomed galaxy simulations \cite[e.g.][]{Jonsson10,
	  Guidi15}, and by \citet{Torrey15} to compute images of galaxies from the {\sc Illustris} simulation \citep{Vogelsberger14}. Note, however, that \cite{Torrey15} did not actually include dust in the {\sc sunrise} models. \skirt{} has been applied to galaxy models \cite[e.g.][]{Gadotti10, Saftly15} but also to dust around AGN \citep{Stalevski12, Stalevski16}. Full MCRT dust models of simulated galaxies in a cosmologically representative volume are yet to be published.

\changes{Previous studies that apply MCRT dust modelling to galaxy zoom simulations can provide insight into this approach. \citet{Scannapieco10} use MCRT to produce representative optical images and decompose them into bulge and disc contributions, but find that dust effects are negligible due to low gas fractions and metallicities in the simulations themselves. Simulations with more realistic gas phase metallicities have also been processed with MCRT to produce mock observables across the UV to sub-mm wavelength range \citep[e.g.][]{Jonsson09, Guidi15, Hayward15}. The influence of galaxy orientation and geometry on attenuation properties and recovered physical quantities are explored by \citet{Hayward15}, showing how the effective attenuation curves vary with orientation and morphological transformation for idealised merger zooms.}
	
In this paper, we generate optical broad-band fluxes and spectra for \eagle{} galaxies using \skirt,
	comparing mock fluxes to \gama\ observations and to the dust-screen model of T15.
	\eagle\ \citep{Schaye15, Crain15} is a suite of hydrodynamical SPH-simulations, with sub-grid parameters calibrated to a small set of low-redshift ($z\sim 0.1$) observables. The simulation reproduces a variety of observations that were not part of the calibration procedure, such as the neutral and molecular contents of $z\sim 0$ galaxies \citep{Bahe16, Lagos15}, and the evolution of galaxy star formation rates and sizes
	\citep{Furlong15b, Furlong15}. There is a hint that the simulation underpredicts specific star formation rates except for the most massive galaxies. 

The simulation does not trace dust explicitly: we describe dust associated with star forming regions using the {\sc mappings} models by \cite{Groves08}, and assume that the ISM dust/gas ratio depends on metallicity. This procedure was developed for this work and for the companion paper of \citet{Camps16}, who looked at the FIR properties of present-day \eagle{} galaxies. This paper compared \skirt\ models to FIR observations of local galaxies to calibrate dust models, showing that observed dust scaling relations can be reproduced. \citet{Camps16} uses dust parameters identical to those used in the present work. The influence of these parameters is discussed in section \ref{sec:template} and the Appendix.

The paper is organised as follows: section \ref{sec:sim} provides a summary of the \eagle{} simulations used in this work, how we define galaxies in our simulated sample and the datasets we compare to. Section \ref{sec:template} details the procedure used to produce observables with \skirt. We investigate the predicted photometric colours in section \ref{sec:testing} and compare the effects of dust to the screen model approach of T15. In section \ref{sec:lines} we present novel measurements of spectral indices for \eagle{} galaxies, and again quantify dust effects. We focus in particular on the star formation proxies of \Ha{} and the extent to which \eagle{} reproduces the statistics of, and the correlation in, the D4000 break. We summarise our findings and conclude in section \ref{sec:conc}. Those only concerned with our main results may want to read
	from section \ref{sec:testing} onwards; outputs of the model are described in section \ref{sec:prod}.

The mock \eagle{} observables used in in this work, and additional data products listed in section \ref{sec:prod}, are to be made available via the public data-base \citep{McAlpine16}. The modelling, described and tested at low redshift ($z \leq 0.1$), is also used to provide these mock observables for galaxies taken from \eagle{} simulations and redshifts that are not considered in this work.

%% file: Simulation.tex
\section{Simulations and Data}
\label{sec:sim}
We provide a brief overview of the \eagle{} simulation suite, see \cite{Schaye15, Crain15}, hereafter S15 and C15, respectively, for full details, review the population synthesis model and dust treatment of T15, and briefly describe the volume-limited sample of galaxies compiled from the \gama{} survey \citep{Driver09}.

\subsection{The \eagle{} simulation suite}
\label{sec:eagle}

\begin{table}
\begin{center}
\caption{Numerical parameters of those simulations of the \eagle{} suite that are used in this paper. From  left to right: simulation identifier, side length of cubic volume $L$ in co-moving Mpc (cMpc), initial mass $m_{\rm g}$ of baryonic particles, Plummer equivalent gravitational softening $\epsilon_{\rm prop}$ scale at redshift $z=0$ in pkpc, where we use \lq pkpc\rq\ to denote proper kiloparsecs.}
\label{tab:sims}
\begin{tabular}{lrrr}
\hline
Name & $L$ & $m_{\rm g}$ & $\epsilon_{\rm prop}$ \\
& cMpc & ${\rm M}_\odot$ & pkpc\\
\hline
\ac{Ref-25} &  25 & $1.81\times 10^6$ & 0.70\\
\ac{RefHi-25} &  25 & $2.26\times 10^6$ & 0.35\\
\ac{Recal-25} &  25 & $2.26\times 10^5$ & 0.35\\
\ac{Ref-100} & 100 & $1.81\times 10^6$ &0.70\\
\hline
\end{tabular}
\end{center}
\end{table}
\eagle{} comprises a suite of cosmological hydrodynamical simulations of periodic cubic volumes performed using a modified version of the {\sc Gadget-3} TreeSPH code (which is an update of the {\sc Gadget-2} code last described by \citealt{Springel05}). Simulations were performed for a range of volumes and numerical resolutions. Here we concentrate on the \lq reference\rq{} model, using simulations at different resolution to assess numerical convergence. In particular, we use models L100N1504, L025N0376 and L025N0752 from table~2, and Recal from Table~3 of S15; in our Table \ref{tab:sims} we refer to these simulations as \ac{Ref-100}, \ac{Ref-25}, \ac{RefHi-25} and \ac{Recal-25} respectively. The \eagle{} suite assumes a $\Lambda$CDM cosmology with parameters derived from the initial {\it Planck} \citep{Planck} satellite data release ($\Omega_{\rm b} = 0.0482$, $\Omega_{\rm dark} = 0.2588$, $\Omega_\Lambda = 0.693$ and $h = 0.6777$, where $H_0 = 100\; h$ km s$^{-1}$ Mpc$^{-1}$). Some simulation details are listed in Table \ref{tab:sims}.

We focus primarily on the \ac{Ref-100} simulation volume. The 100$^3$ Mpc$^3$ volume and mass resolution of $1.2\times10^6 {\rm M_\odot}$ in gas for \ac{Ref-100} provides a sample of $\sim$30,000 galaxies resolved by $>$ 1000 star particles at redshift $z=0.1$, with $\sim$3000 galaxies resolved by $>$ 10,000 star particles. In addition to this primary sample of galaxies, we also use the higher-resolution \ac{Ref-25} and \ac{Recal-25} simulations to test the `strong' and `weak' convergence (see S15 for definition of these terms)
. The 25$^3$ Mpc$^3$ volumes have a factor $2$ ($2^3$) superior spatial (mass) resolution than \ac{Ref-100}. As \ac{Ref-25} uses the same fiducial model at high resolution (with the same initial phases and amplitudes of the Gaussian field), it may be used to test the strong convergence of galaxy properties. The feedback efficiencies adopted by \ac{Recal-25}  were recalibrated to provide better agreement with the $z=0.1$ galaxy stellar mass function at high resolution and to test the weak convergence (see also C15).

The initial conditions of all \eagle{} simulations were generated appropriately for a starting redshift of  $z=127$ using an initial perturbation field generated with the {\sc panphasia} code described by \cite{Jenkins13}. Smoothed particle hydrodynamics (SPH)   is implemented as in \cite{Springel05}, but using the pressure-entropy formulation of \citet{Hopkins13}, including artificial conduction and viscosity \citep{Dehnen12}, a time-step limiter \citep{Durier12}, and the C2 kernel of \citet{Wendland95}. These modifications to the standard {\sc Gadget-3} implementation are collectively termed as {\sc anarchy} (Dalla-Vecchia 2012, in prep., summarised in Appendix A of S15). \citet{Schaller15b} show that these {\sc anarchy} modifications are important in the largest \eagle{} halos, but have minimal effect on galaxies of stellar mass $M_\star$ $\lesssim 10^{11} {\rm M_\odot}$ . To represent important astrophysical process acting on scales below the resolution of \eagle{}, a number of subgrid modules are also employed in the code. Relevant modules include schemes for star formation, enrichment and mass loss by stars, photo-heating, radiative cooling and thermal feedback associated with accreting black holes and the formation of stars, as described below.

Star formation is treated stochastically in \eagle{}. Star formation rates (SFRs) are calculated for individual gas particles using a pressure-dependent formulation of the empirical Kennicutt-Schmidt law \citep{Schaye08}, with a metallicity-dependent density threshold below which star formation rates are zero \citep{Schaye04}. Gas particles thus may have some probabilty of being wholly converted into a star particle at each time step, inheriting the initial element abundances of their parent particle. The gravitational softening scales listed in Table \ref{tab:sims} provide a practical limit on spatial resolution. Cold, dense gas ($T < 10^4$~K, $n_H > 0.1$~cm$^{-3}$) with Jeans lengths below these scales is thus unresolved, and any corresponding gas would artificially fragment in the simulation. To ensure that the Jeans mass of gas is always resolved (albeit marginally), a pressure floor is enforced via a single-phase polytropic equation of state, $P_{\rm EoS} \propto \rho^{\gamma_{\rm EoS}}$. Once formed, star particles are treated as simple stellar populations (SSPs), assuming a universal \citet{Chabrier03} stellar initial mass function (IMF). These SSPs lose mass and enrich neighbouring gas particles according to the prescription of \citet{Wiersma09b}, accounting for type Ia and type II supernovae and winds from massive and AGB stars. Eleven individual elements (H, He, C, N, O, Ne, Mg, Si, S, Ca, and Fe) are followed, as well as a \lq total\rq{} metallicity (the mass fraction in elements more massive than He), $Z$.

Two types of abundances are tracked for the gas in \eagle{}, a {\em particle abundance} that is changed through direct enrichment by star particles and a \textit{smoothed abundance} that smooths particle abundances between neighbours using the SPH kernel \citep[see][]{Wiersma09b}. Diffusion is not implemented in the simulation, therefore no metals are exchanged between gas particles. This may occasionally lead to individual particles exhibiting extreme values as well as large variations in metallicity, even for close neighbours. Although the SPH smoothing is not strictly representative of metal diffusion, it does mitigate extreme values and reduces stochasticity in the metal distribution. For this reason we adopt the \textit{smoothed metallicities} throughout this study, which were also used to compute cooling rates and nucleosynthetic yields during the simulation.

The energy that stellar populations inject into the inter-stellar medium (ISM) through supernovae, stellar winds and radiation is collectively termed \textit{stellar feedback}. Stellar feedback is implemented per star particle (and is separate from enrichment) using the thermal feedback scheme described by \citet{DallaVecchia12}. This implementation sets a temperature change $\Delta T_{\rm SF}$, the temperature by which stochastically sampled gas particle neighbours of stars are heated. The value of $\Delta T_{\rm SF} = 10^{7.5}$K is chosen for the reference model; this is high enough to mitigate catastrophic numerical losses, while low enough to prevent the probability of heating for neighbouring gas particles, $p_{\rm SF}$, from becoming small and leading to poor sampling (see S15). The $p_{SF}$ value depends on both $\Delta T_{\rm SF}$ and the fraction of energy that couples to heat the ISM. The latter fraction is allowed to vary with local gas properties and is calibrated to reproduce observed local galaxy sizes, as detailed by C15.

Black holes are seeded in halos with mass exceeding $10^{10} h^{-1}{\rm M_\odot}$, following \citet{Springel05}. The most-bound gas particle is then converted to a black hole particle with a subgrid mass of $10^5 h^{-1}{\rm M_\odot}$. The black hole grows by subgrid accretion as detailed in S15 and \citet{RosasGuevara15}. A fixed 1.5$\%$ of the rest-mass energy in accreted material provides the energy budget for black hole feedback. This is implemented using a similar stochastic scheme as used for injecting stellar feedback, but with a higher heating temperature ($\Delta T_{\rm BH} = 10^{8.5}$K for the reference models, and $10^{9}$K for \ac{Recal-25}). 

Photo-heating and radiative cooling are implemented as described by \citet{Wiersma09a}, based on the 11 elements traced. This model assumes that gas is in photo-ionisation equilibrium with the cosmic UV+X-ray background as calculated by \citet{Haardt01}. 

We use the friends-of-friends (FoF) algorithm with a linking length of 0.2 times the mean inter particle separation \citep{Davis85, Lacey94} to identify halos. Self-bound substructures within these halos (subhalos) are identified with the {\sc subfind} algorithm \citep{Springel01}. Subhalos may comprise dark matter, stars and gas. Each galaxy is associated with a separate subhalo.

\subsection{Population synthesis and analytic dust treatment}
\label{sec:t15}

\cite{Trayford15} (T15) presented model photometry for \eagle{} galaxies at $z=0.1$. The model adopted here is based on that implementation with some differences as described below. We begin with a brief overview of the T15 model. T15 calculated photometric fluxes in standard $ugrizYJHK$ broad-bands \citep{SDSSfilters, UKIRTfilters}, which included an analytic model for dust obscuration . Stellar emission was calculated using the \galaxev{} population synthesis model \citep{bc03}. This parametrisation uses the stellar ages, smoothed metallicities and initial masses described in section \ref{sec:eagle}. We adopt the same parametrisation in this work where possible. T15 also \lq re-sampled\rq{} the young stellar component at increased mass resolution, to reduce artificial stochasticity in colours resulting from the coarse sampling of star formation in \eagle{}. We use a similar approach here, with full details provided in section \ref{sec:young}. The fiducial dust model employed by T15 was based on the two component dust screen of \citet{CF00}, with the optical depth additionally depending on the gas properties and including a prescription to account for orientation-dependent dust obscuration.

In this paper we adopt many of the same surveying and modelling procedures: individual subhalos are considered potential galaxy hosts, with the \lq galaxy\rq{} comprising the bound material within a 30~pkpc spherical aperture centered on the subhalo's baryonic centre of mass. \changes{This choice was initially made in S15 to reasonably approxiate a Petrosian aperture, and we adopt this galaxy definition for consistency with prior measurements of various galaxy properties. In all but the most massive galaxies, the bound mass excluded by the aperture is negligible (see S15 for details). We find that in $\lesssim 3$\% of cases more than 10\% of the bound baryons lie outside our aperture. While bound material outside of the aperture could modify observable in some galaxies, this effect is insignificant in the majority of galaxies. Primarily, a consistent choice of aperture allows us to isolate aperture effects from other influences.} Each galaxy is processed in isolation, therefore there is no contribution from other structures along the line of sight. We use the same selection of \eagle{} galaxies as T15, processing galaxies with stellar masses of $M_\star \geq 1.81\times 10^8$ M$_\odot$ ($\sim 100$ star particles at standard resolution).

\subsection{GAMA and SDSS survey data}
\label{sec:gama}

The Galaxy and Mass Assembly (\gama) survey \citep{Driver09, Robotham10, Driver11} is a spectroscopic and photometric survey of 5 independent sky fields, undertaken at the Anglo-Australian Telescope, and using the 2dF/AAOmega spectrograph system. The 3 equatorial fields we consider follow up targets from the Sloan Digital Sky Survey (SDSS) Data Release 7 (DR7) \citep{York00, SDSSdr7}, yielding a sample of $\sim 190,000$ galaxies with SDSS $ugriz$ photometry \citep{Hill11, Taylor11} with spectra covering the wavelength range 3700\AA{} to 8900\AA{}, with a resolution of 3.2\AA{} \citep{Sharp06, Driver11}. The \gama{} survey strategy provides high spectroscopic completeness \citep{Robotham10} and accurate redshift determination \citep{Baldry14} for these galaxies, above an extinction-corrected $r$-band Petrosian magnitude limit of 19.8. The galaxy stellar mass estimates and rest frame photometry for the \gama{} sample used in this paper are taken from \citet{Taylor14}. 

Emission line indices in \gama{} were measured assuming single Gaussian profiles, a common redshift for adjacent lines, and a stellar continuum correction simultaneously fit to each spectrum around the measured lines, as described by \cite{Hopkins03} and \cite{Gunawardhana13, Gunawardhana15}. Emission line fluxes are corrected for stellar absorption as described by \cite{Hopkins03}. Dust corrections are obtained using the stellar absorption corrected Balmer emission line flux ratios, also described by \cite{Hopkins03}. The uncertainties associated with correcting Balmer lines for stellar absorption are discussed in both \citet{Hopkins03} and \citet{Gunawardhana13}. 

Derived \Ha{} luminosities and star formation rates are taken from \citet{Gunawardhana13}. Their emission line galaxies (ELGs) are initially selected to have \Ha{} fluxes above the detection limit of $25 \times 10 ^{-20} \; {\rm W m^{-2}}$ and a signal-to-noise ratio of $> 3$, with active galactic nuclei (AGN) identified and removed using standard {\sc [Nii]}$\lambda$6584\AA /\Ha{} and {\sc [Oiii]}$\lambda$5007\AA /\Hb{} diagnostics \citep{Baldwin81}. The \gama{} sample is supplemented with SDSS galaxies with detected \Ha{} emission and signal-to-noise $>$3 from the MPA-JHU catalogue\footnote{http://www.mpa.mpa-garching.mpg.de/SDSS/DR7/}, as the brightest ELGs observed by \sdss{} were not re-observed by \gama{}.

Measurements of the 4000\AA{} break (D4000) are also used in this work. For this, we compare to values measured directly from the \sdss{} DR7 data \citep{Strauss02, SDSSdr7}. We compare a stellar mass-matched sample of \eagle{} galaxies to the publically available \sdss{} D4000 values measured for the MPA-JHU$^1$ catalogue using the code of \citet{Tremonti04}, with the index defined as in \citet{Bruzual83}. For \sdss{} data, we use the mass estimates of \citet{Kauffmann03}.

%% file: SKIRT.tex
\section{Dust Modelling with SKIRT}
\label{sec:template}

\begin{figure}
 \includegraphics[width=0.49\textwidth]{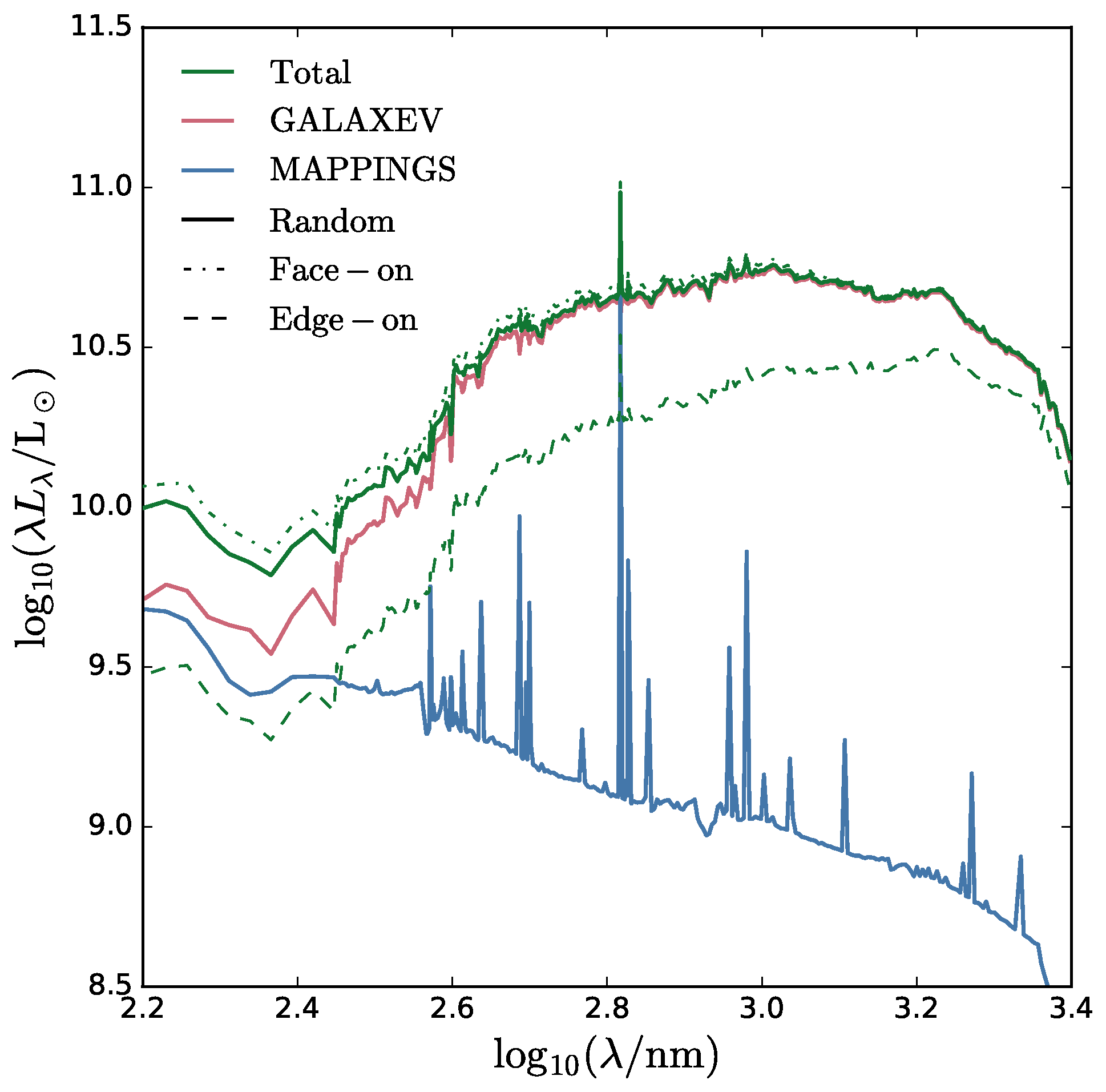}  
 
 \caption{Example integrated rest-frame galaxy SED including ISM dust effects, generated for a late type (S) \eagle{} galaxy taken from simulation \ac{Ref-100} (galaxy ID=15814162 in the \eagle{} database described by \citealt{McAlpine16}). The \textit{green line} indicates the total SED in the presence of dust. \textit{Solid lines} show the integrated SED when the galaxy is observed along the $z$-direction of the simulation volume (an inclination angle of 50.2$^\circ$), while {\em dashed} and {\em dot-dashed} lines denote edge and face-on views, respectively. The \textit{red} and \textit{blue lines} are the ISM dust-free contributions from the old stellar population ($t < 10$~Myr) modelled using GALAXEV \citep{bc03}, and the young population ($t> 10$~Myr) modelled using \starburst{} \citep{Groves08}, respectively. Properties and images of this galaxy can be found in  Table \ref{tab:selgals} and Fig. \ref{fig:pics} respectively. We see from the solid lines that the \starburst{} SED contributes relatively more flux in the UV, and for some strong emission lines.}
 \label{fig:egspec}
\end{figure}

Given a set of sources and a dust distribution, the \skirt{} Monte Carlo code \citep{Baes03, Baes11, Camps15} follows the dust absorption and scattering of monochromatic \lq photon packets\rq{} until they hit a user-specified detector, optionally calculating the heating and re-radiation of the dust grains including non-equilibrium stochastic heating. The position on the detector and wavelength of each arriving photon is registered, allowing a full integral field image of the galaxy to be constructed. Convolving this data cube with a filter yields mock fluxes.

The modular nature of \skirt{} makes it straightforward to implement multiple source components and absorbing media using arbitrary spectral libraries and geometries. The choices and assumptions we make to represent the emissive and absorbing components of \eagle{} galaxies in \skirt{} are detailed in sections \ref{sec:sources} and \ref{sec:dust}. To represent the particle-discretised galaxies of \eagle{} as continuous matter distributions for radiative transfer, particles are smoothed over some scale. For gas particles the SPH smoothing is used, while stellar smoothings are recalculated via a nearest neighbour search of star particles, as explained in Appendix \ref{sec:smooth}. While \skirt{} is capable of very efficient processing, Monte Carlo radiative transfer is fundamentally computationally expensive. We examine the issues of spectral resolution and convergence in Appendix \ref{sec:res}.

\subsection{SKIRT modelling: input SEDs}
\label{sec:sources}

The spectrum of a star particle in the simulation is assigned using spectral energy distribution (SED) libraries. Each particle is treated as a simple stellar population (SSP) with a truncated Gaussian emissivity profile, parametrised by a position, smoothing length and SED. \skirt{} then builds a 3D emissivity profile through the interpolation of these individual kernels. 
	The point of emission of individual photon packets are sampled from these kernels at user-specified wavelengths, and photon packets are launched assuming isotropic emission. For the optical wavelengths considered here, we neglect emission from dust and other non-stellar sources. Our representation of the source component for \eagle{} galaxies, including sub-grid absorption for the youngest stars, is detailed in sections \ref{sec:old} and \ref{sec:young} below, and an example spectrum showing the different SED components is plotted in Fig. \ref{fig:egspec}. Input SEDs and broad band luminosities are stored for \eagle{} galaxies, as described in section \ref{sec:prod}.

\subsubsection{Old stellar populations}
\label{sec:old}

Stellar populations with age greater than 100~Myr are assigned \galaxev{} \citep{bc03} SEDs as described by T15. Briefly, initial masses, (smoothed) metallicities and particle ages are extracted directly from the simulation snapshot. Absolute metallicity values, as opposed to those in solar units, are used to assign SEDs for the reasons given in section 3.1.2 of T15. Stars are assumed to form with a \citet{Chabrier03} IMF covering the stellar mass range of [0.1,100]~M$_\odot$, consistent with what is assumed in \eagle. Star particle coordinates are also taken directly from the simulation output. Smoothing lengths specifying the width of the truncated Gaussian profile are determined using a nearest neighbour search, as detailed in Appendix \ref{sec:smooth}. Note that we do not explore alternative population synthesis models \citep[e.g.][]{sb99, m05, Conroy09} in this paper; differences are expected to be small for the $z=0.1$ galaxies at optical wavelengths studied here \citep[e.g.][]{Gonzalez14}. 

\subsubsection{Young stellar populations}
\label{sec:young}
The treatment of the young stellar component is more involved due to two limitations of the simulation: ({\em i}) the relatively coarse sampling of star formation due to the limited mass resolution (see Table \ref{tab:sims}), and ({\em ii}) the inability of resolving birth cloud absorption associated with recent star formation. Though the diffuse ISM dust can be traced by enriched cold gas, the birth clouds of stellar clusters exhibit structure on sub-kpc scales \citep[e.g.][]{Jonsson10}, below the resolution limit of \eagle{}. Such birth clouds are thought to disperse on timescales of $\sim 10$ Myr \citep[e.g.][]{CF00}. To treat birth cloud absorption, we use the \starburst{} spectral models of \cite{Groves08}, which track stars younger than 10 Myr, and include dust absorption within the photo-dissociation region (PDR) of the star-forming cloud, following the methodology of \citet{Jonsson10}. We therefore have two sources of dust: that associated with birth clouds which is modelled using \starburst{}, and ISM dust whose effects we model using \skirt.

We use an extended version of the re-sampling procedure of T15 to mitigate the effects of coarse sampling. Recent star formation is re-sampled in time over the past 100~Myr, from both star-forming gas particles and existing star particles younger than 100~Myr, as follows. The stellar particle stores the gas density of its parent particle, which is used to compute a star formation rate. We use this rate for young stars, and the star formation rate of star-forming gas particles, to compute how much stellar mass these particles formed on average over the past 100~Myr (assuming a constant star formation rate). We then randomly sample this stellar mass
	in terms of individual star-forming regions, with masses that follow the empirical mass function of molecular clouds in the Milky Way \citep{Heyer01},
\begin{equation}	\frac{{\rm d}N}{{\rm d}M} \propto M^{-1.8} \;\; {\rm with} \; M \in \left[ 700, 10^6 \right] {\rm M}_\odot\,. \label{eq:mfunc} 
\end{equation}
For each particle resampled, sub-particle masses are drawn from this distribution until the mass of the parent particle is exceeded. Rejecting the last drawn sub-particle, the masses of sub-particles are rescaled such that the sum of their masses, $\Sigma m_{\rm r}$, exactly matches that of the parent. Sub-particles are then stochastically assigned formation times using the star formation rate of the star-forming particle. In this way, we replace star-forming gas particles, and young star particles, by a distribution of star-forming molecular clouds with the same total mass and spatial distribution as the original set of particles.

Stellar populations resampled with ages in the range 10~Myr $\leq$ $t_{\rm age}$ $<$ 100 Myr are assigned \galaxev{} spectra, parametrised in the same way as in section \ref{sec:old}. They inherit the position and smoothing length of the particle from which they are sampled. These smoothing lengths are assigned to parent star and gas particles differently, as described in Appendix \ref{sec:smooth}.   

Those resampled to have ages in the age range  $t_{\rm age}$ $\leq$ 10 Myr are assigned the \starburst{} spectra of \citet{Groves08}. One caveat with using these spectral models is that the intrinsic spectra of stars used to model the spectra are specified by the \citet{sb99} (SB99) population synthesis models. This leads to some inconsistency in the modelling of the intrinsic stellar spectra, which come from \galaxev{} for older populations. However, these differences are small in the optical ranges considered here \citep[e.g. see][]{Gonzalez14}. Another caveat is that a \citet{Kroupa01} IMF is assumed for these spectral models rather than that of \citet{Chabrier03}, though again the differences in optical properties are minimal. 

The \starburst{} spectral libraries represent \Hii{} regions, and their emerging spectrum therefore already treats the reprocessing of star light by dust in the star-forming region. In other words, birth cloud dust absorption and nebular emission are included in the \skirt{} input spectra before any \skirt{} radiative transfer is performed. We describe below how we avoid double counting dust in \Hii{} regions.

The \starburst{} SEDs are parametrised as follows:

\begin{itemize}
\item \textbf{Star Formation Rate ($\dot{M}_\star$):} The \starburst{} spectra assume a given (constant) star formation rate between 0 and 10 Myr, the star formation rate assigned to a particle of initial mass $m_{\rm r}$ is given by $\dot{M}_\star = 10^{-7} m_{\rm r} \; {\rm yr}^{-1}$, in order to conserve the mass in stars.
\item \textbf{Metallicity ($Z$):} The metallicities are specified by the SPH-smoothed absolute values of the simulation snapshot. 
\item \textbf{Pressure ($P_0$):} The ambient ISM pressure, $P_0$, is calculated from the density of the gas particle from which the sub-particle is sampled. This is not directly accessible in \eagle{}, but can be estimated using the polytropic equation of state that limits the pressure for star-forming gas \citep{DallaVecchia12}. Because the simulation snapshot contains the density at which each star particle formed, this estimator can be used for re-sampling both star-forming gas and young stellar particles.
\item \textbf{Compactness ($\log_{10}C$):} The compactness, $C$, is a measure of the density of an \Hii{} region. This is calculated using the following equation from \citet{Groves08},
\begin{equation}	\log_{10}C = \frac{3}{5}\log_{10}\frac{M_{\rm cl}}{{\rm M}_\odot} + \frac{2}{5}\log_{10}\frac{P_0 / k_{\rm B}}{{\rm cm^{-3} K}}\,,
\end{equation} where $M_{\rm cl}$ is the star cluster mass, taken to be the re-sampling mass $m_{\rm r}$, $P_0$ is the \Hii{} region pressure taken to be the particle pressure above, and $k_{\rm B}$ is Boltzmann's constant. The parameter $C$ predominately affects the dust temperature and thus the FIR part of the SED, and therefore has little effect on the results presented here, see \cite{Camps16} for a thorough discussion on how $C$ affects FIR colours of \eagle{} galaxies.
\item \textbf{PDR Covering Fraction ($f_{\rm PDR}$): } The photo-dissociation regions (PDRs) associated with \Hii{} regions are influenced by processes well below the resolution of \eagle{}. PDRs disperse over time as O and A stars die out. We assume a fiducial value of $f_{\rm PDR}=0.1$ for the PDR covering factor, which can be compared to the \lq typical\rq{} value of $f_{\rm PDR}=0.2$ used by \citet{Groves08} and \citet{Jonsson10}, following the calibration presented by \cite{Camps16}.  
\end{itemize}

With the parameters of the {\sc starburst} SEDs determined, the \skirt{} source emissivity profile is then set. As explained by \citet{Jonsson10}, the scale of the \Hii{} region emissivity profile should be set so as to enclose a similar mass of ISM to that required to be consistent with the subgrid absorption. Doing so avoids double-counting the dust in the subgrid absorption (which already affects the source spectra) and dust absorption in the diffuse ISM modelled by \skirt{}. To approximate this, we assume a fixed mass for the re-sampled particles of $M_{\rm PDR}= 10 M_{\rm cl}$ \citep[e.g.][]{Jonsson10}, and set the corresponding size of the region to be $r_{\rm HII} = \sqrt[3]{8 M_{\rm PDR} / \pi \rho_g}$ (for a cubic spline kernel), with $\rho_g$ the local gas density, taken from the parent particle. This is taken to be the smoothing length of \starburst{} sources. As noted previously, the \starburst{} model assumes the presence of birth cloud dust and that needs to be accounted for to ensure that the total dust mass is conserved. We budget for this additional dust using the ISM dust distribution, as described in section \ref{sec:dustrep}. \Hii{} region positions are sampled within a kernel of size $r_{\rm s} = \sqrt{r^2_{\rm p} - r^{2}_{\rm HII}}$, with $r^2_{\rm p}$ being the parent kernel smoothing length, and about the parent particle position. This is such that in the infinite sample limit the net kernel of the \Hii{} regions is equivalent to that of the parent. Again, the smoothing lengths of gas and star parent particles are obtained differently, as explained in Appendix \ref{sec:smooth}.
Finally, those sub-particles that are not converted to either a stellar or \Hii{} region source over the re-sampling period are reserved for the absorbing component to ensure mass conservation. Absorption in the (diffuse) ISM is modelled as described in section \ref{sec:dust} below.

\begin{figure*}
{\color{white} \LARGE
\begin{overpic}[width=0.33\textwidth]{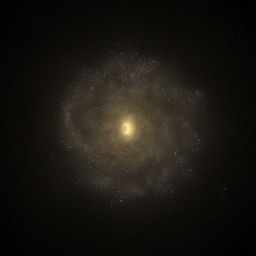}
 \put(5,90) {S}  \put(65,90) {Face-on} \put(5,5) {ID: 15814162}
\end{overpic}  
\begin{overpic}[width=0.33\textwidth]{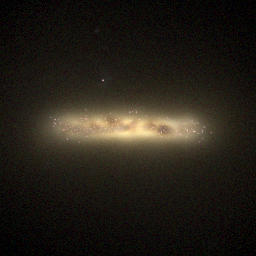}
 \put(5,90) {S}  \put(65,90) {Edge-on} \put(5,5) {ID: 15814162}
\end{overpic}
\begin{overpic}[width=0.33\textwidth]{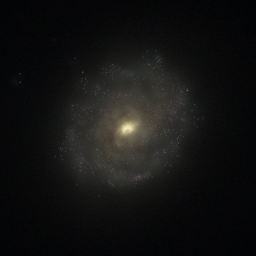}
 \put(5,90) {S}  \put(65,90) {Random} \put(5,5) {ID: 15814162}
\end{overpic}\\
\begin{overpic}[width=0.33\textwidth]{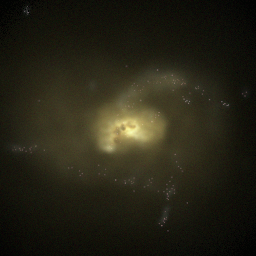}
 \put(5,90) {Irr}  \put(65,90) {Face-on} \put(5,5) {ID: 14318126}
\end{overpic}  
\begin{overpic}[width=0.33\textwidth]{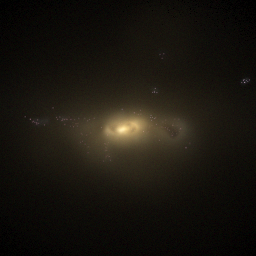}
 \put(5,90) {Irr}  \put(65,90) {Edge-on} \put(5,5) {ID: 14318126}
\end{overpic}
\begin{overpic}[width=0.33\textwidth]{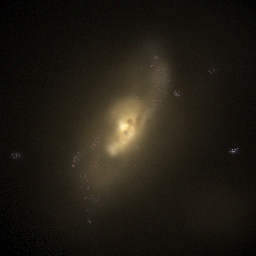}
 \put(5,90) {Irr}  \put(65,90) {Random} \put(5,5) {ID: 14318126}
\end{overpic}\\
\begin{overpic}[width=0.33\textwidth]{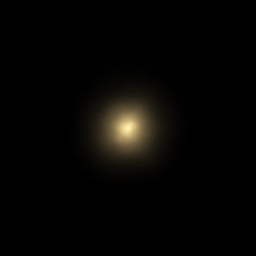}
 \put(5,90) {E}  \put(65,90) {Face-on} \put(5,5) {ID: 19099219}
\end{overpic}  
\begin{overpic}[width=0.33\textwidth]{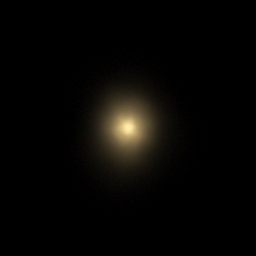}
 \put(5,90) {E}  \put(65,90) {Edge-on} \put(5,5) {ID: 19099219}
\end{overpic}
\begin{overpic}[width=0.33\textwidth]{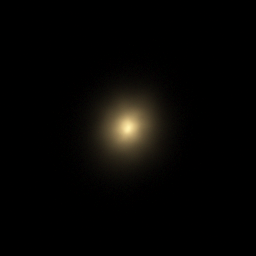}
 \put(5,90) {E}  \put(65,90) {Random} \put(5,5) {ID: 19099219}
\end{overpic}\\
}
 \caption{Observer frame true colour $gri$ images of selected \eagle{} galaxies from \ac{Ref-100} at
   $z=0.1$, including dust obscuration. The galaxy ID can be used to identify
   each galaxy on the \eagle{} database \citep{McAlpine16}, see also Table~\ref{tab:selgals}. Rows from top to bottom show a late type spiral (Hubble type S), an
   irregular type (Irr) and an early type (E) galaxy, with columns from left to right
   showing face-on view (parallel to the stellar angular momentum axis), edge-on view, and the view along the $z$-axis of the
   simulation   volume (\lq random\rq). Images are produced using $4\times 10^6$ photon packages at each of 333
   wavelengths, spaced as described in Appendix~\ref{sec:res}. \Hii{} regions appear pink in the colour scheme, due to strong H$\alpha$
   emission at $\sim$6700 \AA{} red-shifted into the $i$-band, dust lanes are clearly visible in the S and Irr images. The
   kinematic orientation works well for the spiral galaxy, but  the disordered gas distribution caused by an ongoing merger event
   in the Irr case yields more absorption in the face-one view than in the edge-on view. Images are 60 pkpc on a side.}
\label{fig:pics}
\end{figure*}

\begin{table}
\begin{center}
\caption{Properties of galaxies shown in Fig \ref{fig:pics}. All properties are extracted from the \eagle{} database \citep{McAlpine16}, for a 30 pkpc aperture.}
\label{tab:selgals}
\begin{tabular}{llll}
\hline
Type & Galaxy ID. & $M_\star \; [{\rm M_\odot}]$ & SFR [M$_\odot$ yr$^{-1}$]\\
\hline
Late (S) & 15814162 & $9.94 \times 10^{10}$  & 3.96\\ 
Irregular (Irr) & 14318126 & $1.34 \times 10^{11}$  & 4.60\\ 
Early (E) & 19099219 & $2.05 \times 10^{10}$  & $ 0.00$\\ 
\hline
\end{tabular}
\end{center}
\end{table}

\subsection{SKIRT modelling: observed properties}
\label{sec:dust}

Having detailed the parametrisation of the source components, we proceed to describe the modelling of dust in the diffuse ISM. This dust component is mapped to an adaptively refined (AMR) grid, for which the optical depth of each cell is calculated at a given reference wavelength. Neglecting Doppler shifts, this enables the computation of the dust optical depth at any other wavelength once the wavelength-dependence of the dust attenuation is specified. Details of the modelling of the dust and gas contents are given in sections \ref{sec:dustrep} and \ref{sec:grids}, respectively.

\subsubsection{Discretisation of the ISM}
\label{sec:grids}

Dust in galaxies exhibits structure on a range of scales, from galaxy-wide dust lanes to sub-kpc `dark clouds', with significant absorption across the range, down to the scale of molecular clouds \citep[e.g.][]{Hunt09}. We cannot resolve sub-kpc dust structures in \eagle, which is why we include such small-scale dust via the source model of \Hii{} regions, as described in Section~\ref{sec:young}. We use the gas particles in \eagle{} galaxies to estimate how dust is distributed in the diffuse ISM, and use \skirt{} to calculate obscuration by this dust, as follows. 

We discretise the gas density on the AMR grid using the octree algorithm \citep{saftly13}. A cubic root cell of size 60~pkpc is created, centred on the galactic centre of mass, to capture all galactic material (see section \ref{sec:t15}), and is refined based on the interpolated dust density derived from the gas particles, between a specified minimum and maximum refinement level. We increase the refinement level until the photometry is converged. Clearly, the minimum cell size should be smaller than the approximate spatial resolution of \eagle{} to best capture ISM structure in the simulated galaxies. We find that a maximum refinement level of 9 (corresponding to a finest cell of extent $60~{\rm kpc}/2^9=0.11~{\rm kpc}$ or $\approx 1/6$ of the $z=0$ gravitational softening), provides a grid structure that yields converged results when combined with a cell splitting criterion\footnote{This is the maximum fraction of the total dust mass that can be contained within a single dust cell. If the cell contains a larger fraction and is below the maximum refinement level, the cell is subdivided.} of $2\times 10^{-6}$. We therefore adopt a maximum refinement level of 9 for our analysis, together with a minimum refinement level of 4. While we use a minimum cell size twice as large as that of \cite{Camps16}, we have verified that this has a negligible effect on our results in the optical and NIR, while increasing the speed of our \skirt{} simulations.

\subsubsection{Dust model}
\label{sec:dustrep}
Dust traces the cold metal-rich gas in observed galaxies \citep[e.g.][]{Bourne13}. Here we assume that the dust-to-metal mass ratio is a constant, 
\begin{equation}
f_{\rm dust} \equiv {\rho_{\rm dust}\over Z\rho_{\rm g}} = 0.3\,,
\end{equation}
where $Z$ is the (SPH-smoothed) metallicity, and $\rho_{\rm dust}$ and $\rho_{\rm g}$ are the dust and gas density, respectively. The numerical value was determined by calibrating FIR properties of \eagle{} galaxies by \cite{Camps16}, and is consistent within the uncertainties of observationally inferred values \citep[e.g.][]{Dwek98, Draine07b}. The assumption of a constant value of $f_{\rm dust}$ is common and is observed to apply to a wide variety of environments \citep[e.g.][]{Zafar13, Mattsson14}, though there are indications it can vary in some cases \citep[e.g.][]{DeCia13, Feldmann15}. We implement this constant ratio by assigning a dust mass of $m_{\rm dust} = f_{\rm dust} m_g$, where $m_g$ is the particle mass. We use the dust model described by \citet{Zubko04}; a multi-component dust mix tuned to reproduce the abundance, extinction and emission constraints on the Milky Way. Following \citet{Camps16}, gas must be either {\it star-forming} (i.e. assigned a non-zero star formation rate by the simulation or in the re-sampling procedure) or sufficiently cold (with temperature $T < 8000$~K) to contribute to the dust budget.

To account for the dust mass already associated with birth clouds when using the \starburst{} source SEDs, we introduce \lq ghost\rq{} particles that contribute {\em negatively} to the local dust density. These ghost particles are placed at the location of each \Hii{} region, have a mass of $M_{\rm PDR}=10m_{\rm r}$, where $m_r$ is the stellar mass formed in the star-forming region, and a smoothing length equal to three times that of the \Hii{} region. The assumption that the PDR mass, $M_{\rm PDR}$, is ten times that of the stellar mass formed follows the recommendation of \citet{Groves08}, the greater smoothing of the ghost contribution avoids negative dust densities. The creation of \lq holes\rq{} in the dust distribution around \Hii{} regions may seem unphysical, as observed \Hii{} regions are typically embedded in the densest (and dustiest) ISM. However, we have tested an alternative implementation where the dust mass of all contributing particles are downscaled to balance the additional dust invoked by the \starburst{} SEDs, and find little perceptible difference in the results presented here. We will see that ISM dust obscuration is still higher around young stars, even in the presence of these ghost particles.

\subsection{Data products}
\label{sec:prod}
This section describes the data products that are generated by \skirt{}. We reiterate that we do not consider kinematics when using \skirt{}, {\em i.e.} no Doppler shifts are yet accounted for beyond any line broadening present in the input SEDs. We perform a convergence test in Appendix \ref{sec:res} to determine how to best sample the SED, both in terms of wavelength sampling and photon-packet sampling. We construct integrated spectra for all simulated galaxies in three orientations; edge on, face on and randomly orientated with respect to the galactic plane. The calculation of orientations for \eagle{} galaxies is described in section \ref{sec:orientations} below. The data products produced include the following:

\begin{itemize}
\item \textbf{Integrated spectra} capture all the photon packets emanating from the mock galaxy for the fixed list of specified wavelengths, and in a given direction. The standard resolution spectra consist of 333 wavelengths in the range 0.28$-$2.5 ${\rm \mu}$m, chosen to sample the rest-frame $ugrizYJHK$ photometric bands (see Appendix \ref{sec:res} for details). Spectra are produced with and without ISM dust at redshifts $z=0$ and redshift $z=0.1$ (the snapshot redshift from which the galaxies were selected). An example integrated rest-frame SED of a star-forming galaxy and including dust attenuation is plotted in Figure \ref{fig:egspec}.
\item \textbf{Data cubes}, or mock IFU data, consist of 256x256 spatial pixels, each with a spectrum at standard spectral resolution. Given that the field of view corresponds to 60~pkpc on a side, this corresponds to $234\;{\rm pc /pixel}$. Images are produced in both the rest and observed frames, but only for dust attenuated galaxies with $M^{\star} > 10^{10} {\rm M_\odot}$. Again, these do not include kinematic effects, which will be the focus of future work.
\item \textbf{Broad-band photometry} The fluxes through the $ugrizYJHK$ filters and are obtained by convolving the integrated spectra with the filter transmission curves \citep[transmission curves were taken from][]{SDSSfilters, UKIRTfilters}. We compute both rest-frame and observed-frame photometry for the entire galaxy sample both with and without ISM dust attenuation. 
\item \textbf{Broad-band images} are produced by integrating along the wavelength axis of the data cubes. These are generated including dust for the $ugriz$ SDSS bands, \changes{and provided in 3 colour PNG (portable network graphic) format\footnote{\changes{Note that these images are initially intended for illustrative purposes only, as the detailed light distributions are dependent on the somewhat \textit{ad-hoc} choice of stellar smoothing \citep[similarly demonstrated by][]{Torrey14}. While we find the influence of smoothing to be small for the results presented in this paper (see Appendix \ref{sec:smooth}), analysing the influence smoothing has on morphologies is left to a future work.}} via the approach of \citet{Lupton03}.} Figure \ref{fig:pics} shows three-colour $gri$ images  at $z=0.1$ for three different galaxies and three orientations. We picked a late-type, an irregular and an early-type galaxy. Some properties of these galaxies are listed in Table \ref{tab:selgals}. Structural features resembling spiral arms and tidal tails are distinguishable for the late and irregular types, respectively, while the early type exhibits a smooth, featureless light distribution. Star-forming \Hii{} regions appear pink due to \Ha{} emission in the \starburst{} SEDs for these $z=0.1$ galaxies. We also observe scattering and absorption by dust for the late and irregular types.
\end{itemize}

Data products will be made available via the \eagle{} public database \citep{McAlpine16}, with the exception of data cubes, which are available through collaboration with the authors\footnote{To access the database and receive updates on its content, register at {\tt http://icc.dur.ac.uk/Eagle/database.php}}. We will below compare the output from the \skirt\ simulations to those with the same source model but without obscuration by ISM dust. Note, however, that the \starburst\ source model always includes dust associated with the birth cloud. We will refer to the these models as \lq ISM dust-free\rq\ in what follows.

%% file: Orientation.tex
\section{Attenuation Properties of SKIRT galaxies}
\label{sec:orientations}

\begin{figure}
 \includegraphics[width=0.48\textwidth]{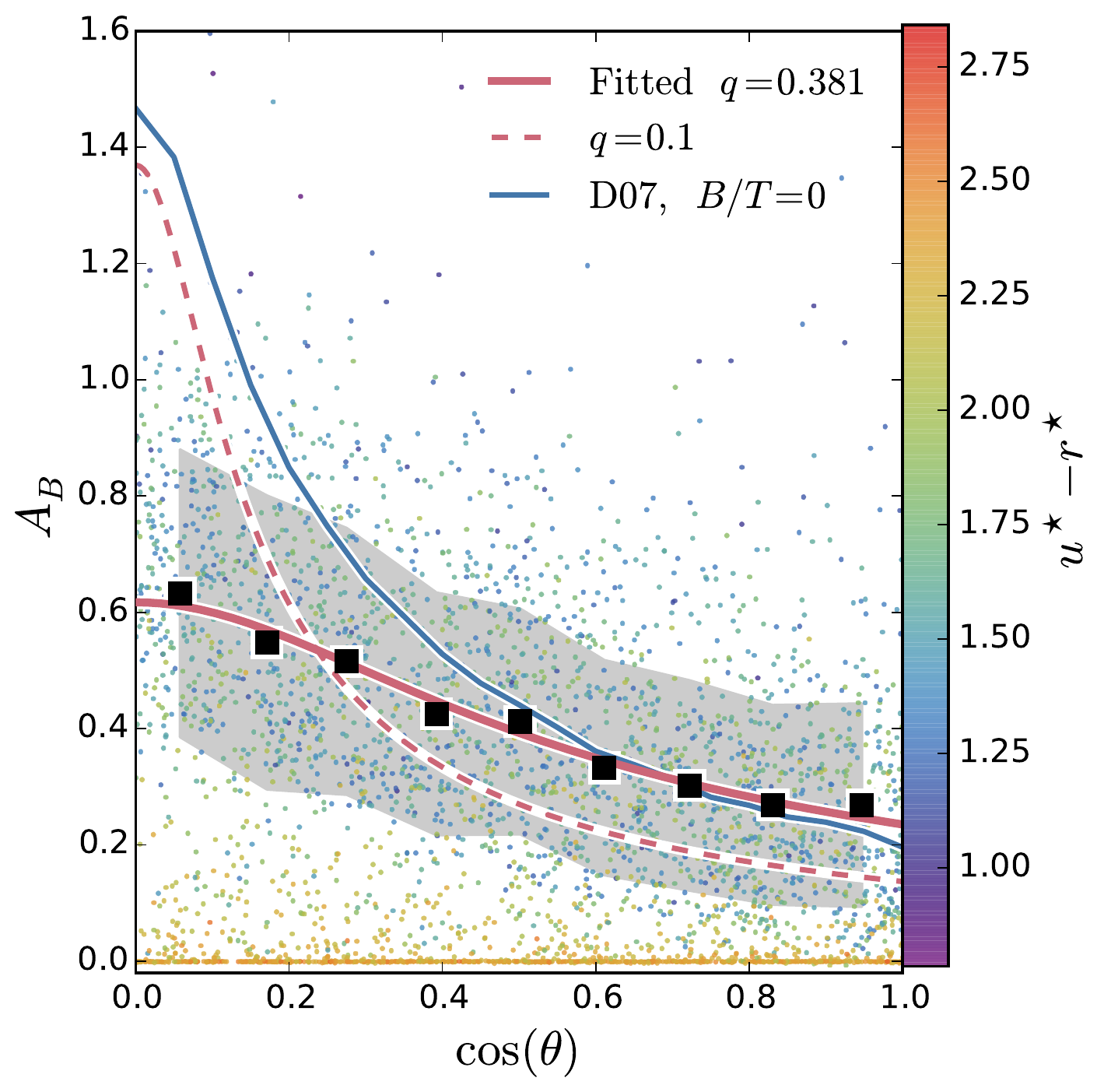}  

 \caption{The Johnson rest-frame $B$-band attenuation, $A_B$, as a function of inclination angle; $\cos\theta=1$ corresponds to face-on.  	\textit{Points} are individual \eagle{} galaxies from simulation \ac{Ref-100} at $z=0.1$, with $M_\star > 10^{10} {\rm M}_{\odot}$, coloured by ISM dust-free $u^\star-r^\star$ colour. \textit{Black squares} indicate the median relation for galaxies with $A_B \geq 0.05$, binned by $\cos(\theta)$. The \textit{grey, shaded} region indicates the 16th-84th percentile range in each bin. The \textit{red, solid line} shows the best fit of the form Eq.~(\ref{eq:ellipse}), which has $q=0.396$. The \textit{red, dashed line} shows the same fit, except with $q=0.1$ to represent a thinner discs, which follows the trend of the intrinsically blue galaxies better. Galaxies with $A_B < 0.05$ are typically red in $u^\star-r^\star$ and occupy a tight distribution in $A_B \approx 0$, independent of orientation. For comparison we also overlay the observed relation from \citet{Driver07} (D07 in the legend) for the median attenuation curve of the disc component only, of galaxies with bulge/total ratio smaller than 0.8.}
  \label{fig:orient}
\end{figure}
\begin{figure*}
  \includegraphics[width=0.99\textwidth]{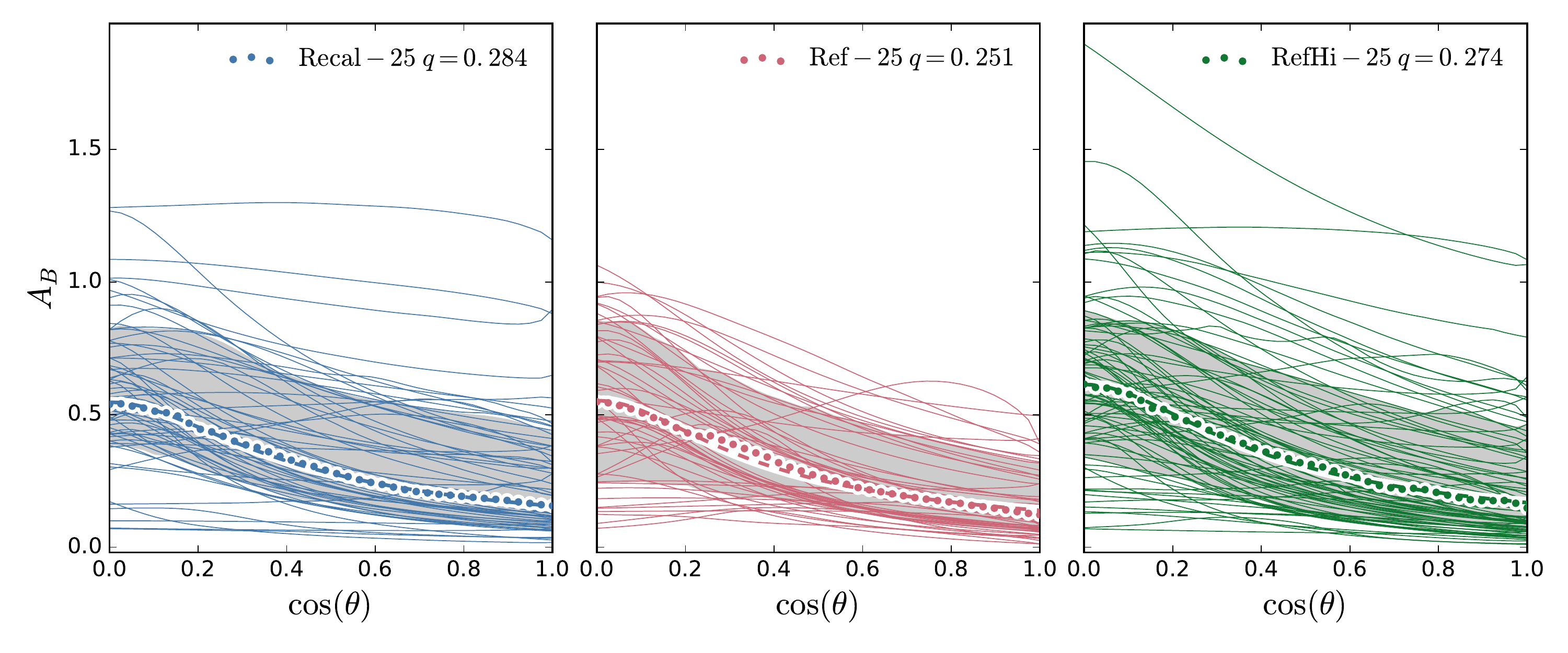}  
 \caption{Johnson rest-frame $B$-band attenuation as function of orientation for \eagle{} galaxies, taken from simulations \ac{Recal-25}, \ac{Ref-25} and \ac{RefHi-25} at $z=0.1$ (left to right panels, see Table~\ref{tab:sims} for details). Thin curves show $A_B$ for individual galaxies as a function of inclination, the shaded circle represent the median trend, and the grey region includes the 16th-84th percentiles. The dashed line approximately overlying the circles represents the fit of Eq.~(\ref{eq:ellipse}); the best-fit value of $q$ is indicated in each panel. We see significant galaxy-to-galaxy variation in the shape of the attenuation-inclination relation, with the highest attenuation values in the higher-resolution galaxies, but little difference between the trend of the median inclination as a function of $\cos(\theta)$.}
 \label{fig:aocurves}
\end{figure*}

\begin{figure*}
 \includegraphics[width=0.99\textwidth]{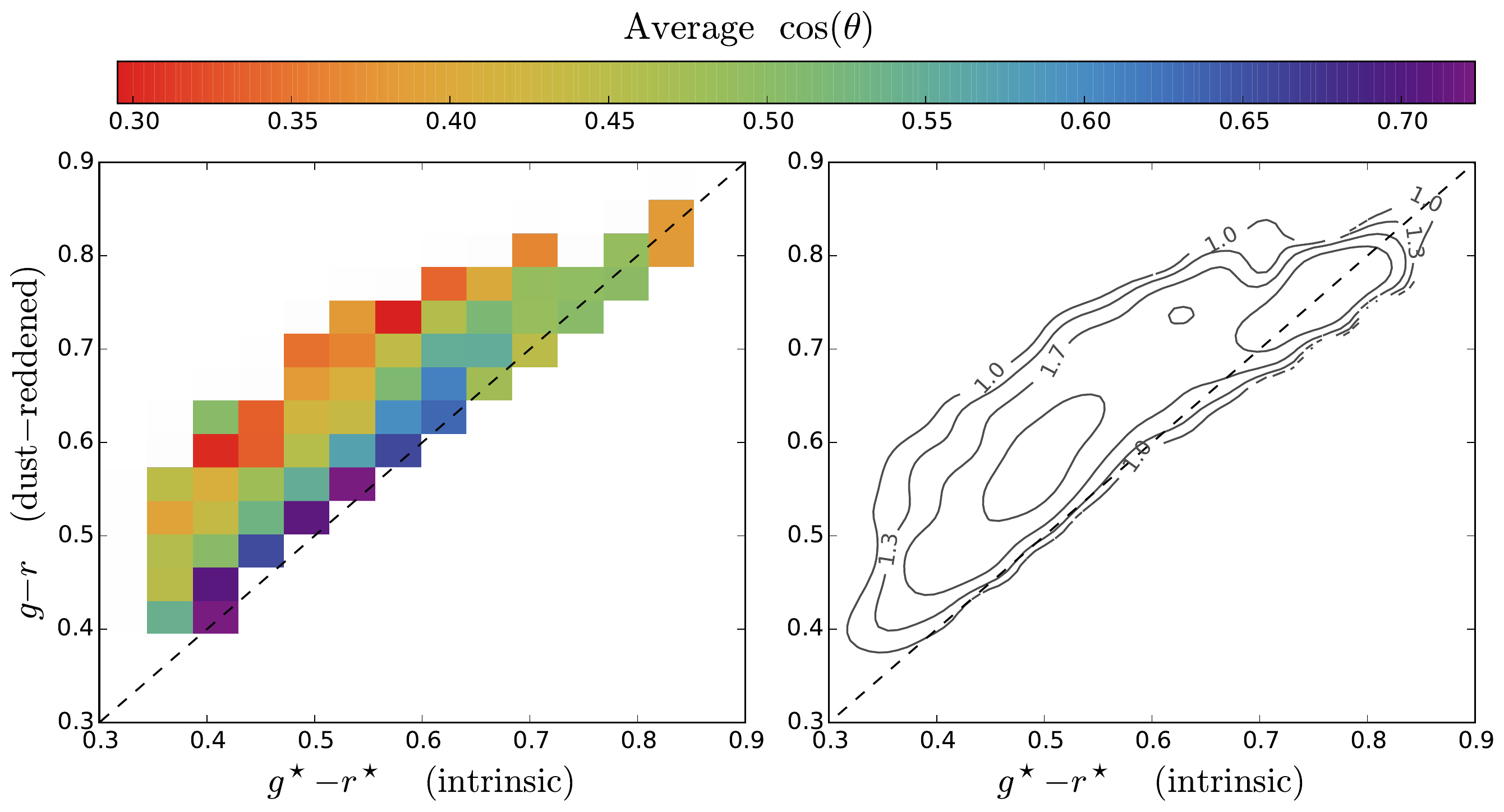}  
 \caption{Comparison of ISM dust-free SDSS $g-r$ colour to $g-r$ including ISM dust for \eagle{} galaxies in the mass range $10 \leq \log(M_\ast / {\rm M_\odot}) < 11$ at $z=0.1$; the 
 	\textit{black dashed line} indicates the 1:1 relation in both panels. In the \textit{left panel}, the median value $\cos(\theta)$ is plotted in regularly spaced bins in colour (only bins with more than 10 galaxies are shown). The \textit{right panel} shows the number density $n$ of galaxies per colour-colour bin as grey contours labeled by $\log(n)$.  There is a clear trend of increased reddening at higher $\cos(\theta)$ as expected. \changes{Some galaxies lie marginally below the 1:1 line, the reasons for which are discussed in the text.}}
 \label{fig:orscat}
\end{figure*}

\begin{figure}
 \includegraphics[width=0.48\textwidth]{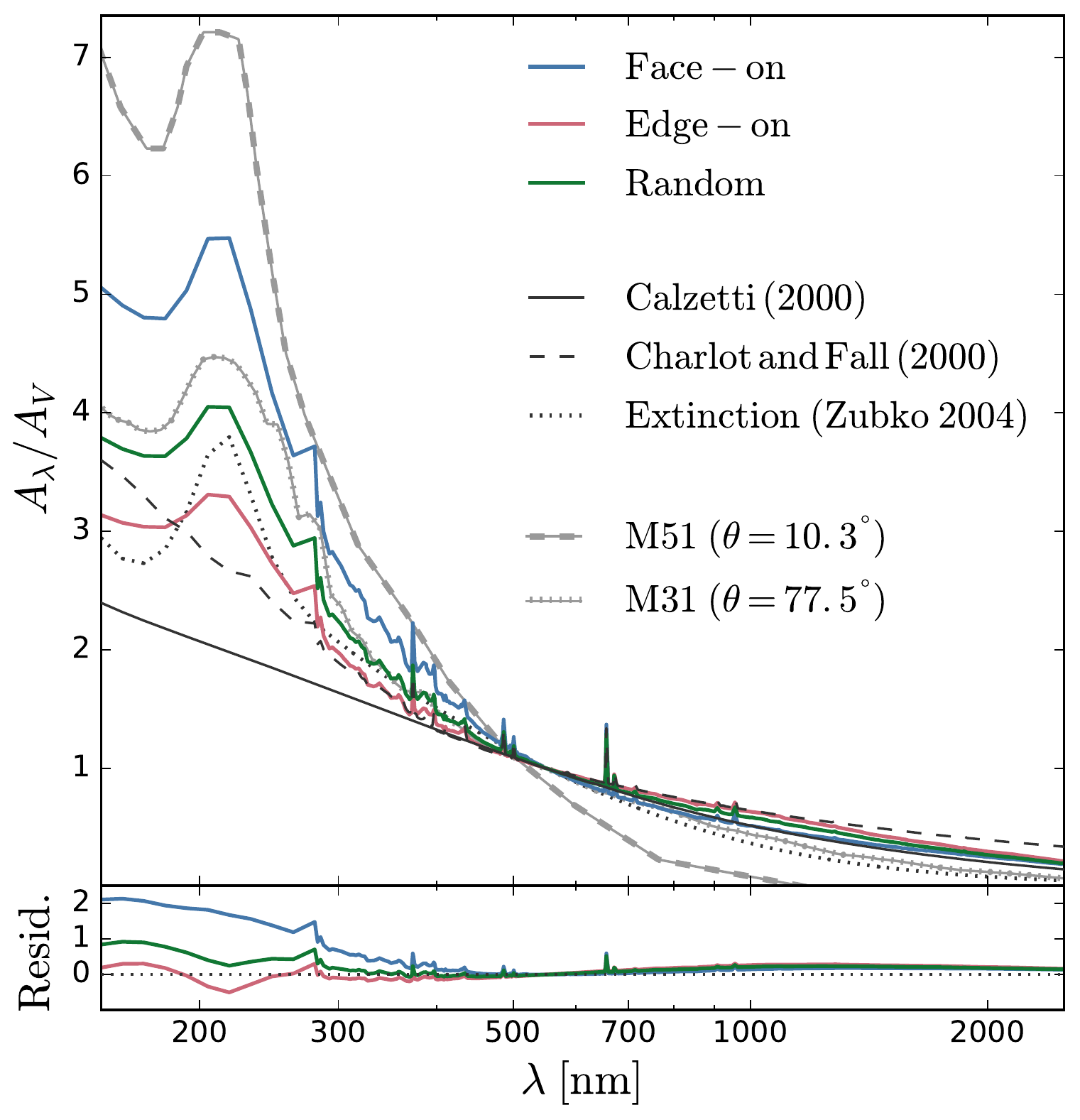}  
 \caption{Attenuation as a function of wavelength, normalised to attenuation in the $V$-band at 550~nm. The luminosity-weighted average attenuation curves of intrinsically blue ($u^\star - r^\star < 2.0$) \eagle{} galaxies for face-on, edge-on and random orientations are plotted as blue, red and green curves, respectively. The attenuation laws of \citealt{Calzetti00} and \citealt{CF00} are plotted as solid and dashed black lines, respectively, and the attenuation curve inferred for the galaxy M51 by \citet{DeLooze14} and for M31 by \citet{ViaeneP}, are plotted as dashed and dotted grey lines, respectively. The inclination of these galaxies is plotted in the legend. \changes{We also plot the intrinsic extinction curve used by \skirt{}, \citet{Zubko04}}. The bottom panel shows the residuals between the three \eagle{} curves and the dust extinction curve of \citet{Zubko04}, assumed by \skirt{}. The shape of the attenuation curve for \eagle{} galaxies varies with orientation, with face-on galaxies exhibiting a stronger wavelength dependence.}
 \label{fig:ecurves}
\end{figure}

In this section we focus on how attenuation depends on galaxy orientation at low redshift ($z=0.1$). This helps us to separate the effects of geometry and dust content, and facilitates the interpretation of a comparison with observations. 

\subsection{Broad-band attenuation}
\label{sec:bbA}

The orientation of a galaxy can profoundly affect its measured colours, particularly in the case of thin spiral galaxies where the edge-on view is much more affected by dust than the face-on view. Indeed, the reddest galaxies observed in the local Universe are often edge-on spirals \citep[e.g.][]{Sodre13}. \changes{The \skirt{} modelling naturally accounts for this effect, as opposed to the two component screen model presented in T15 which relies on simple geometrical arguments to account for this.} To quantify orientation effects in disc galaxies, we use 3 lines of sight: parallel, perpendicular and randomly oriented with respect to the galactic plane. This helps constrain orientation effects on dust extinction for each galaxy individually, as well as providing a set of photometry with random orientations used when comparing to data. 

We assume that the disc of a galaxy is perpendicular to the spin vector, ${\bf S}$, of its stars. We calculate ${\bf S}$ by summing the spin vectors of all star particles within a shell with inner and outer radii of 2.5~pkpc and 30~pkpc, respectively, in the centre-of-mass rest-frame of the galaxy. The outer radius corresponds to the maximum radius of a galaxy assumed in Section~\ref{sec:eagle}, the inner radius was chosen to avoid a significant contribution from a bulge or regions strongly affected by gravitational softening. We found that with this selection, ${\bf S}$ is generally dominated by the dynamically cold rotating disc component, if present. We characterise the orientation of a galaxy by its inclination angle $\theta$, such that face-on galaxies have $\cos(\theta)=1$.

In Fig.~\ref{fig:orient} we plot the attenuation in the $B$-band, $A_B$, as function of orientation, for all \eagle\ galaxies from simulation \ac{Ref-100} with stellar mass $M_\star > 10^{10} {\rm M}_{\odot}$;  points represent individual galaxies coloured by intrinsic $u^\star-r^\star$ ($^\star$ denotes intrinsic photometry). Two sequences are observed: ({\em i}) a broad sequence of intrinsically blue (star-forming) galaxies where $A_B$ increases with decreasing $\cos(\theta)$ and ({\em ii}) a very tight sequence of intrinsically red (passive) galaxies showing no orientation dependence. Such a dichotomy is of course unsurprising: red galaxies typically have low cold gas fractions and therefore negligible dust attenuation, whereas star forming (intrinsically blue) galaxies have relatively high cold gas fractions, with gas distributed in a disc, hence dust obscuration is higher and depends on orientation.

The median $A_B$ of intrinsically blue galaxies with $A_B \geq 0.05$ in bins of $\cos(\theta)$ is plotted as black squares in Fig.~\ref{fig:orient}), with the grey region enclosing the 16th-84th percentiles. The median attenuation $A_B$ of \eagle\ galaxies increases from 0.3 to $\geq 0.6$ from face-on to edge-on orientation but the scatter around the trend is large ($\approx 0.5$~mag). We fit the angular dependence of the median relation using the ellipsoidal dust model discussed by T15,
\begin{equation}
  \label{eq:ellipse}
    A_B(\cos\theta) = a \frac{q}{\sqrt{q^2 + (1 - q^2)\cos^2\theta}}\,,
\end{equation}
where $a$ is the edge-on obscuration and $q$ represents the axial ratio of the galaxy; lower $q$ corresponds to thinner galaxies. We treat $a$ and $q$ as free parameters in the fit, and plot the fitted curve in red. The functional form fits the trend well for an axial ratio of $q=0.381$. The scatter around the median likely originates from both diversity in the ISM distribution in different galaxies, i.e. deviation from an idealised disc, but also from errors in identifying the correct orientation of the disc plane. Indeed, we showed in Figure \ref{fig:pics} that the galactic plane is not always easily defined, as evidenced by the irregular galaxy shown in the middle row.

\citet{Driver07} use a sample of galaxies from the \ac{MGC} with estimated bulge-to-total ($B/T$) light ratios of $B/T < 0.8$ to measure the extent to which the location of the \lq knee\rq\ in the $B$-band luminosity function \citep{Schechter76} depends on inclination. They fit their results with the model of \citet{Tuffs04} to obtain the typical attenuation separately for bulge and disc components. We plot the relation of \citet{Driver07} for a typical disc ($B/T=0$) in Fig.~\ref{fig:orient} as a solid blue line. 

The median \eagle{} $A_B$ values (black squares) and the fitted form of Eq.~(\ref{eq:ellipse}) (red line), are consistent with those obtained by \citet{Driver07} for nearly face-on discs ($\cos(\theta)\geq 0.5$), but are significantly lower for highly inclined discs. While there is uncertainty in the absolute values measured for $A_B$, as discussed by \citet{Driver07}, the difference between typical face- and edge-on $A_B$ values is better constrained\footnote{\citet{Driver07} derive the relative attenuation directly by measuring how the knee position of the luminosity function differs for edge-on and face-on galaxies.} and clearly significantly larger in the data compared to \eagle{}. Note that the \citet{Driver07} data is represented by a pure disc (blue line) for simplicity. While \eagle{} spirals clearly possess bulges (see e.g. Fig. \ref{fig:pics}), the \textit{difference} between face-on and edge-on attenuation found by \citet{Driver07}  varies little with $B/T$. The blue line provides a guide curve to highlight the smaller range in $A_B$ with inclination for \eagle{}. Decomposition of \skirt{} light profiles into bulge and disc contributions is left to a future study. 

The $q=0.1$ curve, which models thinner discs for the same dust content as the fitted curve, demonstrates much better agreement with \citet{Driver07} than the median \eagle{} relation. This suggests that the discrepancy is likely due to \eagle\ galaxies not being as thin as observed galaxies. The fact that \eagle\ galaxies are thicker than observed is not only due to numerical resolution. Indeed, we compare the $A_B(\cos\theta)$ relation for galaxies taken from the three $(25 \; {\rm Mpc})^3$ simulations listed in Table \ref{tab:sims}. These simulate the same volume, but at different resolutions. The number of galaxies with  $M_\star > 10^{10} {\rm M_\odot}$ in this smaller volume is $\lesssim 100$, therefore the two-sequences in the $A_B(\cos\theta)$ relation are not well constrained if we simply use the mock photometry of randomly oriented galaxies as we did in Fig.~\ref{fig:orient}. We therefore calculate $A_B(\cos\theta)$ for all sufficiently massive galaxies ($M_\star > 10^{10} {\rm M_\odot}$) at 40 inclinations for each galaxy, equally spaced in $\cos(\theta)$, and plot the resulting curves in Fig.~\ref{fig:aocurves}. Equation~(\ref{eq:ellipse}) is fit to the median relation and plotted as a dashed coloured line. While higher values of $A_B$ are seen in the higher resolution \ac{RefHi-25} and \ac{Recal-25} samples, the difference with respect to the median values of \ac{Ref-25} is small. Neither the plotted curves for individual galaxies nor the fits using Eq.~(\ref{eq:ellipse}) to the median trend, show strong evidence for $A_B$ being more sharply peaked at improved numerical resolution.

The weaker inclination dependence and lower edge-on values of $A_B$ in \eagle{} are instead likely a consequence of \eagle's subgrid physics, in particular the use of an imposed Jeans-limiting, polytropic relation for star forming gas (section \ref{sec:eagle}). This relation yields a Jeans length at the star formation threshold of the $\sim$1.5~kpc, and \eagle{} discs are unable to be much thinner than this. This relation is imposed to avoid numerical fragmentation below the resolution of the simulation, as explained by S15. Dust discs in observed galaxies, on the other hand, are much thinner, $\sim$100-200~pc \cite[e.g.][]{Xilouris99, DeGeyter14, Hughes15}. In a thin disc seen edge-on, the dust optical depth to young stars will be much higher than if the disc where thick, and this seems to be the main difference between observed and simulated galaxies. 

This comparison demonstrates that the $A_B(\cos\theta)$ dependence displays both strong and weak convergence behaviour, with increased numerical resolution not changing the relation significantly - and not improving the agreement with the data. We show in Appendix \ref{sec:smooth} that reducing all star particles to point sources only boosts the edge-on value of $A_B$ by $\lesssim 0.1$~mag. We conclude from this that the lower values for $A_B$ for edge-on \eagle{} galaxies are likely a result of the the simulations being unable to represent cold gas; the high column densities and clumpy structure of molecular gas observed in real disc galaxies is not reproducible in the \eagle{} simulations without realistic modelling of gas with $T \lesssim 10^4$K. The influence that thicker discs (and thus lower edge-on attenuation) has on our results is discussed further below.

\subsection{Broad-band colour effects}
\label{sec:reddening}

The extent to which inclination affects the optical colour distribution of \eagle\ galaxies with $10^{10} {\rm M}_{\odot} \leq M_\star < 10^{11} {\rm M}_{\odot}$ is illustrated in Fig.~\ref{fig:orscat}, where we plot intrinsic (ISM dust-free) $g^\star-r^\star$ colour against $g-r$ in the presence of dust. In the left panel we shade regularly spaced colour-colour bins by the median value of $\cos\theta$ of galaxies in that bin (provided the bin contains more than 10 galaxies). We see a clear trend in attenuation with inclination, especially for intrinsically blue galaxies of $g^{\star} - r^{\star} \lesssim 0.6$, with galaxies possessing median $\cos(\theta)$ values of $\approx 0.25$ and $\approx 0.65$ for maximal and minimal offsets from the 1:1 relation, respectively. For galaxies with redder intrinsic colours, the trend is less pronounced and the maximal offset is lower, as expected for \changes{less dusty} galaxies. 

In the right panel of Fig.~\ref{fig:orscat} we plot logarithmically spaced contours representing the number of galaxies per colour-colour bin. Intrinsically red ($g^{\star} - r^{\star} \approx 0.75$) galaxies follow the 1:1 relation closely with little offset, whereas intrinsically blue ($g^{\star} - r^{\star} \approx 0.4$) galaxies are offset to redder colours and show a large scatter. Worth noting is the approximately constant median offset to the red of $\approx 0.1$~mag for galaxies with $g^{\star} - r^{\star} \lessapprox 0.6$, implying that a similar average reddening is experienced by star forming galaxies regardless of star formation rate. 

Some galaxies lie marginally {\em below} the 1:1 relation. \changes{In most cases this is due to uncertainty in the photometry (see appendix \ref{sec:res}), particularly in dust-free galaxies where the attenuation is small anyway, and these negative reddening measurements are $\sim 0.001$ mag. However in rare instances, measured for higher redshift \eagle{} galaxies, significant negative reddening is observed. This can be attributed to those galaxies demonstrating heavy obscuration in their central regions, leading to higher contribution of young stars in the outskirts, conspiring to produce bluer colours overall }

\subsection{Attenuation curves}
\label{sec:Acurves}

The \textit{extinction curve} is an intrinsic property of a given dust grain population; combining the wavelength dependent cross-sections of absorption and scattering. Our choice of dust mix thus sets the optical depth of dust cells modelled by \skirt{}. However, the extinction does not provide a direct mapping between the intrinsic and observed SEDs, which additionally depends on the relative distribution of stars and dust and the orientation of the galaxy along the line of sight. This galaxy and line-of-sight specific mapping is referred to as the \textit{attenuation curve}. One example of why the curves may differ significantly is that the young stars that dominate emission at short wavelengths are in general embedded in dusty regions and hence their blue light is more strongly dust-attenuated.

 As a result, the attenuation curves may differ systematically in shape from the extinction curve of the individual dust cells. The shape of the curve is also likely orientation dependent, for example stars in a central bulge may be obscured in edge-on but not face-on projections. As a result, the normalisation, shape and orientation dependence of the attenuation curve are to some extent degenerate in observed integrated spectra when an attenuation proxy such as the Balmer decrement is used. 

 It is typical to assume a fixed shape of the attenuation curve to de-redden observed SEDs. \changes{Using the SEDs we generate for \eagle{} galaxies, we can explore the typical attenuation curves that arise from our MCRT treatment, and how these may vary systematically with orientation. While \eagle{} galaxies appear to have thicker discs than observed (see section \ref{sec:bbA}), we hope to provide an indication of the ways in which real galaxy attenuation curves can vary from basic screen models using the comparatively realistic and diverse morphologies that arise in \eagle{} galaxies.} Studies of variation in galaxy attenuation curves have been performed for observed galaxies assuming idealised geometries by \citep[e.g.][]{Byun94, Baes01, Wild11, Kriek13} and for small samples of zoomed galaxy simulations by \citet[e.g.][]{Natale15}.

We plot the $r$-band luminosity-weighted average attenuation curves of intrinsically blue ($u^\star - r^\star < 2.0$) \eagle{} galaxies, normalised by the attenuation in the $V$-band, in Fig. \ref{fig:ecurves}. Face-on, edge-on and random projections are plotted as blue, red and green curves, respectively. Recall that dust in birth clouds is accounted for in our models by the \starburst{} SEDs of \cite{Groves08} for which we do not have the analytical description of the intrinsic attenuation curve. Therefore, we approximate the attenuation given the modelled dust content assuming a foreground screen. Fortunately, the proportion of optical light attenuated in the \Hii{} or associated PDR regions is small relative to the diffuse component, except at some specific atomic transitions. Nevertheless, we find that the increased attenuation visible in Fig.~\ref{fig:ecurves} at the \Ha{} and \Hb{} wavelengths is still clearly present even when only the diffuse contribution is taken into account: this is because PDR regions are preferentially embedded in \changes{denser regions of the ISM, and it is this diffuse ISM dust that causes the high attenuation. We emphasise that preferential attenuation of young stars due to dust in a birth cloud screen is explicitly built in to both the \skirt{} and the \citet{CF00} model employed by \ac{T15}. The difference is due to additional preferential attenuation of the diffuse ISM represented as a single screen in\citet{CF00}.}
	 	
In all cases, attenuation increases rapidly towards shorter wavelengths with significantly higher attenuation at certain discrete wavelengths and a broad absorption feature at $\approx 220$~nm. For face-on galaxies, the slope is much steeper than the intrinsic dust extinction law. The discrete wavelengths correspond to atomic transitions at which star forming regions dominate emission, their boosted attenuation is due predominately to the increased diffuse dust around these regions\footnote{Constructing ISM attenuation curves for the HII regions alone yields curves similar to the \citet{Calzetti00} and \citet{CF00}, with the \Ha{} feature reduced by $\approx 90$~\%.}. The feature at $\approx 220$~nm is intrinsic to the assumed dust extinction law.

The average edge-on attenuation curve is less steep, or `greyer', than that of face-on galaxies. This is because the dust in \eagle{} galaxies is spatially correlated with star forming gas, therefore intrinsically blue stars are preferentially obscured in the face-on view, while in the edge-on view that dust also acts as a screen for older stars. The $A_\lambda$ curve for the randomly oriented values exhibits an intermediate steepness between the face-on and edge-on curves. 

The extinction curve assumed by \skirt{} when processing \eagle{} galaxies is that of \citet{Zubko04}. \changes{The extinction curve is plotted as a black dotted line in the top panel (again normalised to 1 for the $V$-band), while in the bottom panel we plot the residuals of the \eagle{} attenuation curves once the $V$-band normalised extinction is subtracted, to isolate the influence of geometry and orientation. We see that the \eagle{} curves are steeper than the intrinsic extinction curve, again a manifestation of the preferential obscuration of young stars, most exaggerated for the face-on projection.} We also see that the \eagle{} curves lie above the extinction curve at NIR wavelengths. This can be ascribed to \changes{absorption overtaking scattering as the primary photon-dust interaction at wavelengths longer than optical, leading to a smaller fraction of attenuated light being scattered into the line of sight than at optical wavelengths. As a result, when curves are normalised at optical wavelengths, the NIR attenuation appears boosted relative to the pure extinction curve.}

For comparison, we plot the attenuation curves of the \citet{Calzetti00} and \citet{CF00} screen models. Comparing to the screen-model curves we see that the attenuation curves for \eagle{} are generally steeper at all orientations. The screen models are closest to the edge-on curve at short wavelengths, due to dust behaving as a screen for many stars in an edge-on view. \changes{The \citet{CF00} curve represents a two component screen model, accounting for additional attenuation of young stars associated with stellar birth clouds. This age-dependent attenuation model provides better agreement with the \eagle{} curves than the single screen \citet{Calzetti00} model, laying closest to the edge-on \eagle{} curve. The fact that young stars are also preferentially attenuated by diffuse ISM in our \skirt{} modelling may explain why the \eagle{} attenuation curves are steeper still.}

We also plot attenuation curves derived for local galaxies M31 and M51 (from \cite{ViaeneP} and \cite{DeLooze14}, respectively) M31 is a relatively edge-on galaxy, with an inclination angle of $\theta=77^\circ$ \citep{Brinks84}, whereas M51 is practically face-on at $\theta=10^\circ$ \citep{DeLooze14}. The M31 attenuation curve lies between the face-on and edge-on curves at wavelengths short-ward of the $V$-band, residing closest to the random projection curve. The M51 curve is steeper than any of \eagle{} or screen-model curves. The M31 and M51 curves are both steeper than the \eagle{} curves for comparable galaxy orientations. They also show more difference in slope than between the face-on and edge-on curves. We suggest that this is because \eagle\ galaxy discs are thicker, and smoother than observed discs, both a consequence of limitations in the sub-grid physics. \changes{This could indicate that the orientation dependence we identify in \eagle{} galaxies may become stronger if \eagle{} galaxies possessed more realistic, thinner discs.}

\changes{Observational studies have explored attenuation curve variation through SED fitting of low and high redshift galaxy samples and assuming screen-like attenuation \citep[e.g.][respectively]{Wild11, Kriek13}. \citet{Wild11} find a similar trend between attenuation curve slope and inclination for nearby galaxies as we observe here. However, both \citet{Wild11} and \citet{Kriek13} find a slight weakening of the 2175\AA{} bump feature for face-on galaxies, which is not apparent in \eagle{}. As this feature and its variation is attributed to poorly understood dust grain species that inhabit certain regions of galaxies, and given our modelling does not include spatial variation of the dust mix, this is perhaps unsurprising. A better understanding of the nature of these enigmatic dust grains, and their location in galaxies, would allow us to incorporate this into our modelling.}

%% file: SkirtPhotometry.tex
\section{{\sc skirt} colours of {\sc eagle} galaxies}
\label{sec:testing}

\begin{figure}
 \includegraphics[width={0.48\textwidth}]{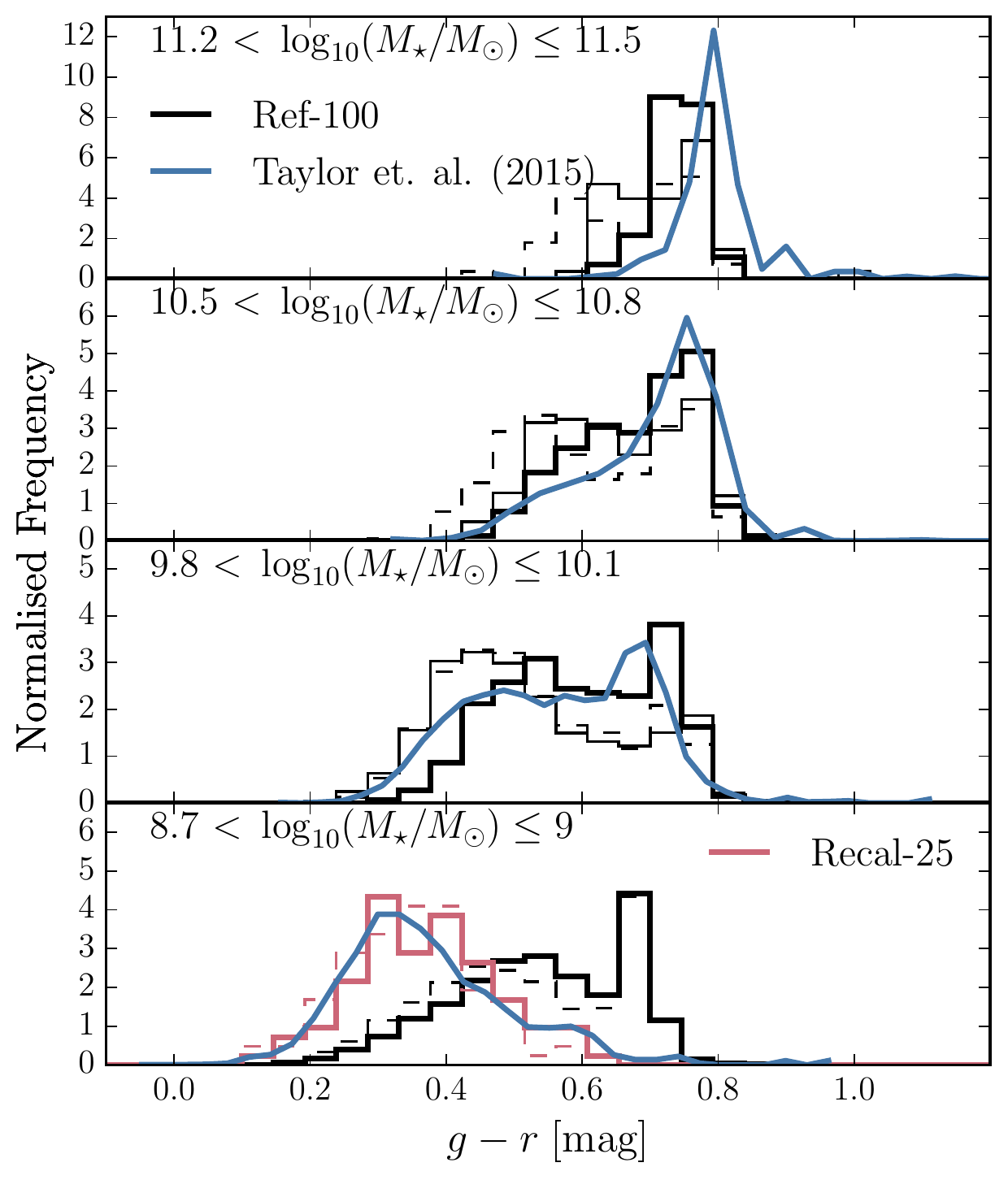}  
 \caption{Rest-frame $z=0.1$ SDSS $g-r$ colour distributions of galaxies in four non-contiguous bins of stellar mass, labelled in each panel. \eagle\ galaxies of simulation \ac{Ref-100} are shown in black, where dashed lines neglect ISM dust, \changes{thin solid lines indicate the \citet{Trayford15} (T15) colours}, thick solid lines include ISM dust modelled using \skirt. Red lines (bottom panel only) are galaxies from the higher resolution simulation \ac{Recal-25}, processed with \skirt. \textit{Blue} lines show the observed distributions of \citet{Taylor14}.}
 \label{fig:coldists}
\end{figure}

In this section we compare colours of \eagle\ galaxies to \gama\ data, as well as to the fiducial model of T15 (their GD+O model, which we will refer to as the \lq T15\rq\ model below). We recall that the models we discuss have two sources of dust, that associated with birth clouds which is modelled by \starburst, and ISM dust taking into account using \skirt\ radiative transfer.
	We will sometimes refer to models without ISM dust as \lq intrinsic\rq\ colours and to these galaxies as \lq dust free\rq\, but note that this only refers to ISM dust, not the dust associated with the \starburst\ source model. 

\subsection{Comparison with observations}
\label{sec:obs}
\subsubsection{Colour distribution at a given stellar mass}
The distribution of rest-frame galaxy $g-r$ colours at $z=0.1$, in narrow (0.3~dex) non-contiguous stellar mass bins, showing both $z=0.1$ \eagle{} galaxies and \gama{} data, is plotted in Fig.~\ref{fig:coldists} \changes{(which can be compared to the simpler dust-screen model of T15, thin solid lines)}. Stepped lines represent the \eagle{} colour histograms, using black to denote simulation \ac{Ref-100} and red to denote simulation \ac{Recal-25}. \changes{These are either thin dashed to indicate colours without ISM dust, thin solid to represent the GD+O model of T15, or thick solid representing those obtained using radiative transfer with \skirt.} Continuous blue lines correspond to rest-frame (volume-limited) \gama{} galaxy colours without dust correction from \citet{Taylor14}. Stellar masses for the \gama{} galaxies, inferred through SED fitting, are taken from \cite{Taylor11}. All distributions are normalised to unit area. Here we compare the \skirt{} to observed distributions, additionally comparing to the T15 model in sections \ref{sec:cmd}-\ref{sec:pass}.

Comparing dashed to solid lines in the top panel demonstrates that dust reddening in massive ($10^{11.2} < {M_\star}/{\rm M_\odot} < 10^{11.5}$) blue galaxies is significant, with blue galaxies redder in \skirt{} compared to intrinsic colours by 0.1-0.2~mag in $g-r$,  changing the bi-modal ISM dust-free colour distribution to a single red peak at $g-r\approx 0.75$. The intrinsically blue colours of massive \eagle{} galaxies is caused by relatively low levels of residual star formation, not completely suppressed by the AGN feedback. The dust content of these star forming regions is high, however, leading to significant dust-reddening when processed with \skirt. At these masses, about half of the galaxies on the \skirt{} red sequence are dust reddened from an intrinsically bluer colour\footnote{\changes{Note that while the dust attenuation in \eagle{} appears to be systematically lower than observed for edge-on galaxies (see section \ref{sec:bbA}), increasing attenuation would not necessarily lead to the \eagle{} red sequence position shifting to even redder colours. Unlike in a screen model where extreme reddening is possible, in our \skirt{} modelling galaxies with high dust content have colours that saturate to that of old stellar populations, as these populations are preferentially unobscured by dust}. If the dust clouds are made optically thick, the galaxy photometry is essentially that of the unobscured population. More realistic attenuation values might, however, lead to more galaxies appearing as members of the red sequence.}.

Dust reddening also affects the $g-r$ colour of galaxies with masses $10^{10.5} < {M_\star}/{\rm M_\odot} < 10^{10.8}$ strongly (second panel from top), shifting blue galaxies to higher $g-r$ by $\approx  0.1 \; {\rm mag}$, to $g-r \approx 0.6$, and changing the bi-modal colour distribution into a single red peak with a tail to bluer colours. This blue tail, due to galaxies with more moderate reddening, hints that the intrinsic colour distribution is in fact bi-modal. At these masses, about a third of the \textit{`green valley'} population with $g-r \approx 0.65$ comprises dust-reddened galaxies. The remaining galaxies have intrinsic colours that puts them in the green valley, and are typically transitioning between the blue and red populations, as discussed in detail by \citet{Trayford16}.

The second-lowest mass bin ($10^{9.8} < {M_\star}/{\rm M_\odot} < 10^{10.1}$) again contains a population of strongly-reddened galaxies. A distinct bi-modality remains after reddening, with the red peak stronger than the blue peak, opposite to the case of intrinsic colours. Intrinsically blue galaxies appear less attenuated on average, with the blue peak shifted by only $\approx 0.05$ mag relative to the ISM dust-free photometry, to $g-r \approx 0.5$. The \textit{`green valley'} population is also boosted relative to the ISM dust-free photometry. Recalling Fig.~\ref{fig:orscat}, we see that the dust-boosted red and green galaxy populations produced by \skirt{} have a tail to significantly bluer colours. The tail consists of galaxies that have little or no ISM dust as well as dusty galaxies seen nearly face-on with ISM dust-free colours typical of the star-forming population.

At the lowest stellar masses, $10^{8.7} < {M_\star}/{\rm M_\odot} < 10^{9}$ (bottom panel), \eagle{} galaxies show very little reddening when processed with \skirt{}. Indeed, comparing the ISM dust-free and \skirt{} $g-r$ distributions separately for the \ac{Ref-100} and \ac{Recal-25} simulations shows that dust effects are minimal. 

In the most massive bin, observed colours from \gama{} conform to a tight red sequence centred at $g-r \approx 0.7$. The \skirt{} distribution is similar but shifted by $\approx 0.05$~mag to the blue. The median stellar metallicity of \eagle{} galaxies agrees well  with the observationally inferred values (S15), and the stars in these galaxies are generally old. It is therefore somewhat surprising that the simulated and observed colours do not agree better, since reddening is not important for these galaxies anyway (either in our model, or in the \gama\ data). \cite{Trayford16} showed that the metallicity distribution of star particles in \eagle{} galaxies is nearly exponential, and it is the lower $Z$ particles that make \eagle{} galaxies bluer than observed. A possible reason for the discrepant colours is thus that massive \eagle{} galaxies have too low metallicities, even though the mass-weighted simulation metallicity agrees well with the luminosity-weighted observed metallicity, see \citet{Trayford16} for more discussion.

The second most massive bin shows striking consistency between \skirt{} and observed colours. The agreement with the data is in fact superior to that obtained with the dust-screen model of \ac{T15}.  In particular, the relative fraction of red and blue galaxies is much closer to the observed ratio when using the \skirt{}. The reason for this is explored further below.

The second lowest mass bin shows similarly good agreement with the observed distribution, with \skirt{} colours systematically shifted to somewhat redder values ($\lesssim 0.05$~mag). Again, the colours conform better to observation than those presented by \ac{T15}, with the latter's dust-reddened colours in fact close to the intrinsic \eagle{} colours.

Finally, in the lowest mass bin, the \ac{Ref-100} colours show poor agreement with observation. \citet{Furlong15} showed that at these lower galaxy masses, numerical effects and poor sampling in \eagle{} cause the star formation rates to be too low and too many galaxies to be quiescent. We therefore also show the colours for the higher-resolution \ac{Recal-25} simulation (red) These. agree well with \gama. The \skirt{} colours of each of the simulations are very similar to those of \ac{T15}, which is understandable as both are very close to the intrinsic colours (i.e. are subject to minimal reddening) in this mass range.

\subsubsection{Colour-mass diagram}
\label{sec:cmd}

\begin{figure}
 \includegraphics[width=0.48\textwidth]{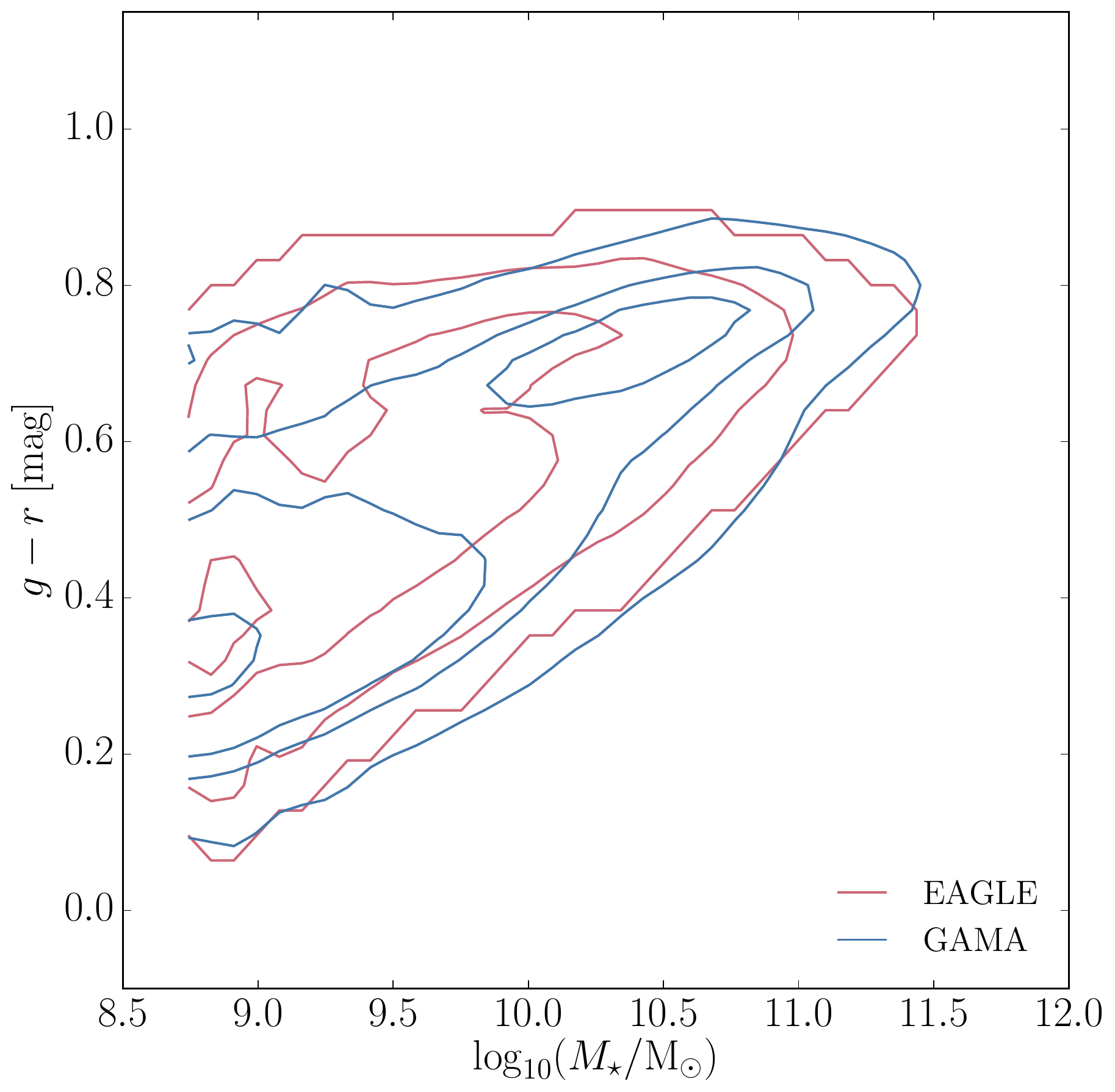}  
 
 \caption{Rest-frame $z=0.1$ SDSS $g-r$ colour as a function of galaxy stellar mass, for qualitative comparison of \eagle{} and \gama{}. \textit{Red Contours} are for randomly-oriented \eagle{} galaxies, processed using \skirt{}. The \eagle\ sample combines galaxies taken from simulation \ac{Ref-100} at high mass with galaxies from simulation \ac{Recal-25} at lower stellar mass, to mitigate the poorer resolution of \ac{Ref-100} (see text). \textit{Blue contours} show the observed distribution taken from \citet{Taylor14}. The contour levels from low to high exclude 3, 18, 48, and 93 percent of the galaxies in each sample.}
\label{fig:cmd}
\end{figure}

The $g-r$ colour versus stellar mass distribution of \eagle\ galaxies is compared to that obtained from \gama\ in Fig.~\ref{fig:cmd}. As explained above, the star-formation rate and hence colour of low-mass galaxies in \ac{Ref-100} is not well-resolved numerically \citep{Furlong15}. We therefore combine the more massive \eagle{} galaxies from \ac{Ref-100} with the low mass galaxies from the higher-resolution \ac{Recal-25} simulation, crossfading between the two in the mass range $10^{10} {\rm M_\odot}$ to $10^{9} {\rm M_\odot}$ as in \ac{T15}. The observed \gama{} contours are based on the analysis of \citet{Taylor14}. Note that the crossfading between two simulations (performed as described in T15) extends the mass range over which galaxies are well resolved, but also introduces some inconsistencies, for example the different simulation volumes probe different environments. As such, this is intended only to provide a qualitative comparison with the observations, with quantitative comparison facilitated by Fig.~\ref{fig:coldists}.

Similar to Fig.~\ref{fig:coldists}, we see that the \eagle{} colours obtained with \skirt{} generally show good agreement with the observed distribution. The blue cloud and red sequence populations in \eagle{} appear to be in approximately the observed position and contain a roughly similar share of the galaxy population across the mass range.  The green valley population is enhanced relative to the \ac{T15} photometry, in better agreement with \gama{} data. The inconsistent surplus of blue galaxies at the high-mass end, $M_\star{} \gtrapprox 10^{10.5}{\rm M}_\odot$, is also largely suppressed with respect to \ac{T15}. This is attributable to the more representative treatment of the spatial distribution of the dust in \skirt{}, with the ISM dust enshrouding young stars, rather than being distributed in a diffuse galaxy-sized disc as assumed by the screen model of \ac{T15}. 

However, there are still some notable discrepancies between \eagle{} and \gama{}. Across all masses the red sequence in \eagle{} is flatter than observed, with slightly bluer colours at high mass and redder colours at low mass. This is consistent with the findings of \ac{T15}, and is symptomatic of the fact that the metallicity of \eagle{} galaxies does not increase with $M_\star$ as steeply as observed. This is at least in part due to insufficient numerical resolution, as shown by S15 (their Fig.~12). A moderate surplus of blue galaxies relative to the observations can also still be seen between $\sim 10^{10} {\rm M_\odot}$ and $\sim 10^{10.5} {\rm M_\odot}$, likely due to a combination of lower passive fractions and lower typical dust attenuation in the \eagle{} galaxies relative to those observed. Differences between the observed and simulated stellar mass functions also contribute to discrepancy: the \eagle{} simulation has a deficiency of galaxies at the knee of the mass function ($M_\star \sim 10^{10.5} {\rm M_\odot}$, S15), such that the contours are skewed to lower masses than in the \gama{} distribution.

\subsection{Comparison of \skirt{} colours to dust-screen models}
\label{sec:fit}
We now turn to comparing the \ac{T15} photometry with that generated using \skirt{}. The screen model presented by \ac{T15} has several parameters, notably $\tau_{\rm BC}$, the dust optical-depth in the birth-clouds of stars, $\tau_{\rm ISM}$, the dust optical depth in the ISM, and $q=b/a$, the axial ratio of the oblate spheroid within which the ISM dust is assumed to be distributed\footnote{Using the standard nomenclature for $a$ and $b$ denoting the primary and secondary axes respectively. Note that in T15 this is erroneously referred to as $q=a/b$.}. The fiducial values of these parameters were informed by observational studies, but they do not necessarily reflect the ISM distribution in \eagle{}. To test whether the radiative transfer photometry is better reproduced with  a different parametrisation of the \ac{T15} model, we fit the \ac{T15} model to the \skirt{} results. The parameter fits are obtained using Bayesian inference, where a \ac{MCMC} method is used to find the maximum-likelihood parametrisation. We simultaneously find the \ac{ML} values of $\tau_{\rm ISM}$ and $q$,  enforcing the constraint that $\tau_{\rm BC} = 2 \tau_{\rm ISM}$ as in the fiducial model of \citet{CF00}. The application of this constraint and full details of the MCMC procedure are given in Appendix \ref{sec:mcmc}. 

The \ac{ML} parameters from fitting the fiducial \skirt{} model using the free parameters of \ac{T15} are given in Table \ref{tab:params}. We find that the \ac{ML} $\tau_{\rm ISM}$ and $\tau_{\rm BC}$ values needed to describe the \skirt{} results for \ac{Ref-100} are $\approx 10\%$ lower than the fiducial values of \ac{T15}. This offset is small, implying similar typical attenuation values in both models. Because the values of $\tau_{\rm ISM}$ and $\tau_{\rm BC}$ used in \ac{T15} come from the original fitting by \citet{CF00} of the SDSS observations, it is encouraging that they are recovered independently by fitting the \skirt{} results: this suggests that our \skirt{} model yields realistic average optical dust attenuations for galaxies of a given metallicity and gas fraction, and also that the relative stellar and ISM geometries and our dust mapping are reasonable.     

However, the $q$ parameter is significantly higher for the \skirt{} model, implying less inclination dependence and lower edge-on attenuation in the model, as is indeed seen in Fig.~\ref{fig:orient}. This most likely reflects the artificially `puffed-up' ISM in simulated galaxies. We repeat that this higher disc thickness is partly set by numerical resolution, but is mostly due to the assumed temperature-density relation for star-forming (disc) gas in the \eagle{} sub-grid model.

Overall, we find that the fiducial \skirt{} model, which is based on physical modelling of the \eagle{} galaxies, produces typical dust attenuations at optical to NIR wavelengths similar to the published model of \ac{T15}. Although the $f_{\rm dust}$ and $f_{\rm PDR}$ parameters of the model were chosen to reproduce FIR observables by \cite{Camps16}, this result is in fact independent of that calibration. To demonstrate this, we apply our \ac{ML} fitting procedure to an `uncalibrated' \skirt{} model; produced using the default literature values of $f_{\rm dust}$ and $f_{\rm PDR}$. Similar levels of agreement between the fiducial $\tau_{\rm ISM}$, $\tau_{\rm BC}$ and $q$ values of \ac{T15} are recovered, as shown in Appendix \ref{sec:uncal}. Given these findings, when comparing models we do not modify the parameters of the \skirt{} and screen model from the published values of \citet{Camps16} and \ac{T15}. The fiducial \skirt{} model photometry and that of the \ac{T15} model are compared below.

\begin{table}
\begin{center}
\caption{Maximum likelihood parameters for the model of \ac{T15} that best describes the fiducial \skirt{} photometry. These values are derived using an \ac{MCMC} approach, as detailed in Appendix \ref{sec:mcmc}.}
\label{tab:params}
\begin{tabular}{llll}
\hline
Model & $\tau_{\rm ISM}$ & $\tau_{\rm BC}$ & $q$ \\
\hline

\ac{T15} values &  0.33 & 0.67 & 0.2\\
\ac{Ref-100} ML &  0.301 & 0.602 & 0.556\\

\hline
\end{tabular}
\end{center}
\end{table}
\subsubsection{Colour-colour distributions}
\label{sec:colcol}

\begin{figure*}

\includegraphics[width=\textwidth]{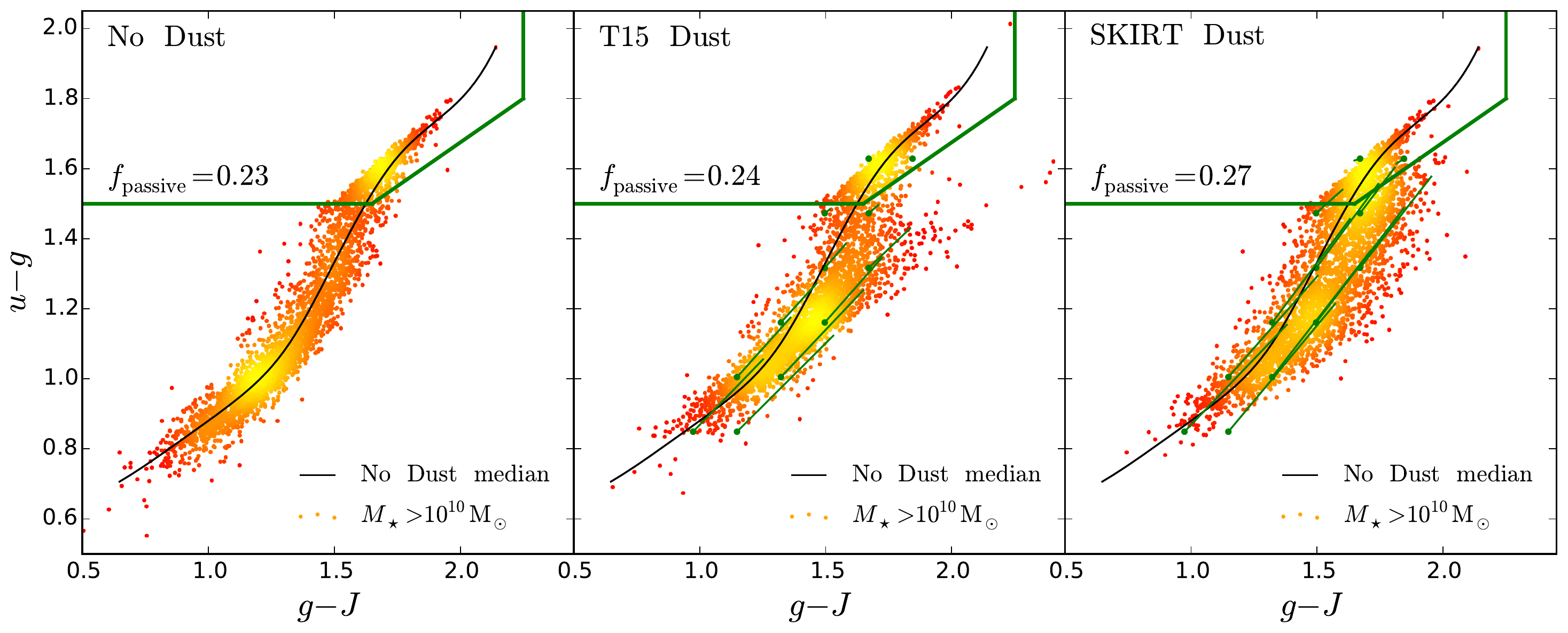}  

\caption{Rest-frame $z=0.1$ $ugJ$ colour-colour diagrams for different photometric models. \textit{Points} are \eagle\ galaxies from simulation \ac{Ref-100} with stellar masses $>10^{10}{\rm M_\odot}$, shaded from red to yellow to indicate local point density. The dust-free photometry of \ac{T15}, the \ac{ML} parametrisation of the dust-screen model of \ac{T15}, and the \skirt{} photometry including ISM dust and \Hii{} regions are shown in the left, middle and right panels respectively. \textit{Thick, green lines} show the passive galaxy cut advocated by \citet{Schawinski14}, the corresponding passive fraction of \eagle\ galaxies using this cut is indicated in each panel. \textit{Black curves} indicate the median $u-g$ values as a function of $g-J$ for the dust-free photometry, these are repeated in each panel to guide the eye. \textit{Thin, green lines} indicate the mean reddening vectors of \eagle{} galaxies in the dusty models relative to the dust-free model. These emanate from green points, specifying the centre of each bin from which a vector is computed. Note that vectors point from bottom-left to top-right. The \skirt{} model gives  higher passive fractions than the intrinsic and \ac{T15} dust photometries, indicating significant pollution of the passive region by star forming galaxies.}
\label{fig:colcol}
\end{figure*}

To further explore the effects of dust-reddening, we compare colour-colour distributions for dust-free photometry of \eagle{} galaxies from T15, the fiducial dust-screen model of \ac{T15} and the full \skirt{} modelling in Fig.~\ref{fig:colcol}. We plot rest-frame $u-g$ against $g-J$ colours, analogous to the $UVJ$ diagram used by \citet{Williams09}, to separate actively star forming but dust reddened galaxies from intrinsically red and passive galaxies. Data points are shaded from red to yellow by point density, to indicate how galaxies are distributed. Thin green lines indicate the mean reddening vectors for the two dust models relative to the intrinsic photometry of \ac{T15}, in regular bins of $ugJ$ intrinsic colour. These are only plotted for bins with $> 10$ galaxies. The colour-colour bin centres from which the vectors emanate are highlighted by green points. 
	We over-plot the $ugJ$ cut used to separate active from passive galaxies by \citet{Schawinski14} in bold green. The passive fractions of \eagle{} galaxies, as inferred from applying this cut, are indicated in each panel.

The three photometric models produce qualitatively similar distributions, but with some important differences. The dust-free model in the left panel exhibits two well-defined peaks, a `blue peak' at $(u-g,\, g-J) \approx (1.2,1.0)$ and a `red peak' at $\approx (1.6,1.6)$. The intrinsic distribution is relatively tight for galaxies with $M_\star > 10^{10} {\rm M}_\odot$, with $\lesssim 0.2$~mag scatter in $u-g$ for a given $g-J$ colour. The passive fraction is $f_{\rm passive} = 0.23$ for galaxies with  $M_\star > 10^{10} M_\odot$.

The middle panel, showing galaxy colours produced by the fiducial dust screen model of \ac{T15}, exhibits a similar distribution in the passive region. From the lack of visible lines at $u-g > 1.4$ , we see that there is minimal reddening of galaxies into or within this region. The recovered passive fraction of $f_{\rm passive} = 0.24$ reveals that indeed the passive region is $< 5\%$ polluted by the galaxies defined as active in the dust-free (left) panel. The active region galaxies, however, exhibit more variation. While the blue peak is similarly well defined relative to the dust-free colours, the position of the peak is shifted to slightly redder colours by $\sim 0.1$~mag and is broadened by scatter to redder colours, with many more star-forming galaxies having $g-J > 1.5$. The mean reddening vectors are small relative to the distance between the peak and the most extremely attenuated active galaxies, which have colours $g-J \gtrsim 2$.  

The right panel, showing the \skirt{} model colours, reveals some consistent features. The passive region galaxies reveal a similar distribution as in the left and middle panels, with a red peak in a very similar position. The active region galaxies also occupy a similar region of the $ugJ$ plane to the \ac{T15} (middle) panel, with a blue peak shifted to slightly redder colours relative to the dust-free model in the left panel. However, in detail there are some notable differences. The blue peak is significantly depleted relative to the other panels, with the red peak exhibiting a tail to bluer colours. Though the \skirt{} model does not possess the extremely reddened galaxies of the \ac{T15} (middle) panel, the magnitude of the mean reddening vectors are generally larger across the intrinsic distribution and show less reduction with redder intrinsic $u-g$ colour relative to \ac{T15}. We also see an enhanced `green' population of galaxies with intermediate colours $1.25 < u-g \lesssim 1.5$ in the active region. The enhanced average reddening across the distribution also results in a higher recovered passive fraction of $f_{\rm passive} = 0.27$. This indicates that the passive region has a $\approx 15\%$ pollution by galaxies defined as active in the dust-free (left) panel. The reddening of galaxies from the blue peak to the red peak in the $ugJ$ diagram corresponds to the significant boost (depletion) of the red (blue) sequence population for the \skirt{} photometry relative to the dust-free colours seen in Fig. \ref{fig:coldists}.

The differences between the \ac{T15} and \skirt{} panels in Fig. \ref{fig:colcol} can be attributed to the nature of the dust modelling. The \ac{T15} reddening vector is close to parallel with the sloped boundary of the passive region, as illustrated by a tail of extremely reddened active galaxies with $g-J > 2$. This is unsurprising, as the screen model of \citet{Calzetti00} is used by \citet{Schawinski14} to define the boundary between active and passive galaxies. This may explain why few dusty galaxies move into the passive region when applying the \ac{T15} screen model. The \skirt{} model, which exploits the 3D distribution of dust around stars, yields generally steeper reddening vectors of higher magnitude, both of which contribute to moving dusty star forming galaxies into the region were galaxies are deemed to be passive when using the \citet{Schawinski14} colour-colour cut. This is because nascent stellar populations embedded in dense ISM are effectively shielded in the \skirt{} model, leading to more active galaxies masquerading as passive. Again, the fraction of galaxies misclassified as passive could be underestimated due to the lack of highly attenuated edge-on \eagle{} galaxies, attributable to the artificially `puffed-up' ISM in the simulation. 

\subsubsection{Passive fractions}
\label{sec:pass}
\begin{figure}
 \includegraphics[width=0.48\textwidth]{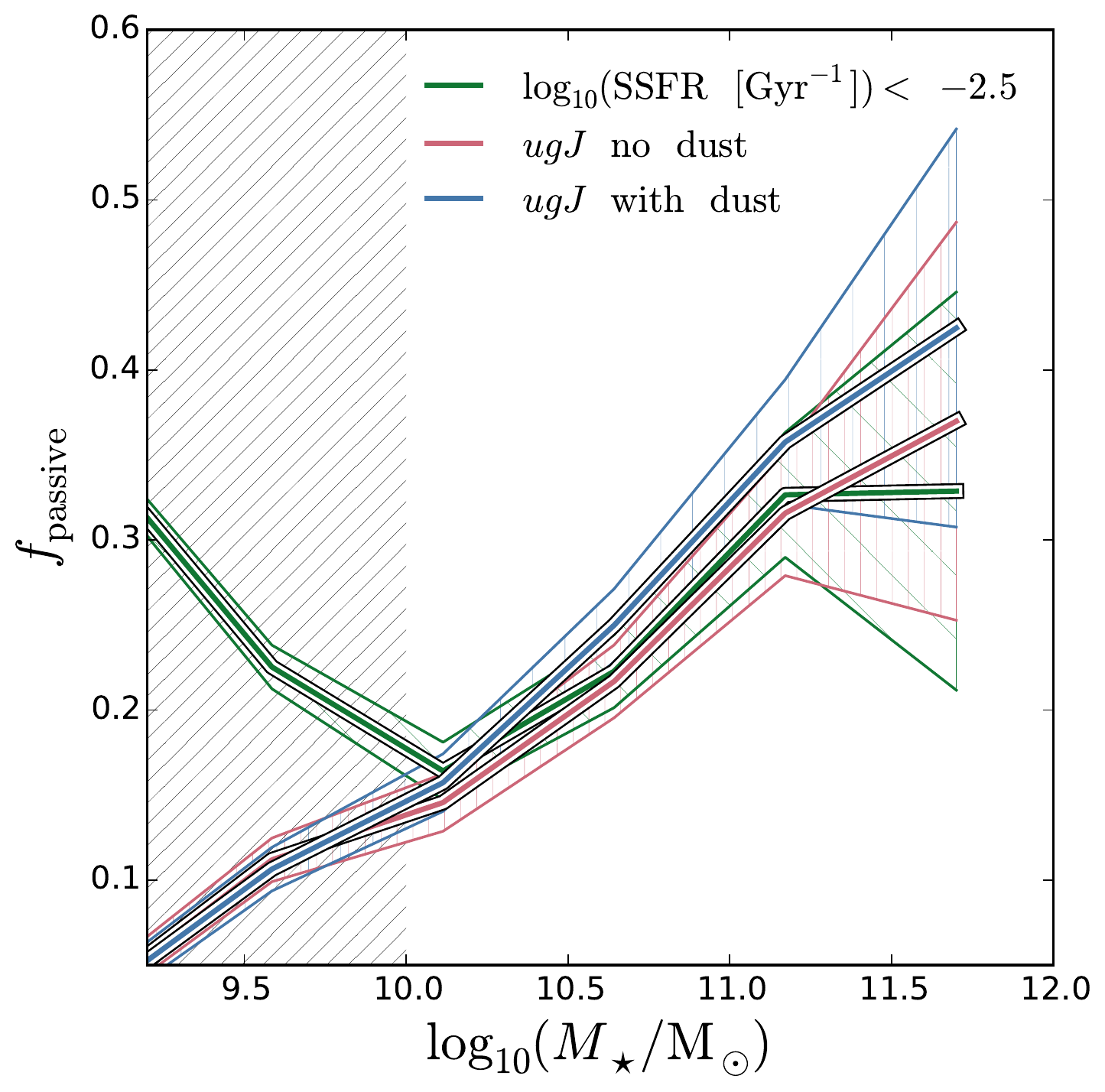}  
\caption{Passive galaxy fractions for the \ac{Ref-100} \eagle{} simulation in bins of $M_\star$, equally spaced in $\log_{10}{M_\star/{\rm M_\odot}}$. The \textit{green} line shows the values obtained using a $\dot M_\star/M_\star$ cut at $10^{-2.5}$ Gyr$^{-1}$. \textit{Red} and \textit{blue} lines show the fractions obtained using the photometry without and with ISM dust, respectively. \textit{Coloured, hatched regions} indicate the uncertainty for each line, corresponding to the fractional Poisson error on the number of galaxies in each bin. The diagonally hatched region $M_\star < 10^{10} {\rm M_\odot}$ indicates where the true passive fraction (green line) decreases with stellar mass, due to resolution and volume effects. We see that the discrepancy between the passive fractions obtained using intrinsic and dust attenuated photometry increases with stellar mass above $10^{10} {\rm M_\odot}$.}
\label{fig:passv}
\end{figure}

We use the colour-colour cut of \citet{Schawinski14} from Fig. \ref{fig:colcol} to calculate passive fractions as a function of mass for both \skirt{} and ISM dust-free photometry. The results are shown in Fig.~\ref{fig:passv}, where we compare to the passive fractions calculated directly from the aperture star formation rates \citep[see S15,][]{Furlong15}. The value of the specific star formation rate ${\rm sSFR}\equiv \dot M_\star/M_\star$ below which a galaxy is deemed passive is somewhat arbitrary. As we are using the \citet{Schawinski14} cut in $ugJ$ for the colour cut, we use a value of $10^{-2.5}$ Gyr$^{-1}$ which yields good agreement with the dust-free photometric estimates for numerically well-resolved galaxies ($M_\star \gtrapprox 10^{10} {\rm M_\odot}$). Note that S15 and \citet{Furlong15} used a higher value of $10^{-2}$ Gyr$^{-1}$.

The differences between the photometric passive fraction estimates with and without ISM dust become apparent at masses $M_\star > 10^{9.5} {\rm M_\odot}$. For better-resolved galaxies, $M_\star > 10^{10} {\rm M_\odot}$, the passive fraction obtained when including dust reddening lies $\approx$ 0.1~dex above the value estimated using intrinsic colours or calculated using the sSFR cut. This offset suggests that, using colours alone, $\approx 15 \%$ of the apparently passive population may be misclassified active galaxies for stellar masses $M_\star > 10^{10} {\rm M_\odot}$. This fraction could be higher still if our mock photometry had levels of attenuation closer to observation (see Fig. \ref{fig:orient}). It is important to note that the $ugJ$ selection used here also represents a particularly stringent passive cut, hence why we compare to a specific star formation rate cut of $10^{-2.5}$ Gyr$^{-1}$ rather than one at $10^{-2}$ Gyr$^{-1}$. We find that by relaxing the $ugJ$ cut (a $-0.1$ shift in $u-g$ to approximate a $10^{-2}$ Gyr$^{-1}$ selection) leads to a higher proportion of active galaxies being misclassified as passive due to dust effects ($\approx38\%$). It seems that the use of a stringent cut minimises misclassification of passive galaxies to the $15\%$ level. 

There is a striking divergence between the sSFR and photometrically defined passive fractions for galaxies with $M_\star < 10^{10} {\rm M_\odot}$ apparent in Fig.~\ref{fig:passv}. This occurs in a region where star formation rates are subject to resolution and volume effects, however it is the metallicities of these galaxies that is likely driving this discrepancy. Indeed, a (passive) $10$~Gyr old stellar population with a metallicity of $Z=0.4 Z_\odot$ will lie {\em below} the $u-g = 1.5$ threshold of \citet{Schawinski14} in the \citet{bc03} model - and hence is too blue to be classified as passive. As the SDSS sample of \citet{Schawinski14} is dominated by galaxies of mass $\gtrsim 10^{10}$ M$_\odot$, it is likely that only a few of the observed galaxies would be affected by this.

%% file: Indices.tex
\section{Spectral Indices}
\label{sec:lines}
\begin{figure*}
 \includegraphics[width=0.99\textwidth]{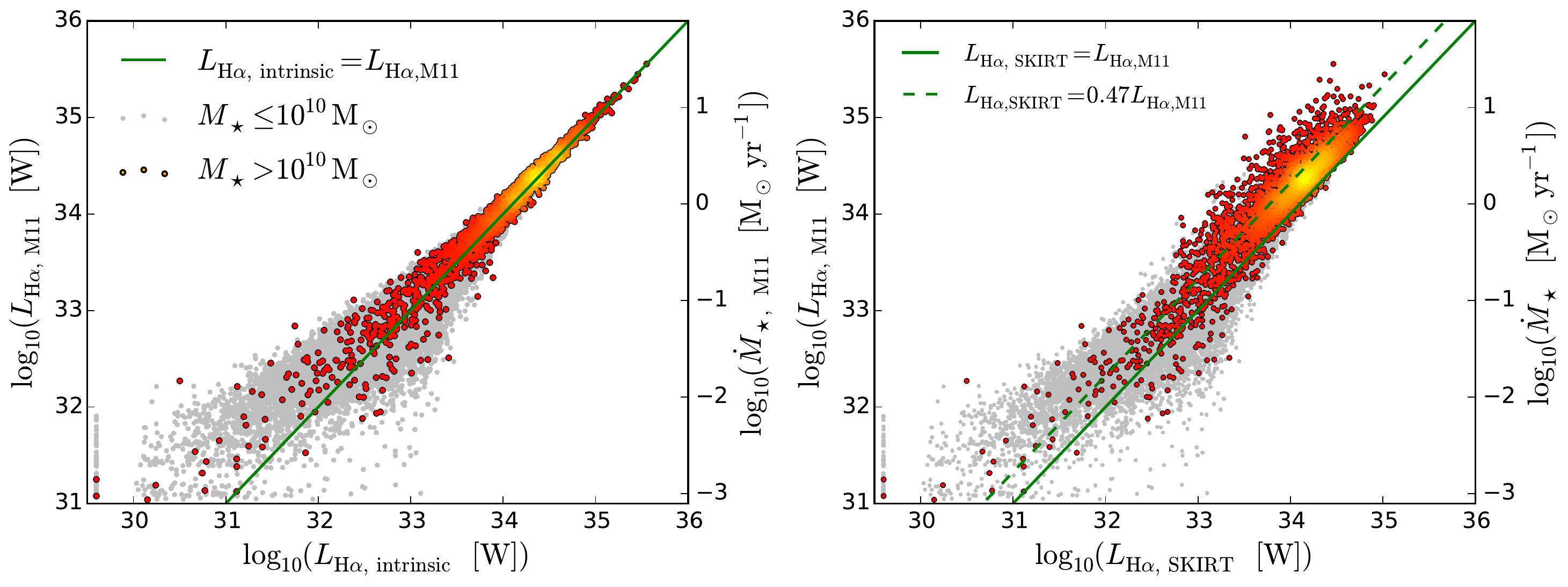}  
 \caption{\Ha{} fluxes of \eagle\ galaxies. \textit{Left panel}: ISM dust-free \Ha{} flux from the \starburst\ model compared to \Ha{} flux computed from Eq.~(\ref{eq:sfr}), where $\dot M_\star$ is the instantaneous star formation rate within a 30~kpc aperture; the value of $\dot M_\star$ is indicated on the right $y$-axis. Coloured dots are individual \eagle\ galaxies with $M_{\star} > 10^{10} \; {\rm M_\odot}$ with colour a measure of the point density of galaxies in the plot; \textit{grey points} represent galaxies of lower stellar mass. Galaxies where \Ha{} is not detected are clipped to $\log_{10}(L_{\rm H\alpha} \; {\rm [W]}) = 29.8$; the \textit{green line} indicates the 1:1 relation to guide the eye. \textit{Right panel}: same as left panel, but showing the \Ha{} fluxes computed using \skirt\, {\em i.e.} including ISM dust; the \textit{dashed} line is the best linear fit to the coloured points (excluding undetected galaxies), and represents the average dust attenuation factor. $L_{{\rm H}\alpha}$ values measured from ISM dust-free spectra generally recover those given by equation \ref{eq:sfr} very well, as expected, with a relatively spread at low values ascribed to shot noise in the sampling of \Hii{} regions. The dust attenuated values show an offset and large ($\sim 1$ dex) scatter. }
 \label{fig:hacomp}
 
\end{figure*}

\begin{figure}
 \includegraphics[width=0.5\textwidth]{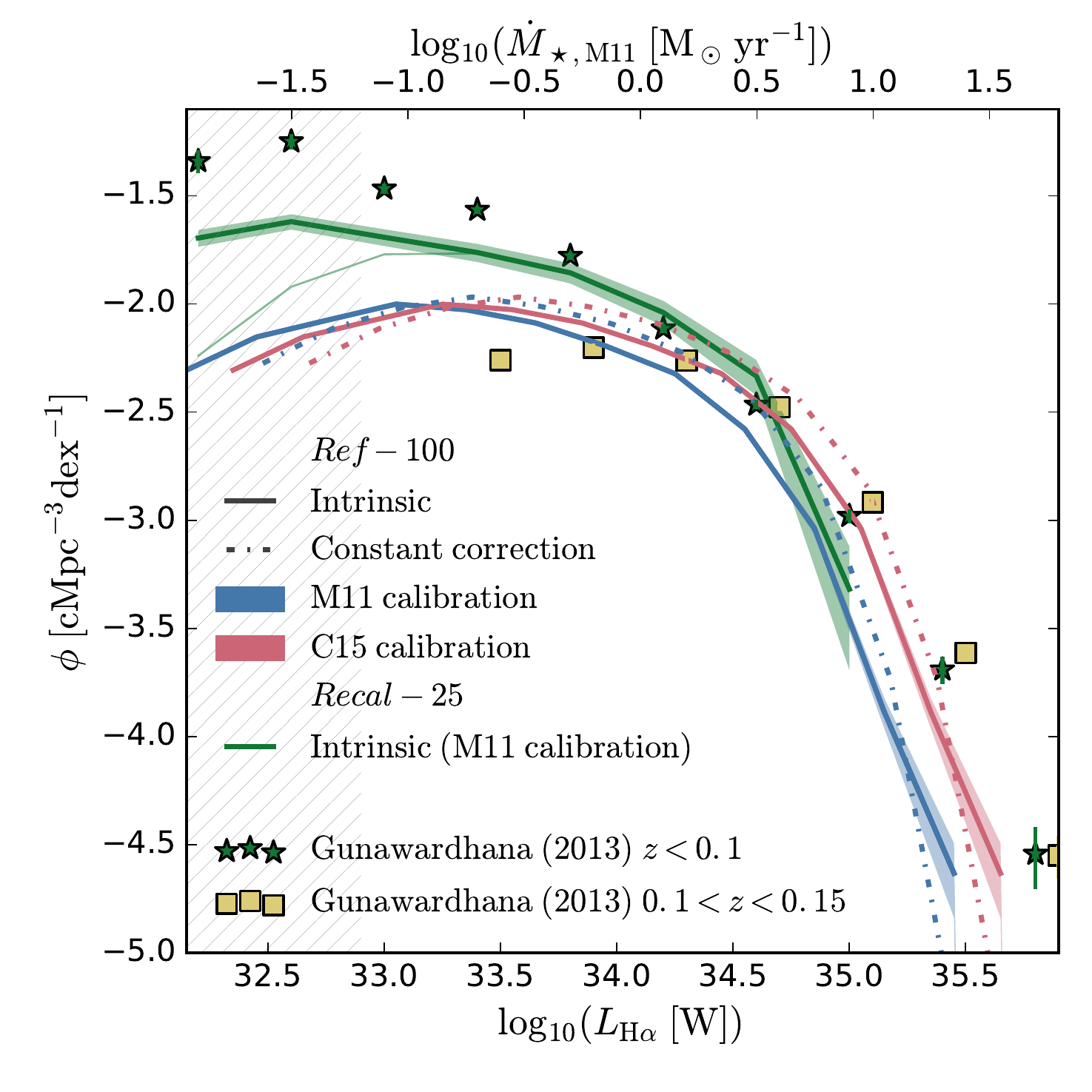}  
 
 \caption{Redshift $z=0.1$ dust corrected \Ha{} luminosity functions of \eagle\ galaxies compared to observations. \textit{Solid coloured lines} are the ISM dust-free \eagle\ LFs for simulation \ac{Ref-100} for randomly oriented \eagle\ galaxies with $M_\star \geq 1.81\times 10^8$ M$_\odot$ using the \citealt{Murphy11} and \citealt{Cheng15} (see text) ratio of  $\dot M_\star/L_{\rm H\alpha}$, respectively, with the shading showing the Poisson error range. The \textit{dot-dashed} curves are the corresponding LFs computed using \skirt, dust corrected using the best-fit constant dust correction to the attenuated spectra
 	($A$, see Fig.~\ref{fig:hacomp}).\textit{Green curves} are the ISM dust free LF for simulation \ac{Recal-25} using the \citealt{Murphy11} ratio of $\dot M_\star/L_{\rm H\alpha{}}$, for galaxies with $M_\star \geq 1.81\times 10^8$ M$_\odot$ and $M_\star \geq 2.26\times 10^7$ M$_\odot$ shown as \textit{thin} and \textit{thick green lines}, respectively. The \textit{hatched region} marks where these deviate by $> 10\%$, indicating luminosities for which incompleteness is important for the $M_\star \geq 1.81\times 10^8$ mass limit. Observational data is taken from \citet{Gunawardhana13}, who use the Balmer decrement dust-corrected \gama{} sample, for redshift ranges $z<0.1$ (star symbols) and $0.1<z<0.15$ (squares). Error bars indicate the statistical uncertainty on observed data points, but are too small to be seen for all but the highest luminosities. The simulated LF using the \citealt{Cheng15} calibration is closer to the observed measurements; the intrinsic LFs possess a broader knee than when a constant dust correction is applied.}
  \label{fig:haLF}
\end{figure}

We apply \skirt{} modelling to compute two spectral indices that are often used as proxies for star formation activity, the \Ha{} luminosity, $L_{{\rm H}\alpha}$ \citep[e.g.][]{Kennicutt98}, and the strength of the 4000~\AA{} break, D4000 \citep[e.g.][]{Kauffmann03}. We compare the indices for \eagle{} galaxies to both theoretical and observed values below, concentrating on the effects of dust, and test the correlation of these indices with the intrinsic star formation rate of \eagle\ galaxies.

\subsection{Dust effects on the \Ha{} flux}
\label{sec:dusteff}

The \Ha{} luminosity of a galaxy is thought to be a reliable proxy for its star formation rate, basically because the recombining gas that emits the \Ha{} photons is thought to be photo-ionised by massive ($>10{\rm M}_\odot$) and hence recently ($<20~$~Myr) formed stars \citep[e.g.][]{Kennicutt98}. However, as we have seen, such star-forming regions are typically dust obscured (not just by the dust in \Hii{} birth-clouds, but by ISM dust as well), and the measured flux therefore needs to be corrected for dust. Here we compare the intrinsic  $L_{{\rm H}\alpha}$ values, emanating only from \Hii{} regions in our model, to those obtained using empirical corrections to observed spectra. When measuring emission lines, we only use the \Hii{} region component of the source spectra (the blue line in Fig.~\ref{fig:egspec}) to minimise the contribution (either in emission or more likely in absorption) from the stellar continuum (a correction that needs to be applied to observational data). This enables us to isolate the effects of dust on inferred line fluxes, which we can investigate with \skirt, from those caused by continuum fitting.

The relation between \Ha{} line flux and star formation rate advocated by \citet{Murphy11} is \begin{equation} 
\label{eq:sfr}
L_{{\rm H}\alpha, \;{\rm M11}}\ = \left( \frac{\dot{M}_\star}{\rm 5.37 \times 10^{-42} \; M_\odot \; yr^{-1}} \right) \; {\rm erg \; s^{-1}}\,,
\end{equation} 
and is a recalibration of the relation from \citet{Kennicutt98}. The recalibration accounts for the different IMFs assumed by \citet{Kennicutt98} (Salpeter) compared to \citet{Murphy11} (Kroupa), and is thus consistent with the (Kroupa) IMF of the \mappings{} models that we use for young stars. For a like-for-like comparison with $L_{{\rm H}\alpha}$ values obtained with \skirt{}, we use the instantaneous $\dot{M}_\star$ within a 30~pkpc aperture from the \eagle{} database \citep{McAlpine16}. This is calculated by summing over the $\dot{M}_\star$ values of gas particles within that aperture.

As a first test of our \Hii{} region prescription and subsequent measurement procedure for calculating $L_{{\rm H}\alpha}$, we compare the values obtained using the ISM dust-free spectra, $L_{{\rm H}\alpha, \;{\rm intrinsic}}$, to $L_{{\rm H}\alpha, \;{\rm M11}}$ in the left panel of Figure \ref{fig:hacomp}. For this comparison we over-plot galaxies with $\log_{10}(M_\star/{\rm M_\odot}) > 10$ coloured by the local point density, with lower-mass galaxies under-plotted in grey. The green line indicates the 1:1 relation to guide the eye. 

The tight 1:1 correlation between the two values for high SFR, $\dot{M}_\star \gtrsim 0.1 \; {\rm M_\odot \; yr^{-1}}$ ($L_{{\rm H}\alpha} \gtrsim 10^{33} {\rm W}$), is reassuring, implying that the resampling technique used to parameterise \Hii{} regions (see section \ref{sec:young}) reproduces the expected \Ha{} measurements. This good agreement is expected: the same population synthesis models and a similar treatment of nebular components are employed in both the \starburst{} model that we use, and the model used by \cite{Murphy11} that yields the conversion factor given in Eq.~\ref{eq:sfr}.

At low SFR, the $L_{{\rm H}\alpha, \;{\rm intrinsic}}/L_{{\rm H}\alpha, \;{\rm M11}} $ ratio exhibits large scatter. This is due to \Hii{} regions being sampled stochastically from the mass function of Eq.~(\ref{eq:mfunc}), yielding increased sampling noise for lower star formation rates. Galaxies without \Hii{} regions, and thus without ${\rm H}\alpha$ in our modelling, are plotted at $L_{{\rm H}\alpha, \;{\rm intrinsic}} = 10^{29.5} \; {\rm W}$ in the figure. 	

The right hand panel shows the corresponding plot for the ISM dust attenuated spectra (without attempting to correct the \Ha{} flux for dust correction). We over-plot the 1:1 relation with a constant dust attenuation factor of $A = 0.47$ (dotted green line, this corresponds to 0.82~mag extinction). This factor is the mean offset between the measured (dust attenuated) and intrinsic \Ha{} luminosities, as determined using a least-squares fit for \ac{Ref-100} galaxies of $\log_{10}(M_\star/{\rm M_\odot}) > 10$. \changes{The convergence of \Ha{} luminosities and $A$ values are tested for the other simulations listed in Table \ref{tab:sims} in Appendix \ref{sec:half_conv}}. The observed average attenuation in local galaxies is $\approx 0.4$ \citep{Kennicutt92} (1~mag of extinction), a factor of $\approx 1.2$ lower, but our value is still within the systematic uncertainty of the average extinction inferred from Balmer decrement measurements at $z \lesssim 0.5$ \citep[e.g.][]{Ly12}.  

\subsection{The \Ha{} luminosity function}
\label{sec:haLF}

Having shown that our implementation of mock \Ha{} emission lines yields line luminosities consistent with the underlying assumed SSP model, we now proceed to compute the \Ha{} luminosity function and compare it to data. It is common practise to apply a constant dust correction to observed \Ha{} fluxes obtained from narrow-band surveys to infer \lq intrinsic\rq, ISM dust-free, luminosities \citep[e.g.][]{Sobral13}. This yields an (intrinsic) \Ha{} line luminosity function with a Schechter form. However, the bright-end slope of the best-fit function that results from applying a constant dust correction is steeper than when the dust correction is performed using Balmer decrements \citep[e.g.][]{Gilbank10, Gunawardhana13}. This systematic difference is partially attributable to the star formation rate dependence of attenuation\footnote{Poor sampling of the strongest \Ha{} emitters, due to the small volume of narrowband surveys, may also contribute a steep bright-end.} \citep{Brinchmann04, Zahid13}. Because \skirt{} provides both the ISM dust-free and dust-attenuated SEDs, we can compare with the `true' dust correction for our simulated galaxies at different orientations and investigate this effect further.

A potential caveat for the realism of \Ha{} luminosities computed for \eagle{} galaxies is that the simulated specific star formation rates may be low compared to observations, both locally and at higher redshifts \citep{Furlong15}. This is in fact somewhat puzzling
	since the stellar mass functions do agree relatively well. The observed SFRs are inferred from various proxies, including
	emission lines, FIR and radio data. However, the calibration may rely on assumptions about the UV continuum, which depends on the assumed IMF and population synthesis model \citep[see][]{Kennicutt98}. Some recent studies suggest that stellar rotation and binary stars may contribute more to the UV continuum than previously thought, affecting this calibration \citep[e.g.][]{Hernandez13, Horiuchi13}. \changes{When instead considering the \Ha{} fluxes (including dust effects) the realism of galaxy attenuation and ISM also becomes important. Discrepancies in star formation rates and ISM attenuation can have degenerate effects on the \Ha{} measurements. It is therefore important to consider the intrinsic \Ha{} and dust attenuated \Ha{} measurements separately to isolate reasons for discrepancy or agreement.}
 
Recently, \citet{Cheng15} performed SED fitting of galaxies with SDSS and WISE photometry and obtained star formation rates for $z \approx 0.1$ star-forming galaxies. Their values are systematically lower by $\approx 0.2$~dex compared to
	previous work such as that of \cite{Kennicutt98}, and in better agreement with those predicted by \eagle{}. 
	If the lower $\dot M_\star$ values of \citet{Cheng15} point to a previous miscalibration of star formation rate indicators, then the $L_{{\rm H}\alpha}$
	values predicted by Eq.~(\ref{eq:sfr}), and thus present in the intrinsic \eagle{} SEDs (see Fig. \ref{fig:hacomp}), may be
	too low. To test the effect of such a change in normalisation when comparing to the \Ha{} luminosity function, we plot $L_{{\rm H}\alpha}$ both with
	and without a $+0.2$~dex shift, referring to this as the \citet{Cheng15} conversion\footnote{Note that \citet{Cheng15} do not explicitly advocate such a correction, rather, this conversion represents the case that the discrepancy they find exists due to previous miscalibration of absolute SFR.}. 

Figure \ref{fig:haLF} compares mock \Ha{} luminosity functions to data. Comparing thick and thin green lines, which use the same $\dot M_\star$ to \Ha{} flux conversion, but correspond to imposing a mass limit of $M_\star{} \geq 2.26\times 10^7$ M$_\odot$ (100 star particles at high resolution) and $M_\star \geq 1.81\times 10^8$ M$_\odot$, respectively, enables us to estimate the level of numerical convergence. A hatched region marks where these differ by more than 10$\%$, ie. where incompleteness effects become important for the higher mass cut. We therefore focus on the model \Ha{} luminosities above $10^{33}$~W. 

Up to luminosities of $\sim 10^{34}$~W, the solid green and solid blue curves (that both use Eq.~\ref{eq:sfr}) differ by $\approx 0.5$~dex (a factor of three), indicating that the \ac{Recal-25} simulation predicts significantly higher values of $\phi$ then \ac{Ref-100}, which is in better agreement with the data. As seen in Fig \ref{fig:hacomp}, below $L_{{\rm H}\alpha} \sim 10^{34} {\rm W}$ the poor sampling of \Hii{} regions contributes to this resolution effect. For the hatched region, $L_{{\rm H}\alpha} \lesssim 10^{32.9}$~W, the discrepancy is driven by incompleteness due the imposed mass limit for the \ac{Ref-100} simulation. However, at intermediate luminosities this is due to real differences between the properties of the galaxies in the high- and standard-resolution runs. Indeed, we recall from Fig.~\ref{fig:coldists} that these two simulations differ substantially in the lowest mass bin: \ac{Recal-25} galaxies tend to be intrinsically blue and star forming, whereas a significant fraction of \ac{Ref-100} galaxies are intrinsically red and passive. This contributes to the boost in the \ac{Recal-25} luminosity function at intermediate $L_{{\rm H}\alpha}$ ($ 10^{32.9} \lesssim L_{{\rm H}\alpha} \lesssim 10^{34}$~W), yielding better agreement with the observed luminosity functions. The higher number density of galaxies at these intermediate \Ha{} luminosities is due to the similar contribution of volume and resolution effects. Volume effects and convergence are discussed further in Appendix \ref{sec:half_conv}. The offset between the blue and green curves at lower luminosities is therefore, in part, a measure of numerical convergence. 
	 
We next compare solid lines (intrinsic luminosities) to dot-dashed lines (\skirt{} dust-attenuated luminosities corrected using a constant dust correction), for either blue or red lines. \changes{These agree well at the faint end ($L_{{\rm H}\alpha}\lessapprox 10^{34}$~W) where the luminosity functions are close to flat, but significant differences can be seen for brighter galaxies, with the number of bright sources higher at the knee of the constant dust correction luminosity function. A constant dust correction tends to overestimate the true level of dust attenuation around the knee, and underpredict it for the most \Ha{} bright galaxies, resulting in a steeper bright end slope.} Note that the constant dust correction we use is essentially indistinguishable from the common observational assumption of 1~mag in this plot.

Finally, we compare the intrinsic \Ha{} luminosity function (solid lines) to the observed result corrected for dust using the Balmer decrement (symbols). Although similar at faint fluxes, $L_{{\rm H}\alpha}\lessapprox 10^{35}$~W, the simulated luminosity function is significantly below the observations at the bright-end when using Eq.~\ref{eq:sfr} (blue), with the difference much reduced when using the \cite{Cheng15} conversion (red curve). The shape at the bright-end is so steep that even a small error in the observed luminosity determination can make \eagle{} and \gama{} consistent. In addition, the \eagle{} stellar mass function is lower than observed around the knee (S15), which will also contribute to the deficit at brighter luminosities.

In summary: the \eagle{} \Ha{} luminosity function is in relatively good agreement with observations when applying the \citet{Cheng15} inspired$^7$ conversion between $\dot M_\star$ and \Ha{} luminosity. The effects of insufficient numerical resolution are apparent at lower luminosities ($L_{{\rm H}\alpha} \lesssim 10^{34}$~W). Applying a constant correction to the \eagle{} dust-attenuated \Ha{} values does not reproduce the shape of the underlying `true' \eagle{} \Ha{} luminosity function well, over-estimating the star formation rates at higher \Ha{} luminosities ($L_{{\rm H}\alpha} \gtrsim 10^{34}$~W). \changes{This comes about due to the higher attenuation in more \Ha{} luminous galaxies, shown directly in Appendix \ref{sec:half_conv}.}

\subsection{D4000 Index}
\label{ref:d4000}

\begin{figure}
 \includegraphics[width=0.49\textwidth]{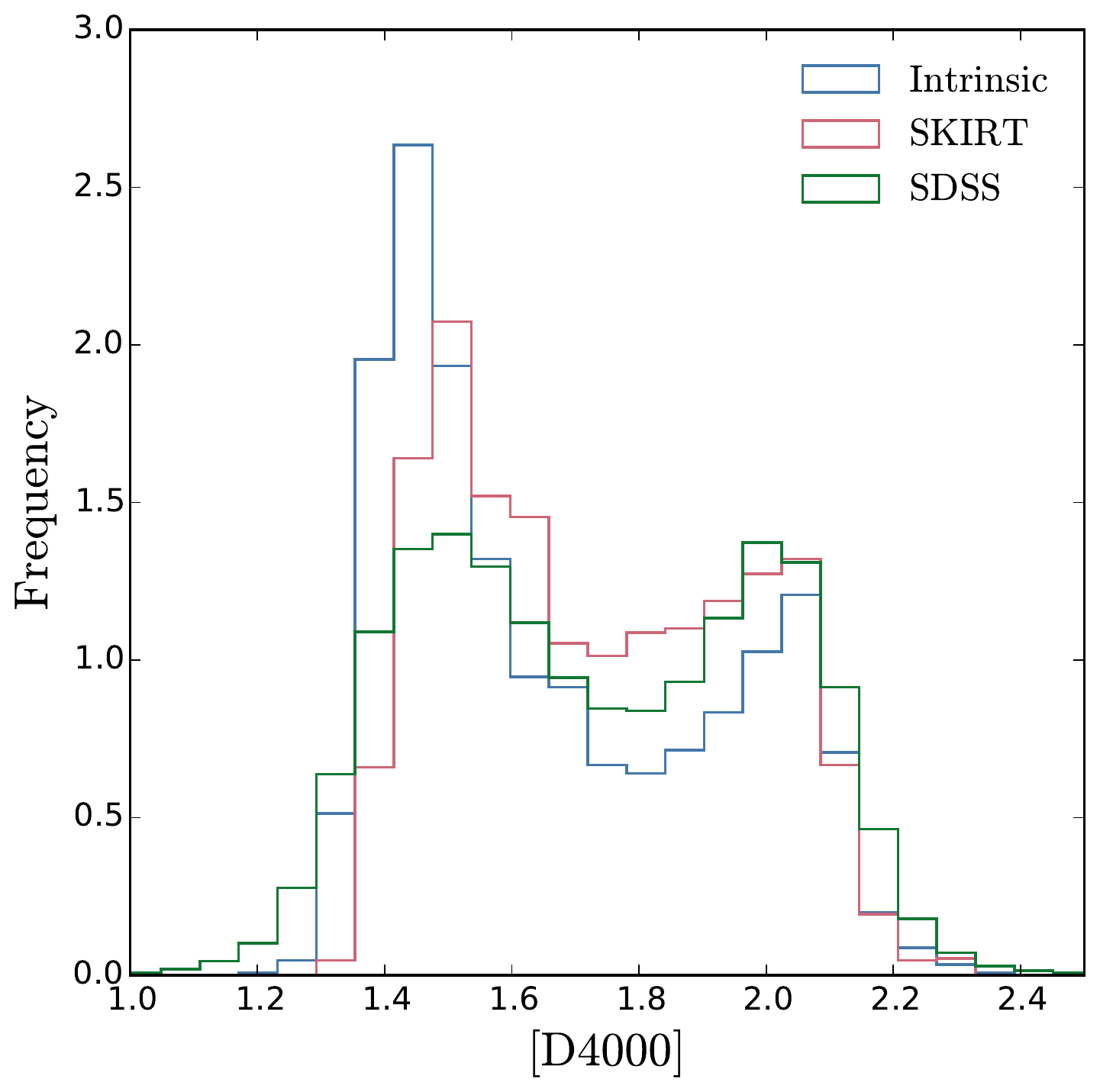}  
 
 \caption{Histogram of D4000 values for intrinsic and radiative transfer (\skirt{}) photometry of $z=0.1$ galaxies with $10^{10} {\rm M_\odot} < M_\star < 10^{11} {\rm M_\odot}$. The low and high D4000 peaks are taken to represent active and passive populations respectively. \textit{Blue} indicates the intrinsic values, \textit{red} after dust is applied. Values of D4000 measured for a mass-matched sample of SDSS galaxies are plotted in green. A considerable difference in the active and passive peaks recovered with this technique is observed when dust is applied. }
 \label{fig:d4000}
\end{figure}

\begin{figure}
 \includegraphics[width=0.49\textwidth]{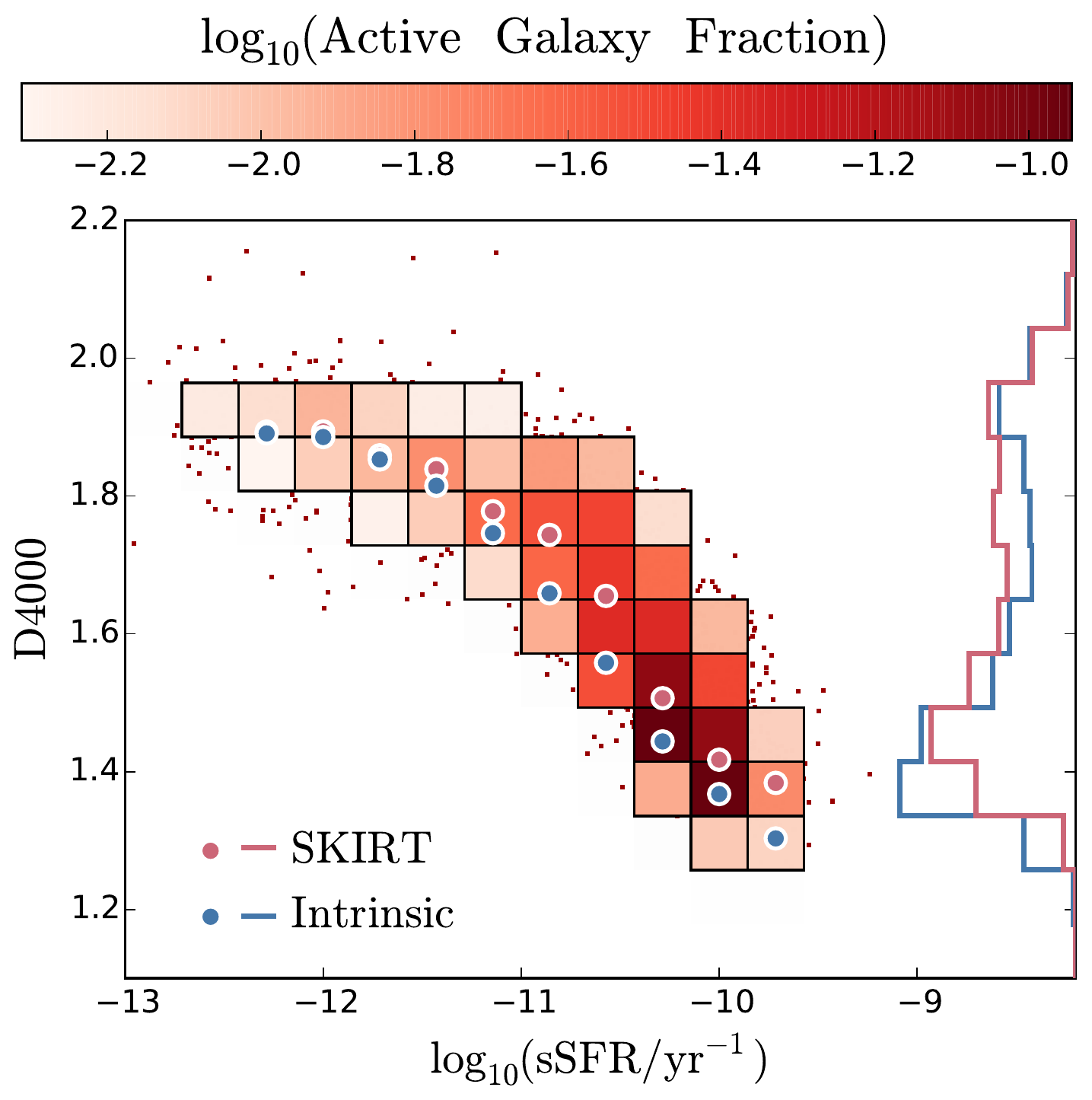}  
 
 \caption{D4000 index as a function of specific star formation rate (sSFR) for $z=0.1$ \eagle{} galaxies with $M_\star > 10^{10} {\rm M_\odot}$. \textit{Circles} indicate the median D4000 value in bins of $\log_{10}$sSFR, with \textit{blue} representing the intrinsic values and \textit{red} including dust (\skirt{}). \textit{Shaded squares} represent evenly-spaced bins in D4000 and log sSFR. These are shaded by the fraction of active galaxies (SFR$>0$) they contain for the \skirt{} values of D4000, and are only displayed within the 95\% contour of the active galaxy distribution. Galaxies outlying the 95\% contour are represented individually as \textit{red points}. D4000 histograms are also displayed as in \ref{fig:d4000}, again \textit{blue} represents intrinsic values and \textit{red} \skirt{} values. We see that, despite the trend of D4000 with sSFR for $\log_{10}({\rm sSFR / yr^{-1}}) > -12$, a significant number of galaxies are scattered toward the high D4000 population when dust is included. A few galaxies also exhibit relatively low D4000 despite low sSFRs ($\log_{10}({\rm sSFR / yr^{-1}}) < -12$)}
 \label{fig:ssfr}
\end{figure}

The 4000 \AA{} break (D4000) was used by \citet{Kauffmann03} to analyse a large sample of galaxies from the SDSS. In older stellar populations, the opacity due to several ions in stellar atmospheres combines to induce a noticeable discontinuity in the flux around 4000 \AA{}, which is mostly absent in hot stars. The size of the discontinuity is hence a measure of the relative contribution of young hot stars to the flux - and hence of the specific star formation rate of the galaxy \citep[e.g.][]{Bruzual83}. The D4000 index is the logarithm of the ratio of the red and the blue continuum, measured using narrow-band top-hat filters. We use $\left[4050, 4250 \right]$ \AA{} and $\left[3750, 3950 \right]$ \AA{} to sample the red and blue continuum respectively, as defined by \citet{Bruzual83}. The D4000 index is commonly used to distinguish between active (star forming) and passive populations, similar to the colour-colour plots of Fig.~\ref{fig:colcol}. Indeed, D4000 can be thought of as a colour index (analagous to e.g. $u-r$), but is generally considered robust against dust effects due to it being a differential measure with a relatively small wavelength separation between bands.

The D4000 distribution used by \citet{Kauffmann03} and \citet{Kauffmann03b} are from a flux-limited selection of SDSS galaxies, and shows strong bi-modality where massive galaxies have a high value of D4000, low-mass galaxies have low values, with the transition mass around $M_\star \sim 3\times 10^{10}{\rm M}_\odot$, see Fig.~1 of \citet{Kauffmann03b}. Even the largest, 100$^3$~Mpc$^3$, \eagle{} volume does not have enough massive galaxies to allow for a direct comparison with these measurements. 

To make a comparison to data meaningful, we therefore create a sample of SDSS galaxies mass-matched to the \eagle{} population over the range $10^{10} {\rm M_\odot} < M_\star < 10^{11} {\rm M_\odot}$, using the observed masses from \citet{Kauffmann03}. The corresponding broad-band \citep{Bruzual83} D4000 values are taken from the MPA-JHU catalogue, released for SDSS DR7 \citep{SDSSdr7}. We then compare the D4000 distribution to that of \eagle{} for both the intrinsic and dust-reddened spectra in Fig. \ref{fig:d4000}. Each distribution is normalised to integrate to unity. We employ a cut at D4000=1.8 to separate the active and passive population, which is near the minima of each histogram. We recover passive fractions of $32\%$ and $41\%$ for \eagle{} galaxies using intrinsic and \skirt{} spectra, as compared to $55\%$ for the SDSS sample. 

We first compare the \lq intrinsic\rq, ISM dust-free, \eagle{} distribution (blue) to that produced using \skirt{} (red). Both distributions exhibit a clear bi-modality, with low and high D4000 peaks at $\approx 1.5$ and $\approx 2$ respectively. A stark difference between the two distributions is that the population with low D4000, a common proxy for the star formation, is significantly depleted when ISM dust is included with \skirt{}. There is also a shift of the low D4000 peak to higher values. The relative depletion of the low D4000 population in the \skirt{} distribution corresponds to a relative boost at intermediate and high D4000. This boost reduces in significance for higher D4000 values, falling below $10\%$ near the high D4000 peak of $\approx 2$. Although there is little difference in overall attenuation between the two D4000 bands, including the preferential obscuration of light from young stars by dust attributed both stellar birth clouds and the diffuse ISM via \skirt{} leads to some star-forming galaxies registering higher D4000 values, and even appearing completely passive in this proxy. This is the same effect seen for broad-band colours in Fig. \ref{fig:coldists} and \ref{fig:colcol}. We find that $\approx 20\%$ of \eagle\ galaxies deemed to be passive according to the D4000$>$1.8 criterion, are star forming.

We next compare the \eagle{} distributions to the \sdss{} distribution (green). A clear bi-modality can be seen in the \sdss{} distribution, exhibiting similar peak positions to \eagle{} at D4000~$\approx 1.5$ and $\approx 2$. The \sdss{} peak positions agree more closely with the \skirt{} model distribution, which should be a fairer comparison. The fraction of galaxies in the low D4000 (active) population for the \sdss{} sample is smaller than for either \eagle{} distributions, but agrees better with the \skirt{} histogram. The high D4000 (passive) fraction is larger for \sdss{} than \eagle{}, and also closer to the dust reddened values of \skirt{}. The inferred passive fraction for \sdss{} galaxies is $70\%$ higher than inferred for the intrinsic \eagle{} spectra, and $35\%$ higher than for \skirt{}. While the frequency of \sdss{} galaxies at intermediate D4000 values ($\approx 1.8$) is under-predicted by the intrinsic \eagle{} distribution, the boost in the population seen for the \skirt{} distribution over-predicts the number density of galaxies by a similar factor. Generally the agreement with observation is improved by the inclusion of \skirt{} ISM dust modelling, but remains slightly discrepant. While modelled dust effects can improve the inferred passive fractions, an excess of high-mass active galaxies persists relative to observation. This could reflect a genuine overproduction of active galaxies in \eagle{}, as suggested by \citet{Furlong15}.

Additionally we see that the \sdss{} distribution is broader than for \eagle{}, with tails extending to more extreme high and low values. These tails might be due to outliers with unusually large photometric errors. The limited parameter coverage of the populations synthesis models could also prevent the occurrence of the most extreme values in the simulation.

We used the D4000 continuum band definitions of \citet{Bruzual83} rather than the narrower band definition of \citet{Balogh99} employed by \citet{Kauffmann03}. The reason for this choice is that D4000 is better converged at our standard spectral resolution for \skirt{}. A caveat is that the broad band definition is observed to be significantly more susceptible to dust effects than narrow bands \citep{Balogh99, Kauffmann03}, potentially leading to larger dust uncertainties.

To test how well the measured D4000 predicts star formation activity for the \eagle{} spectra, we plot D4000 against specific star formation rate (sSFR) for $z=0.1$ `active' ($\dot{M}_\star > 0$) \eagle{} galaxies in Fig. \ref{fig:ssfr}. At $\log_{10}({\rm sSFR / yr^{-1}}) > -12$, and despite a clear negative trend of D4000 with sSFR, we see that the median D4000 is higher for the \skirt{} spectra (red circles) than for the spectra without ISM dust (blue circles). The distribution of active galaxies in this plane also shows a significant tail to high D4000 when ISM dust is included (shaded squares), a significant number of galaxies with $\log_{10}({\rm sSFR / yr^{-1}}) > -11.5$ have D4000 consistent with the passive population. This suggests that dust effects are significant for this index when applied to the \skirt{} spectra, and can be ascribed to star formation being hidden by the dense, dusty ISM of some galaxies. Conversely, a few outlying galaxies (red points) show relatively low D4000 at $\log_{10}({\rm sSFR / yr^{-1}}) < -12$. This is due to these galaxies being relatively metal poor, analagous to the anomalously blue passive galaxies identified in \citet{Trayford16} using $u-r$ colour (their Fig. 1d). 

%% file: Summary.tex
\section{Summary and Conclusion}
\label{sec:conc}
We have made mock optical observations of galaxies simulated within the \eagle{} suite of cosmological, hydrodynamical simulations, including the effects of dust, utilising the public \skirt{}\footnote{\tt www.skirt.ugent.be} code. \skirt{} calculates three-dimensional radiative transfer on each galaxy, and we used the cool, enriched gas as a tracer of dust in the diffuse interstellar medium (ISM). To ameliorate limitations resulting from limited numerical resolution, we apply a subgrid model to represent unresolved \Hii{} regions and the associated dust attenuation by using the \starburst\ model of \cite{Groves08}. The full procedure is presented in section \ref{sec:template}, and was developed in this work and in the companion work of \citet{Camps16} (C16). The dust modelling introduces extra parameters, such as the dust-to-gas ratio in the ISM, and parameters of the \starburst\ model such as the covering factor and compactness of H{\sc ii} regions. These were chosen by comparing far-infrared mock observations of \eagle\ galaxies with observations of local galaxies, as described by \cite{Camps16}. We apply no additional calibration in the optical regime.

To enable a detailed comparison of numerically resolved \eagle\ galaxies to observations from the \gama\ survey \citep{Driver11}, we selected galaxies at redshift $z=0.1$ above a stellar mass cut of $M_\star \geq 1.81\times 10^8$ M$_\odot$ ($\sim 100$ star particles at standard resolution). We focus on the largest fiducial \eagle{} simulation, a 100$^3$ Mpc$^3$ volume, which provides a sample of $30145$ galaxies that satisfy this selection criterion. Galaxies are defined and selected in the same way as in \citet{Trayford15}, to enable direct comparison with their dust-screen model. Integrated spectra, broad-band magnitudes, broad-band images and mock IFU data were produced for each galaxy for three orientations (face-on, edge-on and along the $z$-axis of the simulation volume), with properties detailed in section \ref{sec:prod}. These will be made available through the public \eagle{} database \citep{McAlpine16}.

In section~\ref{sec:orientations} we studied the attenuation in the $B$-band as a function of inclination, comparing to the observationally inferred relation of \citet{Driver07}. We found that:

\begin{itemize}
\item The $B$-band attenuation-inclination relation for \eagle{} galaxies, Fig.~\ref{fig:orient}, exhibits large galaxy-to-galaxy scatter, on top of a smooth trend of increasing  median attenuation for galaxies seen more edge-on. This trend is weaker than observed, as is the level of the edge-on attenuation. We attribute this to the fact that the stellar and gas disc in \eagle\ galaxies is thicker than in observed galaxies, a consequence of the artificial pressure floor imposed on the simulated ISM (as opposed to being purely limited by numerical resolution, see Fig.~\ref{fig:aocurves}). 
\item Stacked attenuation curves of \eagle{} galaxies at different orientations, Fig.~\ref{fig:ecurves}, reveal different profiles. Face-on galaxies show the steepest frequency dependence as the youngest stars are preferentially dust obscured by the diffuse ISM component. Conversely, the curve for edge-on galaxies shows a weaker (or \lq greyer\rq{}) frequency dependence as both young and old stellar populations are obscured by the diffuse dust disc, closer to a screen model. We also see that nebular emission features (such as H$\alpha$) suffer from strongly increased \textit{ISM attenuation} (over and above the attenuation of the stellar birth clouds), due to star-forming regions being embedded in denser and dusty gas.   
\end{itemize}

Comparing optical \skirt\ photometry to ISM dust-free models, the dust-screen model of \cite{Trayford15}, and colours from the \gama\ survey, we find that

\begin{itemize}
\item  Optical \skirt{} galaxy colours match the data remarkably well, Fig.~\ref{fig:coldists}. In fact, they match significantly better than the dust-screen model galaxies of \cite{Trayford15}. The \skirt{} colours exhibit a mass-dependent boost of the {\it \lq green valley\rq{}} and red populations compared to either intrinsic or dust-screen colours.
\item The improved agreement with observation demonstrated by \skirt{}, relative to the screen model of \citet{Trayford15}, is attributable to the spatial distribution of dust compared to that of young stars, and the better capturing of orientation effects as compared to a screen model. Fig.~\ref{fig:orscat} shows how highly-inclined galaxies with intrinsically blue colours may scatter to the reddest colours.
\item Dusty galaxies can confound simple star-formation activity proxies, such as colour-colour cuts. Fig.~\ref{fig:colcol} shows that a $ugJ$ colour-colour cut can recover passive fractions well when dust is modelled as a screen, but with the more realistic \skirt{} dust modelling a significant fraction of active galaxies masquerades as passive. We find that for the relatively stringent cut used in this paper, approximately 15$\%$ of galaxies with $M^\star > 10^{10} {\rm M_\odot}$ are classified as passive due to dust reddening. More relaxed cuts may yield significantly higher misclassification rates (see section \ref{sec:pass} for details). 
\end{itemize}

Finally, we investigated spectral indices often used as proxies for star formation activity, such as the H$\alpha$ line flux and the 4000\AA{} break (D4000). We plotted the H$\alpha$ luminosity functions ($\phi$) and compared directly to the function computed for the \gama{} sample by \citet{Gunawardhana13} in Fig.~\ref{fig:haLF}. Given that the star formation rates of \eagle{} galaxies are typically 0.2~dex lower than reported for observations using the \citet{Kennicutt98} calibration \citep{Furlong15}, and that the conversion from $\dot{M}^\star$ to H$\alpha$ follows the standard \citet{Kennicutt98} relation (see Fig.~\ref{fig:hacomp}), it is unsurprising that the \eagle{} luminosity function is systematically low at the bright end. However, the recent study of \citet{Cheng15} argued that absolute $\dot{M}^\star$ values should be normalised $\approx 0.2$~dex lower, so we also plotted $\phi$ using H$\alpha$ boosted by this factor. The D4000 distributions were compared to a mass-matched sample of \sdss{} galaxies in the range $10^{10} {\rm M_\odot} < M^\star < 10^{11} {\rm M_\odot}$. From studying these indices, we found that:
\begin{itemize}
\item The bright end of the recalibrated, intrinsic H$\alpha$ luminosity function ($L_{\rm H\alpha} > 10^{34.5} {\rm W}$) agrees reasonably with observations, particularly when using the \citet{Cheng15} calibration. Applying a constant dust correction to the \skirt{} H$\alpha$ predictions distorts the shape of the luminosity function, appearing more Schechter-like with a steeper bright end than seen in either the intrinsic or observed LFs.
\item D4000 values produced by \skirt{} show similar distributions to a mass-matched \sdss{} sample over the range $10^{10} {\rm M_\odot} < M_\star < 10^{11} {\rm M_\odot}$, Fig.~\ref{fig:d4000}, but with somewhat fewer galaxies at high D4000. Dust reddening significantly boosts the high D4000 population relative to the intrinsic distribution, analogous to the red sequence boosting observed in Fig.\ref{fig:coldists}. This leads to higher inferred passive fractions, with the light from young stars preferentially extinguished by dust. The scattering of galaxies to high D4000 by dust, for a given specific star formation rate, is characterised in Fig. \ref{fig:ssfr}.
\end{itemize}

Including dust radiative transfer effects with \skirt{} allows us to model the inhomogeneous dust distribution and how that is correlated with regions of recent star formation, while improving the level of agreement of mock fluxes with observations compared to the dust screen model of \citet{Trayford15}. However, there are clear limitations. In particular, the edge-on attenuation of disc galaxies is lower than observed. Improving this would likely require higher-resolution simulations with an explicit cold phase, allowing us to resolve thin molecular gas discs on scales $\lesssim 100$~pc

It is perhaps surprising that despite these limitations, the attenuation computed using \skirt{} profoundly influences colours, improving agreement with data. This can be ascribed to the effects that geometry and scattering have on the optical attenuation of galaxies that cannot be captured by screen models. The localised nature of the dusty ISM around young stars is effective at hiding their blue light, leaving the older populations to contribute relatively more to the fluxes. Despite invoking similar mean attenuation values (see section \ref{sec:fit}, appendix \ref{sec:uncal}), a screen model yields more apparently blue galaxies because their star forming regions are relatively much less shielded by dust. 

\changes{Throughout this paper we focus on using a forward modelling approach to compare simulated and real galaxies in the observable domain. While we discuss many benefits of this approach, we note that forward modelling has the potential to obscure the reasons for discrepancy (or agreement) if the influence of individual physical properties are not well understood. An example where such caution is needed would be in our \Ha{} measurements, influenced by both the ISM structure and star formation rates of \eagle{} galaxies. We emphasise that it is therefore important to analyse our forward modelling results in the context of comparison studies in the physical domain \citep[such as][for \eagle{} galaxies]{Schaye15, Furlong15}. We have attempted to provide such context in this work.}

We hope that the more realistic observables produced by \skirt{} provide a useful resource, opening new avenues of investigation for comparing observations to the simulations. Mock observational data presented in this paper will be made accesible via the public \eagle{} database \citep[for updates register at {\tt http://icc.dur.ac.uk/Eagle/database.php}]{McAlpine16}. Our mock observables are also provided for additional redshifts and \eagle{} simulations that are not discussed in this work. 
 
 \section*{Acknowledgements}
JWT would like to acknowledge Peder Norberg for insightful comments at the early stages of this project. \changes{We also thank the anonymous referee, who's comments and suggestions lead to a significant improvement of the manuscript}. The work was supported by the Science and Technology Facilities Council [grant number ST/F001166/1], by the Interuniversity Attraction Poles Programme initiated by the Belgian Science Policy Office ([AP P7/08 CHARM]) by ERC grant agreement 278594 - GasAroundGalaxies and by Netherlands Organisation for Scientific Research (NWO), through VICI grant 639.043.409. We used the DiRAC Data Centric system at Durham University, operated by the Institute for Computational Cosmology on behalf of the STFC DiRAC HPC Facility (www.dirac.ac.uk). This equipment was funded by BIS National E-Infrastructure capital grant ST/K00042X/1, STFC capital grant ST/H008519/1, and STFC DiRAC is part of the National E-Infrastructure. RAC is a Royal Society University Research Fellow. The data used in the work is available through collaboration with the authors. GAMA is a joint European-Australasian project based around a spectroscopic campaign using the Anglo-Australian Telescope. The GAMA input catalogue is based on data taken from the Sloan Digital Sky Survey and the UKIRT Infrared Deep Sky Survey. Complementary imaging of the GAMA regions is being obtained by a number of independent survey programmes including GALEX MIS, VST KiDS, VISTA VIKING, WISE, Herschel-ATLAS, GMRT and ASKAP providing UV to radio coverage. GAMA is funded by the STFC (UK), the ARC (Australia), the AAO, and the participating institutions. The GAMA website is http://www.gama-survey.org/.

%% file: Appendices.tex
\appendix
\section{Smoothing Lengths}
\label{sec:smooth}

\begin{figure*}
 \includegraphics[width=0.99\textwidth]{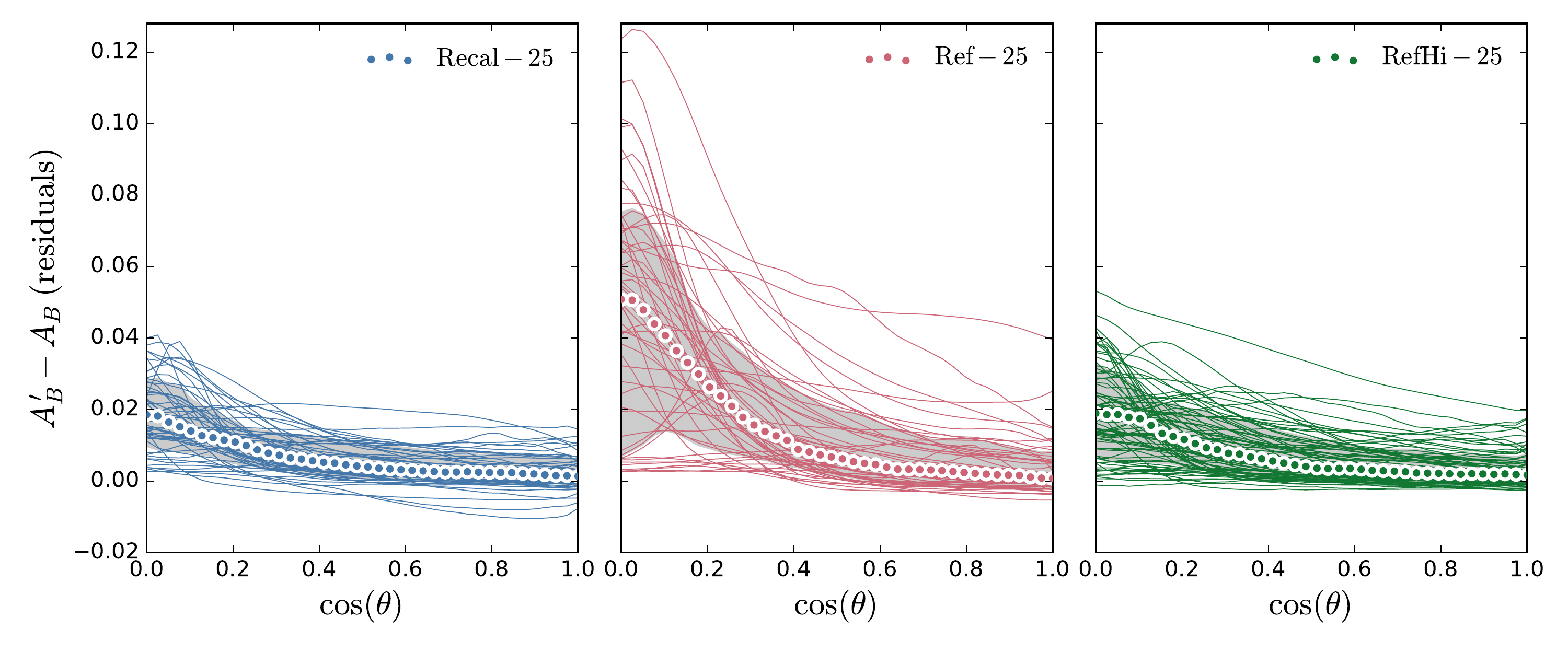}  
 \caption{The difference between $r$-band attenuation in \eagle{} galaxies when using no smoothing ($A^\prime_r$) and employing the standard nearest neighbours smoothing used in this work ($A_r$) plotted as a function of inclination, $\cos(\theta)$. The panels from left to right represent simulations \ac{Recal-25}, \ac{Ref-25} and \ac{RefHi-25}, and are also coloured blue red and green respectively. \textit{Thin Lines} represent $A^{\prime}_r - A_r$ values of individual galaxies taken from their respective simulations at 40 inclinations evenly spaced in $\cos(\theta)$. \textit{Shaded circles} represent the median $A^{\prime}_r - A_r $ value of galaxies at each $\cos(\theta)$ value, with \textit{grey shading} illustrating the 16th-84th percentile range. We see that not using stellar smoothing leads to a slightly stronger inclination dependence for orientation, with increased attenuation for edge-on galaxies. The maximal difference is seen for edge-on galaxies in the \ac{Ref-25} simulation, $\Delta A_B\approx 0.1$, with typical values for edge-on galaxies of $\Delta A_B\approx 0.05$.}
 \label{fig:aosmooth}
\end{figure*}

Star and gas particles in \eagle{} function as tracers of the baryonic mass. Because particles are the smallest resolution elements in the simulations, the distribution of the material represented by a single particle is unresolved. However, some 3D form for the traced material needs to be assumed to facilitate radiative transfer with \skirt{}. A kernel distribution is thus used to set the density profile of the stars and gas. A truncated Gaussian distribution is used to approximate the cubic spline kernel used by the \eagle{} simulations in \skirt{} \citep{Altay13, Baes15}. As this is isotropic, it is parametrised solely by a position and a smoothing length. 

Smoothing lengths are tracked by \eagle{} for baryonic particles on the fly (see S15, appendix A). These values are derived using the distance to the weighted $N$th nearest neighbouring gas particle, to facilitate SPH interaction between gas and chemical enrichment of gas by stars. As such this kernel size represents the spatial smoothing of gaseous material well, but implies that for star particles the smoothing is entirely dependent on their proximity to gas. 

For resolved disc galaxies in \eagle{} this stellar smoothing is reasonable as the galaxies have high gas fractions, with star and gas particles being well mixed. However for a minority of gas-poor elliptical galaxies, the smoothing values may become extremely large (up to $\sim 70$ kpc). This distorts the surface brightness profiles to become kernel-shaped, and renders them much more extended than the actual stellar surface density. 

To alleviate this problem, we re-compute more appropriate smoothing lengths for \eagle{} star particles within each galaxy. There is no unique smoothing scale for star particles that can be defined, as they do not interact with each other using an SPH kernel. Using the same smoothing length calculation between star particles as between gas particles also results in significantly smaller smoothing lengths, due to a higher fraction of galaxy mass being in stars than gas, and such small smoothing lengths yields unrealistic granularity in galaxy images. For this reason we use a somewhat \textit{ad-hoc} method of `morphological convergence', observing galaxy images smoothed on a variety of scales, in a similar vein to \citet{Torrey14}. 

We use a $kd$-tree algorithm \citep{Maneewongvatana01} to identify nearby star particles. This is performed for each galaxy as it is extracted from the simulation data. We find that using as a smoothing length the distance to the 64th nearest neighbouring star particle works well, in the sense that this yields reasonable galaxy images, avoiding both unrealistic granularity and over-smoothing. Using the re-computed or simulation smoothing lengths make only marginal difference to the scientific results presented in this work. Intrinsic properties (i.e. without dust effects) are of course unaffected, as all light emitted by material within the 30 pkpc aperture is measured for consistency with \ac{T15}. The effect on dust attenuated properties is small because the smoothing lengths differ most in large galaxies where there is minimal gas and thus minimal attenuation.

To constrain the effect of stellar smoothing on the attenuation measured for \eagle{} galaxies, we compare attenuation measured for \eagle{} galaxies without any smoothing of sources (i.e. treating star particles as point sources) to those measured using the re-computed smoothing lengths. Fig.~\ref{fig:aosmooth} is set out in a similar way to Fig. \ref{fig:aocurves}, except we plot the \textit{difference} between the $r$-band attenuation without smoothing ($A^\prime_r$) and with smoothing ($A_r$) on the $y$-axis. The thin coloured lines show the residuals for individual galaxies at different orientations, and the data points show the median residuals. Again, all galaxies of mass $M_\star > 10^{10} {\rm M}_{\odot}$ are included for each of the \ac{Ref-25} \ac{Recal-25} and \ac{RefHi-25} simulations.

We see that, in general, treating stellar particles as point sources leads to a stronger dependence of attenuation on inclination, with more attenuation for edge on inclinations. This can be understood as a higher fraction of the stellar emission emanating from near the disc plane when no smoothing is applied, as the smoothing effectively thickens the emmisivity distribution of the stellar disc. Although this effect is measurable, the difference is $\lessapprox 10\%$ for the galaxies in this sample, suggesting that even an extreme choice in stellar smoothing has only a marginal effect on the integrated dust reddening for these galaxies. The difference is most pronounced for the low resolution \ac{Ref-25} galaxies, because this has lower stellar particle resolution and larger smoothing lengths for the same mass range of $M_\star > 10^{10} {\rm M}_{\odot}$.

\section{Resolution and Convergence}
\label{sec:res}

\begin{figure*}
 \includegraphics[width=0.99\textwidth]{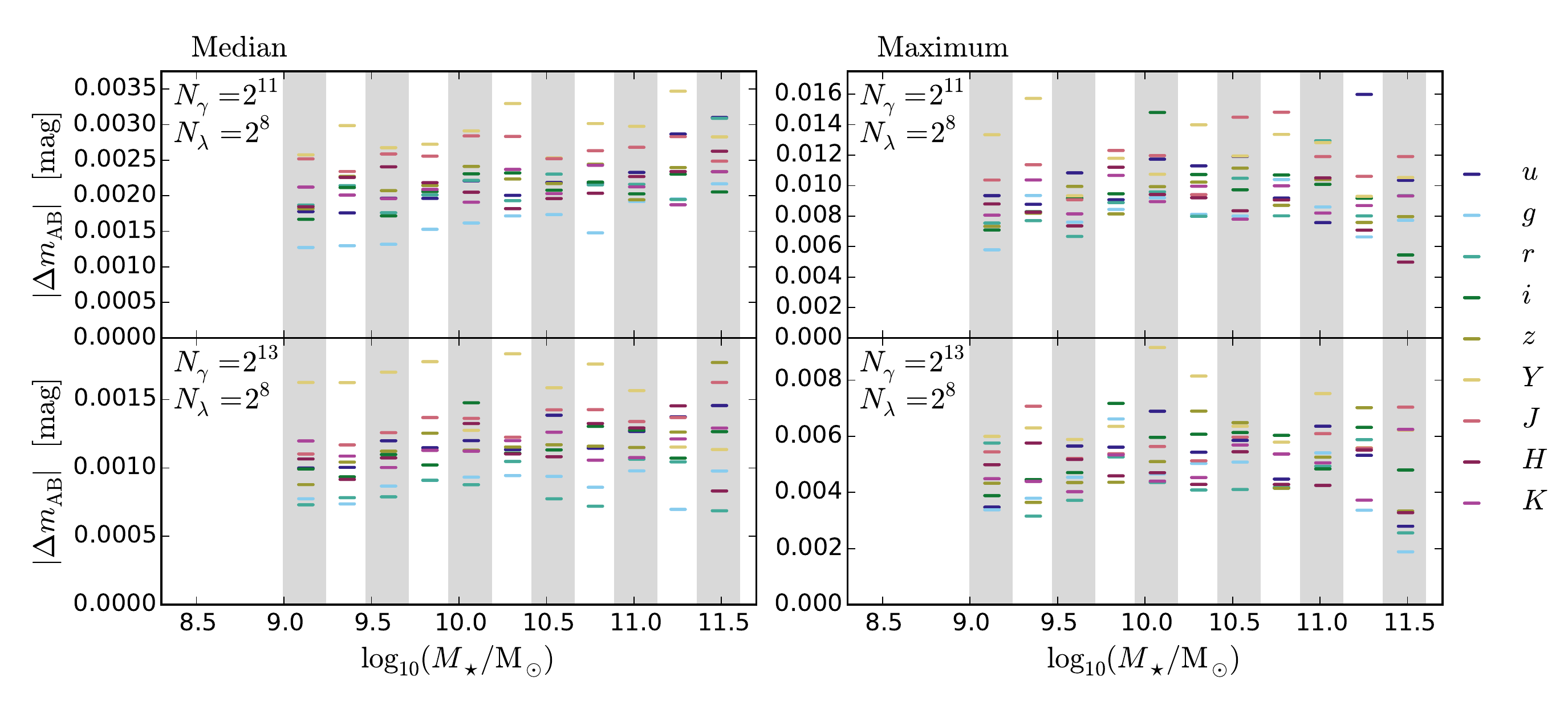}  
 
 \caption{SDSS $ugriz$ and UKIRT $YJHK$ absolute magnitude convergence properties for sample \eagle{} galaxies (section \ref{sec:res}) varying  the number of photon packets launched per wavelength bin, $N_\gamma$, at a constant spectral resolution. \skirt{} simulations with low, medium and high  $N_\gamma$ values of  $2^{11}$, $2^{13}$ and $2^{15}$ respectively are run for each galaxy. The top and bottom panels of each column then show the comparison of the low and medium $N_\gamma$ simulations with the high value respectively. The left panel shows the median absolute magnitude difference in each bin for the 9 photometric bands, the right panel shows the maximum difference.}
 \label{fig:photconv} 
\end{figure*}

\begin{figure*}
 \includegraphics[width=0.99\textwidth]{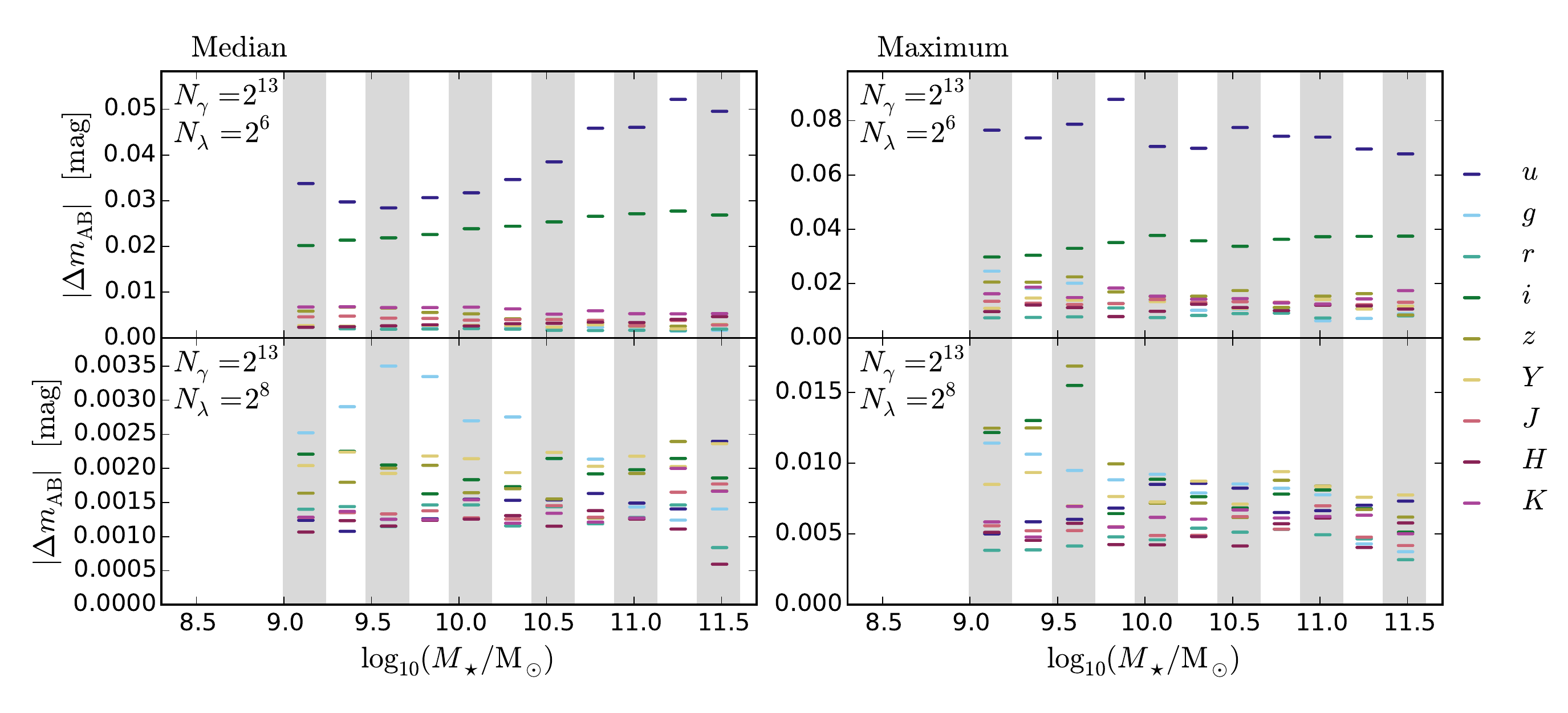}  
 
 \caption{As Figure \ref{fig:photconv}, but now exploring convergence as a function of continuum spectral resolution, $N_\lambda$, at fixed  $N_\gamma$. \skirt{} simulations with low, medium and high  $N_\lambda$ values of $2^{7}$, $2^{8}$ and $2^{9}$ respectively are run for each galaxy. Again, the low and medium resolutions are compared to the high resolution in the top and bottom panels of each column respectively.}
  \label{fig:specconv}
\end{figure*}

\subsection{Radiative Transfer}
\label{sec:rtconv}

Here we test the convergence of photometry results respect to the sampling of photon packages. We look at both the spectral resolution used to represent SEDs, and the number of photon packets sampled at each wavelength. In order to test this efficiently for our fiducial model, we randomly select a test sample of galaxies in 9 contiguous and logarithmically spaced mass bins, over the range $9 < \log_{10}(M_\star / {\rm M}_\odot) \leq 11.6$. 200 galaxies are randomly selected from each mass bin, except in the two highest mass bins where all the galaxies are sampled due to insufficient counts. The highest and second highest bins contain 23 and 64 galaxies respectively.

\subsubsection{Sampling Noise}
\label{sec:photconv}
\skirt{} tracks an equal number, $N_\gamma$, of photon packets per wavelength bin. The stochastic emission and absorption of these photon packets introduces Poisson noise into the photometric measurements. The degree of this variation depends not only on $N_\lambda$, but also on the emissivity at each wavelength, and the complex distribution of sources and dust in the galaxy. A natural target level of convergence is variation comparable to photometric errors in SDSS observations, on the order of $\sim$ 0.01 mag for $griz$ and $\sim$ 0.02 mag for the $u$-band, dominated by uncertainty in the un-modelled atmospheric effects at Apache Point \citep{Padmanabhan08}.

We test photometric convergence by running separate \skirt{} simulations launching $N_\lambda=2^{11}$ $2^{13}$ and $2^{15}$ photon packets per wavelength bin on each of our test sample galaxies. We then compare the variation in $ugrizYJHK$ photometry between the $N_\lambda=2^{15}$ run and the lower $N_\lambda$ runs. Figure \ref{fig:photconv} shows this level of variation. Both the median and maximum variations are below 0.01 mag for $griz$ and 0.02 mag for the $u$-band for $N_\gamma=2^{13}$, the number we used in the paper.

\subsubsection{Spectral Resolution}
\label{sec:specres}

To sample SEDs, \skirt{} performs radiative transfer for a grid of wavelengths. The number of wavelengths we choose is a trade-off between spectral resolution and computational expense. For our purposes we want to resolve the continuum well enough to capture the overall shape of the SED and produce accurate photometry in arbitrary optical broad-bands, as well as focussing on certain spectral indices of interest. After some initial experimentation, we begin with a superposed grid of wavelengths:
\begin{enumerate}
\item $2^8$ (256) logarithmically spaced wavelengths to sample the continuum between 280 nm and 2500 nm
\item 33 wavelengths to sample the peak and continuum either side of 11 prominent spectral lines in emission line galaxies \citep{Stoughton02}
\item 22 evenly spaced wavelengths to better sample the \Ha{} and \Oii{} line profiles 
\item 10 logarithmically spaced wavelengths from 150 nm to 280 nm to sample the UV slope
\item 12 additional wavelengths about the 4000\AA{} break.
\end{enumerate}

We test numerical convergence by measuring the variation between individual galaxies when different spectral resolutions are used. We vary the continuum wavelength grid resolution, (i), using the standard value of $2^8$ as medium resolution and $2^7$ and $2^9$ as low and high resolution respectively and comparing the standard and low resolutions to the high resolution in each plot. Figure \ref{fig:specconv} shows this level of variation.	

We find that the median variation is $<$ 0.01 mag for all bands at standard resolution, $N_\lambda=2^8$. When looking at the most extreme outliers in each bin we see that the most extreme differences are $\sim$ 0.015 mag in the lowest mass bin.  We decide this to be sufficient resolution for our purposes.

\subsection{Covergence of H${_\alpha}$ Luminosity and Flux}
\label{sec:half_conv}

\begin{figure}
\includegraphics[width=0.47\textwidth]{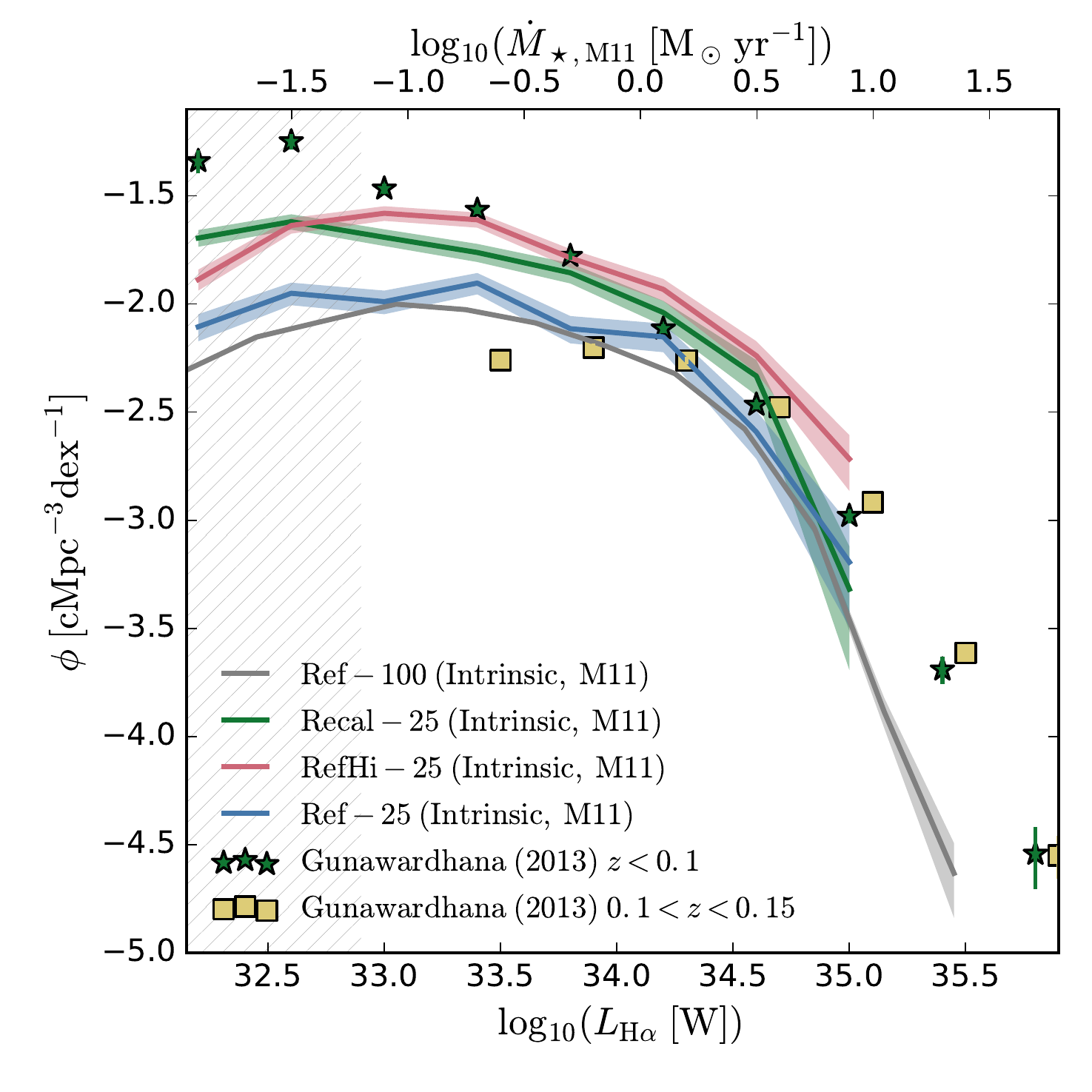}  
 \caption{As Fig. \ref{fig:haLF}, but excluding C15 recalibrated lines and including the \ac{Ref-25} and \ac{RefHi-25} \Ha{} luminosity functions. The \ac{Ref-100}, \ac{Recal-25}, \ac{RefHi-25} and \ac{Ref-25} luminosity functions are represented as \textit{grey}, \textit{green}, \textit{red} and \textit{blue} lines respectively, with Poisson error indicated by the shaded regions of the same colour. }
 \label{fig:HaLF_conv}
\end{figure}

Here we investigate the convergence properties of the \Ha{} line fluxes, by comparing between all the simulations listed in Table \ref{tab:sims} in Fig.~\ref{fig:HaLF_conv}. Comparing \ac{Ref-100} and \ac{Ref-25} simulations, plotted as grey and blue lines respectively, isolates the effects of volume because the sub-grid calibration and resolution are the same. We see that the \ac{Ref-25} LF agrees better with \ac{Ref-100} than any of the other $25^3$~Mpc$^3$ boxes. However the \ac{Ref-25} LF is still between 0 and $0.3$ dex higher at all luminosities sampled. The higher number density of \Ha{}-emitting galaxies is likely due to a $25^3$~Mpc$^3$ being too small to represent large scale modes in the density distribution, and thus does not sample massive halos. As was shown in \citet{Trayford16}, star formation is significantly suppressed within these environments in the \eagle{} simulation. This could lead to the lower normalisation of the LF in the \ac{Ref-100}.

Comparing the \ac{Ref-25} and \ac{Recal-25} LFs instead tests `weak' convergence (defined by S15) with resolution. The \ac{Recal-25} LF is in general higher still, typically by $\sim 0.2$~dex. The higher normalisation of the \ac{Recal-25} LF is attributable to the effect of resolution on passive fractions of galaxies. As was discussed in T15, coarse sampling of feedback events paired with the resolution of the star forming component of galaxies contributes to the surplus of passive galaxies seen at $(M_\star / {\rm M}_\odot) \lesssim 10$ in the lower resolution simulations. Because these resolution effects become significant at a factor of $\sim 8$ lower mass at higher resolution, the proportion of star forming galaxies at relatively low mass is more realistic. This contributes to boosting the \ac{Recal-25} LF closer to observations.

\changes{We also investigate the attenuation of the \Ha{} line by the diffuse dust component, $A({\rm H}\alpha, \; {\rm ISM})$, using the simulations listed in Table \ref{tab:sims} in Fig. B3. We look specifically at the ISM attenuation here, as this is controlled by the ISM structure that may vary with resolution. Taking
\begin{equation}
\label{eq:AHa}
A({\rm H}\alpha, \; {\rm ISM}) = 2.5\log_{10}(L_{{\rm H}\alpha, \; {\rm SKIRT}} / L_{{\rm H}\alpha, \;{\rm intrinsic}})
\end{equation}
we plot $A({\rm H}\alpha, \; {\rm ISM})$ as a function of star formation rate for individual galaxies with $\log_{10}(M_\star/{\rm M_\odot}) > 8.5$ from each simulation as grey points, overplotting the average attenuation in bins of star formation rate for each simulation in different colours, and indicating the scatter using the 16th-84th percentile ranges. We see that attenuation generally increses with star formation rate but exhibits a large scatter, skewed to high values. This is consistent with general observational trends. \citep[e.g.][]{Gunawardhana15}. We notice that the typical attenuation values are similar for all the simulations, and consistent with each other within the scatter, suggesting $A({\rm H}\alpha)$ converges with resolution. This is consistent with what we find for Fig. \ref{fig:aocurves}, that attenuation is limited by the artificial pressurisation of the \eagle{} ISM rather than resolution, and attenuation would likely increase if gas were able to cool to lower temperatures.   }

\begin{figure}
\includegraphics[width=0.47\textwidth]{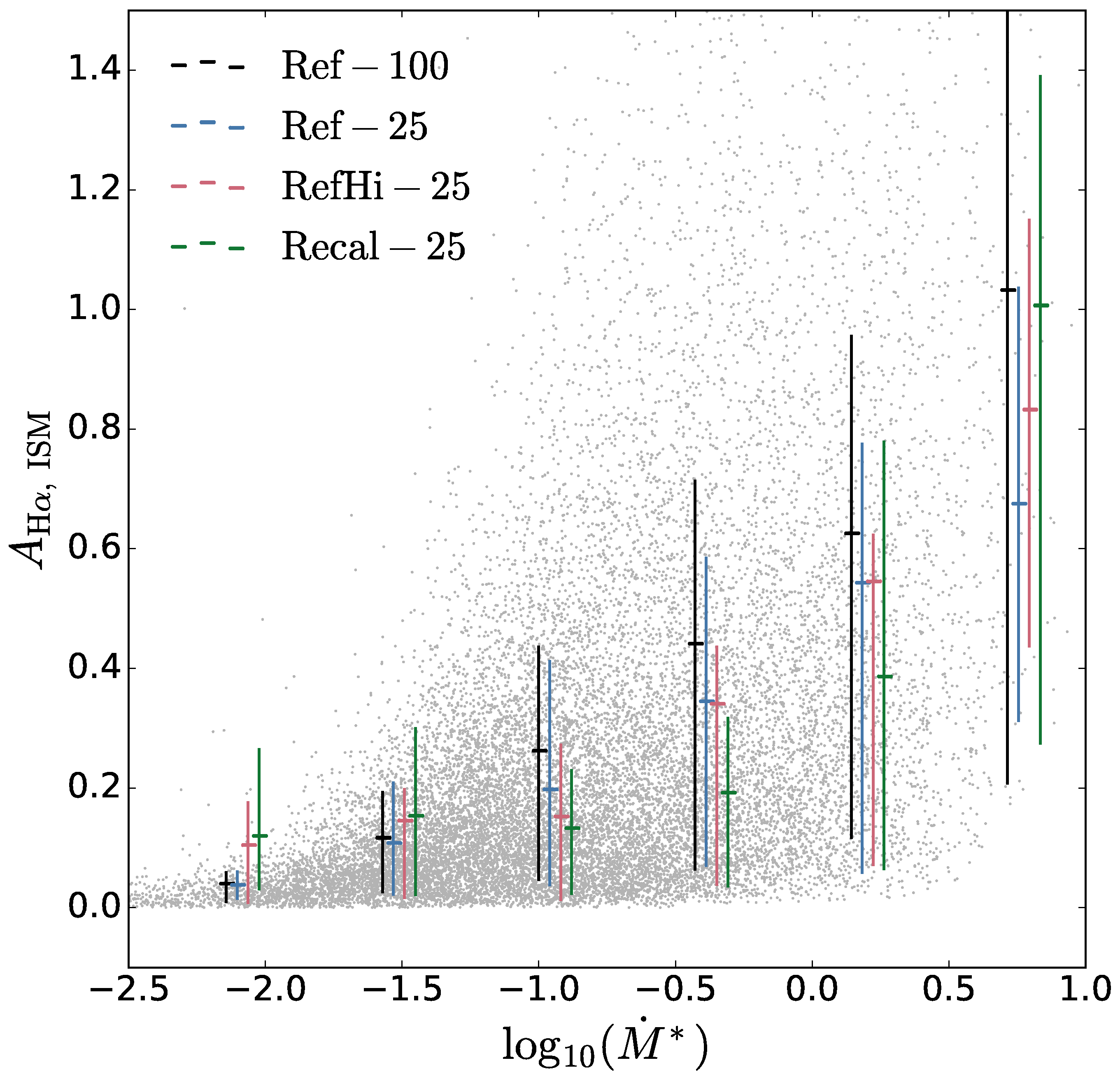}  
 \caption{\changes{Plot of $A({\rm H}\alpha, \; {\rm ISM})$ (Eq. \ref{eq:AHa}) as a function of star formation. Individual galaxies with $\log_{10}(M_\star/{\rm M_\odot}) > 8.5$ from each simulation of Table \ref{tab:sims} are plotted as \textit{grey points}. Average values for each simulation are overplotted as coloured points, with error bars indicating the 16th-84th percentile ranges. We see that the attenuation generally increases as a function of star formation rate, exhibiting large scatter skewed towards high values. The average relations agree between simulations within errors, suggesting that $A({\rm H}\alpha, \; {\rm ISM})$ is well converged in our modelling.}}
 \label{fig:Aha}
\end{figure}

\section{fitting SKIRT results using the T15 model}
\label{sec:mcmc}

To make a detailed comparison between the radiative transfer photometry presented in this work and that of \ac{T15}, we calibrate the dust-screen parameters used by T15 to the results obtained with \skirt. Comparing these parameters to their fiducial \ac{T15} values helps us understand how the models differ. This can be achieved by finding the \ac{ML} parameter values using an MCMC approach. 

The flux density for an \eagle{} galaxy measured in a certain band using the \ac{T15} model, $F_{\rm T15}$, can be expressed
as
\begin{equation}
F_{\rm T15}   \;\;\; = \left( F_{\rm o} + F_{\rm y}e^{-\hat{\tau}_{\rm BC}} \right)e^{- \hat{\tau}_{ISM}  O(\theta \vert q)} \nonumber
\end{equation}
where
 \begin{eqnarray}
\hat{\tau}_{\rm ISM} &= \tau_{\rm ISM} f_{\rm M} f_{\rm Z} \left( \frac{\lambda_{\rm eff}}{550 {\rm nm}}\right)^{-0.7},  \\
\hat{\tau}_{\rm BC} &= \tau_{\rm BC} f_{\rm M} f_{\rm Z} \left( \frac{\lambda_{\rm eff}}{550 {\rm nm}}\right)^{-0.7},  
\end{eqnarray}

Here, $f_{\rm M}$ and $f_{\rm Z}$ are the star forming gas mass and metallicity in units of the Milky Way value, respectively, $\lambda_{\rm eff}$ is the effective wavelength of the filter, and $F_{\rm o}$ and $F_{\rm y}$ are the intrinsic flux densities for star particles older and younger than 10 Myr, respectively.  The $O(\theta \vert q)$ term accounts for galaxy orientation, as detailed in section 3.2.3 of \ac{T15}. The free parameters of the T15 model which we fit for, are $\tau_{\rm ISM}$, $\tau_{\rm BC}$ and $q$, representing the typical ISM optical depth, birth cloud optical depth and dust disc axial ratio. 

The attenuation applied in \ac{T15} is deterministic, apart from the randomised orientation term, $O(\theta \vert q)$, where the disc inclination angle is sampled uniformly in $\cos(\theta)$. However, as we store the inclination angle of each \skirt{} image, the expected $F_{\rm T15}$ value corresponding to a particular \skirt{} galaxy observation, $F_S$, is fully deterministic.  

Clearly, no parametrisation of the simple \ac{T15} model can provide perfect agreement with the \skirt{} photometry. The inclination also has an associated uncertainty. So that any possible $F_S$ has a finite likelihood of being observed with a particular \ac{T15} parametrisation, we build in a Gaussian observational tolerance. This contributes to the log-likelihood as:
\begin{equation}
\ln\mathcal{L}(F \, \vert \, F_{\rm T15}, \sigma) = C - \frac{1}{2\sigma^2}\sum^n_{i=1} (F_i - F_{{\rm T15},i})^2\,,
\end{equation}
where $C$ is a constant, $\sigma$ is a fixed uncertainty, and we sum over each randomly oriented $ugrizYJHK$ observation of an \eagle{} galaxy sample.  The  constant $\sigma$ value means that better resolved galaxies generally provide stronger constraints on the model, so the likelihood is effectively luminosity weighted. The exact value of $\sigma$ we use is unimportant as we aim to maximise $l$, but should be comparable to the observations to avoid numerical errors.  We use the $5^{\rm th}$ percentile of all $F_S$ values as $\sigma$. For simplicity, we do not explicitly incorporate an additional uncertainty on our input inclination angles.

With an imperfect model, the galaxies we use to constrain our fit will affect the recovered \ac{ML} parameters. However, using all galaxies may not provide the best parametrisation for those where dust is effective. For insufficiently resolved galaxies, the \skirt{} dust modelling is itself unreliable, and dust effects are anyway minimal.  For the most luminous galaxies the simple geometric assumptions of \ac{T15} are inappropriate and do not help constrain the parameters.  For this reason we select a sample in stellar mass .  To capture sufficiently resolved galaxies over which the disc geometry assumption is appropriate, we select galaxies in the range $10^{9.75} < M_\star / {\rm M_\odot} < 10^{11} $.

Initially we assume uniform prior distributions for each parameter and fit simultaneously, with the conservative ranges  $q \in [0,1)$, $\tau_{\rm BC} \in [0,10)$ and $\tau_{\rm ISM} \in [0,10)$. However, we found that the $\tau_{\rm BC}$ parameter tends to $\sim 10$. Indeed, relaxing the prior to $\tau_{\rm BC} \in [0,1000)$ yields a median $\sim 800$, such that $F_{\rm y}$  contributes effectively nothing to $F_{\rm T15}$. The reason for this may be that the spectra representing stellar populations younger than 10 Myr are intrinsically different in the models, and the clearing timescale for birthclouds is longer in \citet{CF00} (30 Myr). While \ac{T15} uses the \galaxev{} spectra \citep{bc03}, the \skirt{} model uses the MAPPINGS-III spectra \citep{Groves08}. While both models account for birth-cloud reddening, the MAPPINGS-III spectra include emission lines and a different ionising spectrum \citep[from][]{sb99} than \galaxev{}.

As a second approach, we try fixing $\tau_{\rm BC} = 2\tau_{\rm ISM}$, as suggested in the fiducial \citet{CF00} model and \ac{T15}, while assuming the same priors for  $\tau_{\rm ISM}$ and $q$. The  $\tau_{\rm BC}$ parameter has only marginal influence on the photometry, as it only affects a small fraction of stars, and the \ac{T15} value of $\sim 0.67$ already significantly reduces their contribution. We therefore expect that fixing $\tau_{\rm BC} = 2\tau_{\rm ISM}$ rather than allowing it to freely vary over the range $[0,1000)$ has minimal influence on the other parameters, such that all parameter values remain physically plausible. We find that including this constraint yields \ac{ML} $q$ and $\tau_{\rm ISM}$ values only $\sim 8\%$ and $\sim 11\%$ higher respectively. We therefore use this second approach as our default procedure.

\begin{figure}

\includegraphics[width=0.47\textwidth]{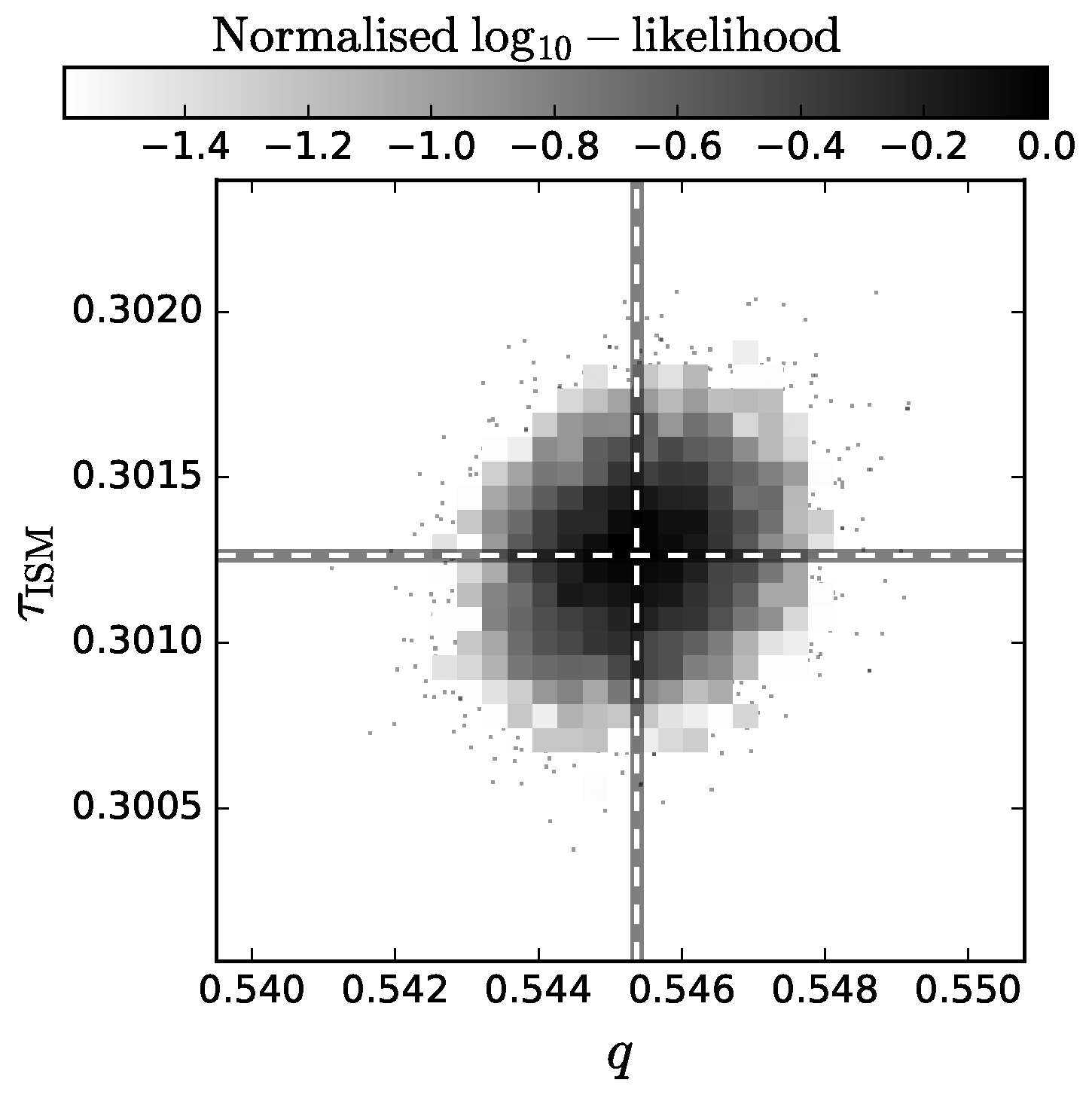}  
\includegraphics[width=0.47\textwidth]{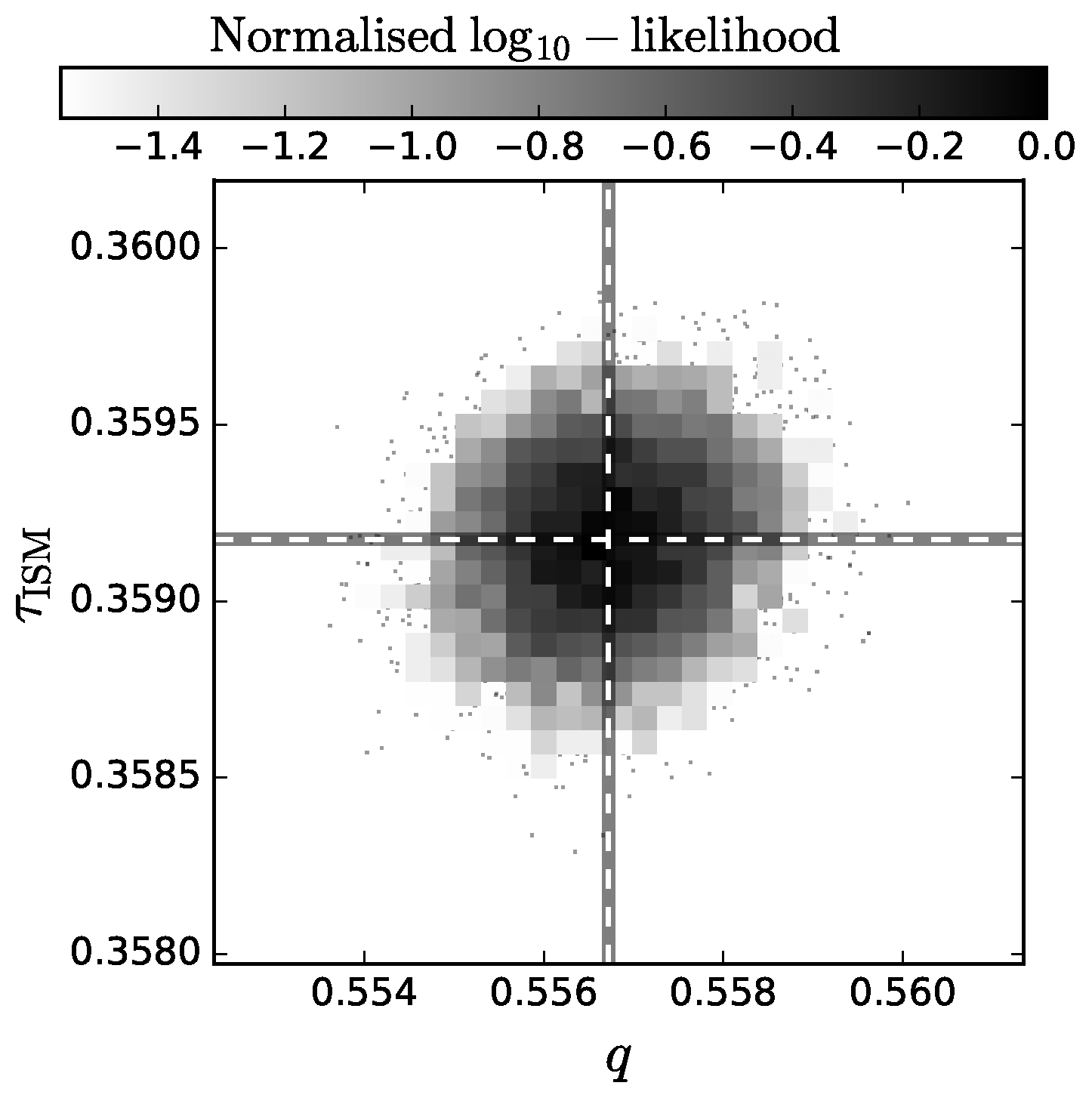}  

 \caption{Shaded maps of likelihood for \ac{T15} $\tau_{\rm ISM}$ and $q$ values fitting the \skirt{} photometry of galxies from the \ac{Ref-100} simulation. The $\tau_{\rm BC}$ parameter is fixed to be $2\tau_{\rm ISM}$. Where the normalised $\log_{10}$-likelihood falls below -1.5, the individual Markov-chain samplings are plotted. The top panel represents the fiducial \skirt{} model, while the bottom panel represents the \lq uncalibrated\rq \skirt{} model discussed in Appendix \ref{sec:uncal}. These are generated by constructing the posterior distribution using an MCMC method, as described in section \ref{sec:fit}. The white dotted lines indicate the median value for each parameter. The ML parameter values are taken from the three dimensional distribution in parameter space, and listed in table \ref{tab:params}.
 }
 \label{fig:contours}
\end{figure}

 With this set-up we recover \ac{ML} values for the three parameters, which are encouragingly very close to the values assumed in \ac{T15}, as discussed in sections \ref{sec:fit} and \ref{sec:conc}. Figure \ref{fig:contours} shows the posterior distribution constructed for the \ac{T15} model parameters given the \skirt{} data, with the \ac{ML} listed in table \ref{tab:params}. We use $1 \times 10^{5}$ samples, employing a burn-in of $1 \times 10^{4}$ samples and a thinning factor of 5. The results of this are presented in Table~\ref{tab:params}, and discussed in section~\ref{sec:fit}. The top panel of Fig. \ref{fig:contours} shows the constructed posterior distribution.

\section{Comparing fiducial and uncalibrated SKIRT models}
\label{sec:uncal}

\begin{figure*}
 \includegraphics[width={0.98\textwidth}]{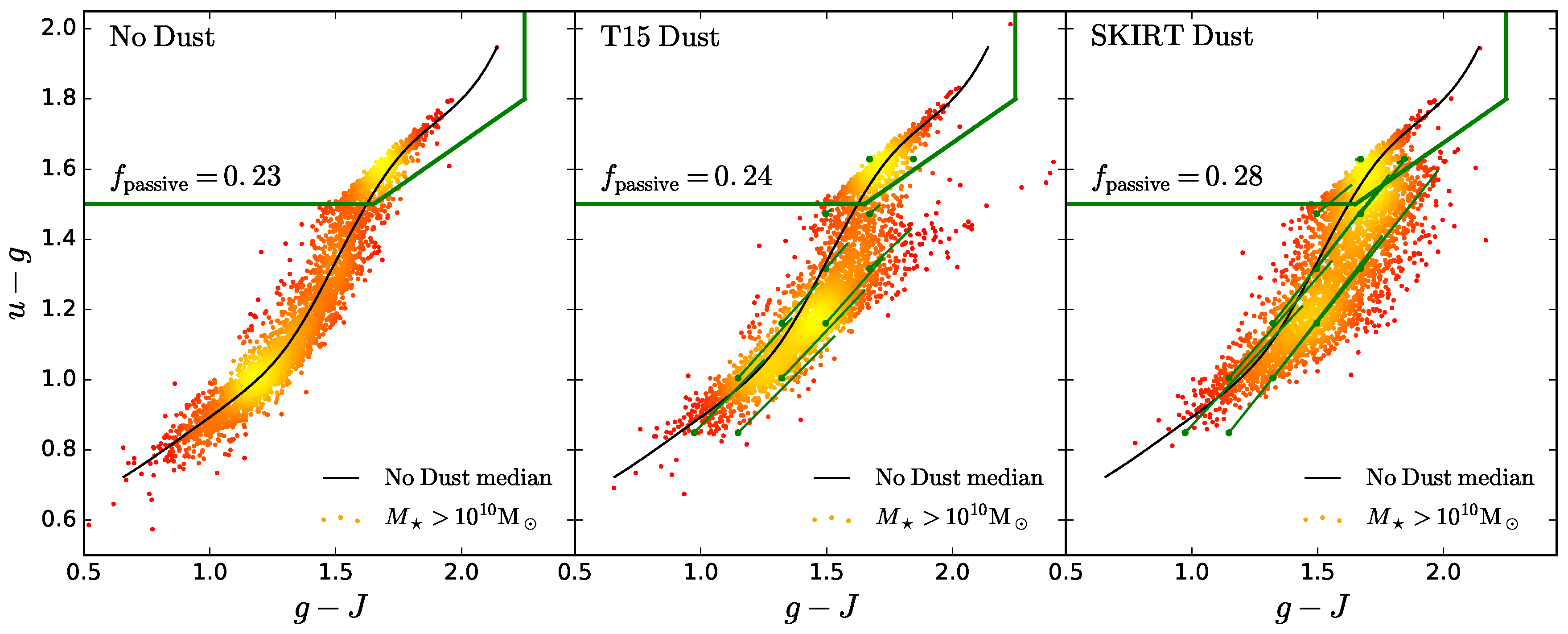}  
 \caption{As Fig. \ref{fig:colcol}, except using the uncalibrated \skirt{} model with values of $f_{\rm dust}=0.4$ and $f_{\rm PDR}=0.2$. We see that this model marginally improves agreement with the observed blue and red fractions for masses $M_\star > 10^{10} M_\odot$.}
 \label{fig:coldists_alt}
\end{figure*}

The $f_{\rm dust}$ and $f_{\rm PDR}$ parameters are the two values in our dust model that are not assigned using the simulation output. These effectively scale the optical depths in the diffuse and birth cloud components respectively. The parameters of our fiducial \skirt{} model were calibrated using local dust scaling relations, indicative of dust mass and temperature, presented in the companion study of \citet{Camps16} (C16). This is desirable as it allows the same model to be consistent with observations over a large spectral range, from optical to FIR wavelenths. FIR measurements also provide a more direct measurement of dust mass than optical attenuation, which depends strongly on star formation histories and the geometries of stellar and ISM distributions. The C16 calibrated values used in our fiducial model are $f_{\rm dust} = 0.3$ and $f_{\rm PDR} = 0.1$. Without FIR constraints, we would default to the best literature values for our assumed dust parameters.

In order to test the effect our choice of dust parameters has on our results, we also generated results using literature values for the parameters of $f_{\rm dust} = 0.4$ \citep{Draine07} and $f_{\rm PDR} = 0.2$ \citep{Groves08}, hereby referred to as the \lq uncalibrated\rq{} \skirt{} model. For comparison we apply the \ac{ML} fitting procedure presented in Appendix \ref{sec:mcmc} to the uncalibrated model and plot the posterior distribution as the bottom panel in Fig. \ref{fig:contours}.

Comparing the posterior distributions of the fiducial and uncalibrated \skirt{} models reveals that the \ac{ML} values for $\tau_{\rm ISM}$ are $\sim 10\%$ below and above their \ac{T15} values respectively. The $q$ values for both models are $\sim 0.4$~dex higher than \ac{T15}, with the uncalibrated model giving a $q$ value $\sim 10\%$ higher than fiducial\footnote{This comparison also highlights incompatibility between the screen and \skirt{} models; the $q$ parameter that inependently describes geometry in \ac{T15} has different \ac{ML} values for two \skirt{} models with identical geometries but re-scaled dust optical depths. However the dependence of $q$ on optical depth is relatively weak. The change in $\tau_{\rm ISM}$ is $\sim 10$ times larger in terms of the marginalised parameter uncertainty.}. Overall, a similar level of agreement with \ac{T15} is achieved for both the uncalibrated and fiducial \skirt{} models.

To exhibit the effects of the calibration on the overall photometry, we also plot $ugJ$ colour-colour distributions in Fig. \ref{fig:coldists_alt}. This is the same plot as Fig. \ref{fig:colcol}, except using the uncalibrated rather than fiducial \skirt{} model. We see that the higher optical depth normalisation has a small effect on the colours, shifting galaxies to marginally redder colour in both indices. It seems that the effects on passive fractions are relatively minor, with only a $\sim 7\%$ change in the passive fractions compared to \ref{fig:colcol}.